\documentclass[english]{article}
\usepackage[latin9]{inputenc}
\usepackage{geometry}
\usepackage{color}
\usepackage{float}
\usepackage{amsmath}
\usepackage{amsthm}
\usepackage{amssymb}
\usepackage{graphicx}

\makeatletter

\providecommand{\tabularnewline}{\\}

  \theoremstyle{definition}
  \newtheorem{defn}{\protect\definitionname}
\theoremstyle{plain}
\newtheorem{thm}{\protect\theoremname}

\@ifundefined{showcaptionsetup}{}{%
 \PassOptionsToPackage{caption=false}{subfig}}
\usepackage{subfig}
\makeatother

\usepackage{babel}
  \providecommand{\definitionname}{Definition}
\providecommand{\theoremname}{Theorem}

\begin{document}

\title{Distributed event-triggered control for \\
 multi-agent formation stabilization\\
and trajectory tracking}

\author{Christophe Viel, Sylvain Bertrand, Helene Piet-Lahanier, Michel Kieffer}

\date{14/05/2018}
\maketitle

\paragraph{\emph{Abstract -} This paper addresses the problem of formation control
and tracking a of desired trajectory by an Euler-Lagrange multi-agent
systems. It is inspired by recent results by Qingkai \emph{et al.}
and adopts an event-triggered control strategy to reduce the number
of communications between agents. For that purpose, to evaluate its
control input, each agent maintains estimators of the states of its
neighbour agents. Communication is triggered when the discrepancy
between the actual state of an agent and the corresponding estimate
reaches some threshold. The impact of additive state perturbations
on the formation control is studied. A condition for the convergence
of the multi-agent system to a stable formation is studied. The time interval between two consecutive communications
by the same agent is shown to be strictly positive. Simulations show the effectiveness of
the proposed approach.}

\paragraph{\emph{Index Terms -} Communication constraints, event-triggered control,
formation stabilization, multi-agent system (MAS).}

\section{Introduction}

Distributed cooperative control of a multi-agent system (MAS) usually
requires significant exchange of information between agents. In early
contributions, see, \emph{e.g.}, \cite{Olfati2007,Ren2008}, communication
was considered permanent. Recently, more practical approaches have
been proposed. For example, in \cite{wen2012,Wen2012_2,Wen2013},
communication is intermittent, alternating phases of permanent communication
and of absence of communication. Alternatively, communication may
only occur at discrete time instants, either periodically as in \cite{Garcia2014J2},
or triggered by some event, as in \cite{dimarogonas2012,fan2013,zhang2015,viel2016}.

This paper proposes a strategy to reduce the number of communications
for displacement-based formation control while following a desired
reference trajectory. Agent dynamics are described by Euler-Lagrange
models and include perturbations. This work extends results presented
in \cite{Qingkai2015} by introducing an event-triggered strategy,
and results of \cite{Qingchen2015,sun2015,tang2011} by addressing
systems with more complex dynamics than a simple integrator. To obtain
efficient distributed control laws, each agent uses an estimator of
the state of the other agents. The proposed distributed communication
triggering condition (CTC) involves the inter-agent displacements
and the relative discrepancy between actual and estimated agent states.
A single \emph{a priori} trajectory has to be evaluated to follow
the desired path. Effect of state perturbations on the formation and
on the communications are analyzed. Conditions for the Lyapunov stability
of the MAS have been introduced. The time interval between two consecutive communications
by the same agent is shown to be strictly positive.

This paper is organized as follows. Related work is detailed in Section~\ref{sec:RelWork}.
Some assumptions are introduced in Section~\ref{subsec:Problem-statement}
and the formation parametrization is described in Section~\ref{sec:Formation problem}.
As the problem considered here is to drive a formation of agents along
a desired reference trajectory, the designed distributed control law
consists of two parts. The first part (see Section~\ref{sec:Formation problem})
drives the agents to some target formation and maintains the formation,
despite the presence of perturbations. It is based on estimates of
the states of the agents described in Section~\ref{subsec:Estimator-dynamic-and}.
The second part (see Section~\ref{sec:Tracking problem}) is dedicated
to the tracking of the desired trajectory. Communication instants
are chosen locally by each agent using an event-triggered approach
introduced in Section~\ref{sec:Event-triggered-communication}. A
simulation example is considered in Section~\ref{sec:Example} to
illustrate the reduction of the communications obtained by the proposed
approach. Finally, conclusions are drawn in Section~\ref{sec:Conclusion}.

\section{Related work}

\label{sec:RelWork}

Event-triggered communication is a promising approach to save energy.
It is well-suited to applications where communications should be minimized,
\emph{e.g.}, to improve furtivity, reduce energy consumption, or limit
collisions between transmitted data packets. Application examples
with such constraints are exposed in \cite{linsenmayer2014,linsenmayer2015}
for the case of a fleet of vehicles, or in \cite{aragues2011} where
agents aim at merging local feature-based maps. The main difficulty
consists in determining the CTC that will ensure the completion of
the task assigned to the MAS, \emph{e.g.}, reaching some consensus,
maintaining a formation, \emph{etc}. In a distributed strategy, the
states of the other agents are not permanently available, thus each
agent usually maintains estimators of the state of its neighbours
to evaluate their control laws. Nevertheless, without permanent communication,
the quality of the state estimates is difficult to evaluate. To address
this issue, each agent maintains an estimate of its own state using
only the information it has shared with its neighbours. When the discrepancy
between this own state estimate and its actual state reaches some
threshold, the agent triggers a communication. This is the approach
considered, \emph{e.g.}, in \cite{zhu2014,garcia2014J,seyboth2013,garcia2015,viel2016,dimarogonas2012,viel2017}.
These works differ by the complexity of the agents' dynamics \cite{zhu2014,garcia2014J,seyboth2013},
the structure of the state estimator \cite{dimarogonas2012,garcia2015,viel2016,viel2017},
and the determination of the threshold for the CTC \cite{seyboth2013,viel2017}.

Most of the event-triggered approaches have been applied in the context
of consensus in MAS \cite{dimarogonas2012,seyboth2013,garcia2015}.
This paper focuses on distributed formation control, which has been
considered in \cite{Qingchen2015,sun2015,tang2011}. Formation control
consists in driving and maintaining all agents of a MAS to some reference,
possibly time-varying configuration, defining, $e.g.$, their relative
positions, orientations, and speeds. Various approaches have been
considered, such as behavior-based flocking \cite{Reynolds1999,Vicsek1995,olfati2006,rochefort2011,moreno2014},
or formation tracking \cite{do2008,chao2012,arboleda2013,mei2011,ren2004}. 

Behavior-based flocking \cite{Reynolds1999,Vicsek1995,olfati2006,rochefort2011,moreno2014}
imposes several behavior rules (attraction, repulsion, imitation)
to each agent. Their combination leads the MAS to follow some desired
behavior. Such approach requires the availability to each agent of
observations of the state of its neighbours. These observations may
be deduced from measurements provided by sensors embedded in each
agent or from information communicated by its neighbours. In all cases,
these observations are assumed permanently available. In addition,
if a satisfying global behavior may be obtained by the MAS, behavior-based
flocking cannot impose a precise configuration between agents.

Different formation-tracking methods have been considered. In leader-follower
techniques \cite{do2008,chao2012,arboleda2013,mei2011}, based on
mission goals, a trajectory is designed only for some leader agent.
The other follower agents, aim at tracking the leader as well as maintaining
some target formation defined with respect to the leader. A virtual
leader has been considered in \cite{chao2001,chao2012,Yohan} to gain
robustness to leader failure. This requires a good synchronization
among agents of the state of the virtual leader. Virtual structures
have been introduced in \cite{ren2004,wang2014}, where the agent
control is designed to satisfy constraints between neighbours. Such
approaches also address the problem of leader failure. In distance-based
control, the constraints are distances between agents. In displacement-based
control, relative coordinate or speed vectors between agents are imposed.
In tensegrity structures \cite{Qingkai2015,nabet2009} additional
flexibility in the structure is considered by considering attraction
and repulsion terms between agents, as formalized by \cite{alfakih2013}.
In addition to constraints on the structure of the MAS, \cite{sun2009}
imposes some reference trajectory to each agent. In most of these
works, permanent communication between agents is assumed. 

Some recent works combine event-triggered approaches with distance-based
or displacement-based formation control \cite{Qingchen2015,sun2015,tang2011}.
In these works, the dynamics of the agents are described by a simple
integrator, with control input considered constant between two communications.
The proposed CTCs consider different threshold formulations and require
each agent to have access to the state of all other agents. A constant
threshold is considered in \cite{sun2015}. A time-varying threshold
is introduced in \cite{Qingchen2015,tang2011}. The CTC depends then
on the relative positions between agents and the relative discrepancy
between actual and estimated agent states. These CTCs reduce the number
of triggered communications when the system converges to the desired
formation. A minimal time between two communications, named inter-event
time, is also defined. Finally, in all these works, no perturbations
are considered.

LBC techniques have been introduced in \cite{rego2013,xu2006,aguiar2007,yook2002}
to reduce the number of communications in trajectory tracking problems.
MAS with decoupled nonlinear agent dynamics are considered in \cite{rego2013,aguiar2007}.
Agents have to follow parametrized paths, designed in a centralized
way. CTCs introduced by LBC lead all agents to follow the paths in
a synchronized way to set up a desired formation. Communication delays,
as well as packet losses are considered. Nevertheless, if input-to-state
stability conditions are established, absence of Zeno behavior is
not analyzed.

\section{Notations and hypotheses}

\label{subsec:Problem-statement}

Table~\ref{tab:Main-notations-tracking} summarizes the main notations
used in this paper.

Consider a MAS consisting of a network of $N$ agents which topology
is described by an undirected graph $\mathcal{G}=(\mathcal{\mathcal{N}},\mathcal{E})$.
$\mathcal{\mathcal{N}}=\left\{ 1,2,...,N\right\} $ is the set of
nodes and $\mathcal{E}\subset\mathcal{\mathcal{N}}\times\mathcal{\mathcal{N}}$
the set of edges of the network. The set of neighbours of Agent~$i$
is $\mathcal{N}_{i}=\{j\in\mathcal{\mathcal{N}}|\left(i,j\right)\in\mathcal{E},\:i\neq j\}$.
$N_{i}$ is the cardinal number of $\mathcal{N}_{i}$. For some vector
$x=\left[\begin{array}{cccc}
x_{1} & x_{2} & \ldots & x_{n}\end{array}\right]^{T}\in\mathbb{R}^{n}$, we define $\left|x\right|=\left[\begin{array}{cccc}
\left|x_{1}\right| & \left|x_{2}\right| & \ldots & \left|x_{n}\right|\end{array}\right]^{T}$ where $\left|x_{i}\right|$ is the absolute value of the $i$-th
component of $x$. Similarly, the notation $x\geq0$ will be used
to indicate that each component $x_{i}$ of $x$ is non negative,
$i.e.$, $x_{i}\geq0$ $\forall i\in\left\{ 1\ldots n\right\} $.
A continuous function $\beta\left(r,\,s\right)\,:\,\left[0,\,a\right)\times\left[0,\,\infty\right)\rightarrow\left[0,\,\infty\right)$ is said to belong to class\textbf{ $\mathcal{KL}$} if for each fixed $s$, the function $\beta\left(.,\,s\right)$ is strictly increasing and $\beta\left(0,\,s\right)=0$, and for each fixed $r$, the function $\beta\left(r,\,.\right)$ is decreasing and $\lim_{s\rightarrow\infty}\beta\left(r,\,s\right)=0$.
{\color{black}
A continuous function $\Omega:\left[0,\,a\right)\rightarrow\left[0,\,\infty\right)$ is said to belong class $\mathcal{K}$ if it is strictly increasing and $\Omega\left(0\right)=0$.
 A continuous function $\Phi:\left[0,\,a\right)\rightarrow\left[0,\,\infty\right)$ is said to belong class $\mathcal{K}_{\infty}$ if it belongs to class $\mathcal{K}$, with $a=\infty $ and $\lim_{r\rightarrow\infty}\Phi\left(r\right)=\infty$. }

\begin{table}
\centering
\begin{tabular}{|c|l|}
\hline 
$q_{i}$ & vector of \emph{coordinates} of Agent~$i$ in some global fixed reference
frame $\mathcal{R}$\tabularnewline
\hline 
$q$ & vector $\left[\begin{array}{cccc}
q_{1}^{T} & q_{2}^{T} & \ldots & q_{N}^{T}\end{array}\right]^{T}\in\mathbb{R}^{N.n}$, \textit{configuration }of the MAS\tabularnewline
\hline 
$x_{i}$  & state vector $\left[q_{i}^{T},\dot{q}_{i}^{T}\right]^{T}$ of Agent~$i$ \tabularnewline
\hline 
$\hat{q}_{i}^{j}$  & estimate of $q_{i}$ performed by Agent~$j$.\tabularnewline
\hline 
$\hat{q}^{j}$  & estimate of $q$ performed by Agent~$j$.\tabularnewline
\hline 
$\hat{x}_{i}^{j}$  & estimate of $x_{i}$ performed by Agent~$j$.\tabularnewline
\hline 
$e_{i}^{j}$  & estimation error between $q_{i}$ and $\hat{q}_{i}^{j}$. \tabularnewline
\hline 
$r_{ij}$ & relative coordinate vector $r_{ij}=q_{i}-q_{j}$ between agents $i$
and $j$.\tabularnewline
\hline 
$r_{ij}^{*}$ & desired value for $r_{ij}$.\tabularnewline
\hline 
$q_{0}$ & reference trajectory\tabularnewline
\hline 
$q_{i}^{*}$ & reference trajectory for Agent~$i$, $q_{i}^{*}=q_{0}+r_{i1}^{*}$ \tabularnewline
\hline 
$\varepsilon_{i}$ & trajectory error for Agent~$i$, $\varepsilon_{i}=q_{i}-q_{i}^{*}$\tabularnewline
\hline 
$t_{j,k}$  & time at which the $k$-th message is sent by Agent $j$.\tabularnewline
\hline 
$t_{j,k}^{i}$  & time at which the $k$-th message sent by Agent~$j$ is received
by Agent~$i$. \tabularnewline
\hline 
\end{tabular}\caption{Main notations\label{tab:Main-notations-tracking}}
\end{table}

Let $q_{i}\in\mathbb{R}^{n}$ be the vector of \emph{coordinates}
of Agent~$i$ in some global fixed reference frame $\mathcal{R}$
and let $q=\left[\begin{array}{cccc}
q_{1}^{T} & q_{2}^{T} & \ldots & q_{N}^{T}\end{array}\right]^{T}\in\mathbb{R}^{N.n}$ be the \textit{configuration }of the MAS. The dynamics of each agent
is described by the Euler-Lagrange model
\begin{equation}
M_{i}\left(q_{i}\right)\ddot{q}_{i}+C_{i}\left(q_{i},\,\dot{q}_{i}\right)\dot{q}_{i}+G=\tau_{i}+d_{i},\label{eq: System dynamique Lagrange}
\end{equation}
where $\tau_{i}\in\mathbb{R}^{n}$ is some control input described
in Section~\ref{subsec:Command-control}, $M_{i}\left(q_{i}\right)\in\mathbb{R}^{n\times n}$
is the inertia matrix of Agent~$i$, $C_{i}\left(q_{i},\dot{q}_{i}\right)\in\mathbb{R}^{n\times n}$
is the matrix of the Coriolis and centripetal term on Agent~$i$,
$G$ accounts for gravitational acceleration supposed to be known
and constant, and $d_{i}$ is a time-varying state perturbation satisfying
$\left\Vert d_{i}\left(t\right)\right\Vert \leq D_{\max}$. The state vector of Agent~$i$ is $x_{i}^{T}=\left[q_{i}^{T},\dot{q}_{i}^{T}\right]$.
Assume that the dynamics satisfy the following assumptions, \textcolor{black}{where
Assumptions A1, A2 and A3 have been previously considered, }\textcolor{black}{\emph{e.g.}}\textcolor{black}{,
in \cite{mei2011,liu2016,makkar2007}:}
\begin{description}
\item [{A1)}] $M_{i}\left(q_{i}\right)$ is symmetric positive and there
exists $k_{M}>0$ satisfying $\forall x$, $x^{T}M_{i}\left(q_{i}\right)x$ $\leq k_{M}x^{T}x$.
\item [{A2)}] $\dot{M}_{i}\left(q_{i}\right)-2C_{i}\left(q_{i},\dot{q}_{i}\right)$
is skew symmetric or negative definite and there exists $k_{C}>0$
satisfying $\forall x$, $x^{T}C_{i}\left(q_{i},\dot{q}_{i}\right)x\leq k_{C}\left\Vert \dot{q}_{i}\right\Vert x^{T}x$ and $\lambda_{\max}\left(C_{i}\left(q_{i},\dot{q}_{i}\right)\right)\leq k_{C}\left\Vert \dot{q}_{i}\right\Vert $.
\item [{A3)}] The left-hand side of (\ref{eq: System dynamique Lagrange})
can be linearly parametrized as 
\begin{equation}
M_{i}\left(q_{i}\right)x_{1}+C_{i}\left(q_{i},\,\dot{q}_{i}\right)x_{2}=Y_{i}\left(q_{i},\,\dot{q}_{i},\,x_{1},\,x_{2}\right)\theta_{i}\label{eq:Prop. Yi}
\end{equation}
for all vectors $x_{1},\,x_{2}\in\mathbb{R}^{n}$, where $Y_{i}\left(q_{i},\,\dot{q}_{i},\,x_{1},\,x_{2}\right)$
is a regressor matrix with known structure and $\theta_{i}$ is a
vector of unknown but constant parameters associated with the $i$-th
agent.
\end{description}
Moreover, one assumes that
\begin{description}
\item [{A4)}] For each $,i=1,\dots,N,$ $\theta_{i}$ is such that $\theta_{\min,i}<\theta_{i}<\theta_{\max,i}$,
with known $\theta_{\min,i}$ and $\theta_{\max,i}$.
\item [{A5)}] Each Agent~$i$ is able to measure without error its own
state $x_{i}$, 
\item [{A6)}] There is no packet losses or communication delay between
agents.
\end{description}
In what follows, the notations $M_{i}$ and $C_{i}$ are used to replace
$M_{i}\left(q_{i}\right)$ and $C_{i}\left(q_{i},\dot{q}_{i}\right)$.

\section{Formation control problem}

\label{sec:Formation problem}

This section aims at designing a decentralized control strategy to
drive a MAS to a desired target formation in some global reference
frame $\mathcal{R}$, while reducing as much as possible the communications
between agents. The target formation is first described in Section~\ref{subsec:Formation-parametrization}.
The potential energy of a MAS with respect to the target formation
is introduced to quantify the discrepancy between the target and current
formations. The proposed distributed control, introduced in Section~\ref{subsec:Command-control},
tries to minimize the potential energy. To evaluate the control input
of each agent despite the communications at discrete time instants
only, estimators of the coordinate vectors of all agents are managed
by each agent, as presented in Section~\ref{subsec:Estimator-dynamic-and}.
The presence of perturbations increases the discrepancy between the\textcolor{blue}{{}
}state vector and their estimates. A CTC is designed to limit this
discrepancy by updating the estimators as described in Section~\ref{sec:Event-triggered-communication}. 

\subsection{Formation parametrization}

\label{subsec:Formation-parametrization}

Consider the relative coordinate vector $r_{ij}=q_{i}-q_{j}$ between
two agents $i$ and $j$ and the target relative coordinate vector
$r_{ij}^{*}$ for all $\left(i,\,j\right)\in\mathcal{N}$. A target
formation is defined by the set $\left\{ r_{ij}^{*},\left(i,\,j\right)\in\mathcal{N}\right\} $.
Consider, without loss of generality, the first agent as a reference
agent and introduce the target relative configuration vector $r^{*}=\left[\begin{array}{ccc}
r_{11}^{*T} & \ldots & r_{1N}^{*T}\end{array}\right]^{T}$. Any target relative configuration vector $r_{ij}^{*}$ can be expressed
as $r_{ij}^{*}=r_{1i}^{*}-r_{1j}^{*}$. 

The \emph{potential energy} $P\left(q,\,t\right)$ of the formation,
introduced for tensegrety formations in \cite{nabet2009,Qingkai2015},
represents the disagreement between $r_{ij}$ and $r_{ij}^{*}$ 
\begin{equation}
P\left(q,\,t\right)=\frac{1}{2}\sum_{i=1}^{N}\sum_{j=1}^{N}k_{ij}\left\Vert r_{ij}-r_{ij}^{*}\right\Vert ^{2}\label{eq:PotEnergy}
\end{equation}
where the $k_{ij}=k_{ji}$ are some spring coefficients, which can be be positive or null. The values of the $k_{ij}$s that make a given $r^{*}$ an equilibrium formation may be chosen using the method developed in \cite{Qingkai2015}. Moreover, we take $k_{ii}=0$ \textcolor{black}{and $k_{ij}=0$ if $\mathcal{E}_{ij}=0$, \emph{i.e.}, if $i$ and $j$ are not neighbors. Since $\mathcal{G}$ is connected,
the minimum number of non-zero coefficients $k_{ij}$ to properly
define a target formation is $N-1$.}  A number of non-zero $k_{ij}$
larger than $N-1$ introduces robustness in the formation, in particular
with respect to the loss of an agent. 
\begin{defn}
\label{Def-The-potential-energy}\cite{Qingkai2015} The
MAS asymptotically converges to the target formation with a bounded
error iff there exists some $\varepsilon_{1}>0$ such as 
\begin{equation}
\lim_{t\rightarrow\infty}P\left(q,\,t\right)\leqslant\varepsilon_{1}.\label{eq:Def Bounded convergence}
\end{equation}
\end{defn}
A control law designed to reduce the potential energy $P\left(q,\,t\right)$
allows a bounded convergence of the MAS. To describe the evolution
of $P\left(q,\,t\right)$, one introduces as in \cite{Qingkai2015}
\begin{eqnarray}
g_{i} & = & \frac{\partial P\left(q,\,t\right)}{\partial q_{i}}=\sum_{j=1}^{N}k_{ij}\left(r_{ij}-r_{ij}^{*}\right)\label{eq:Calcul_gi}\\
\dot{g}_{i} & = & \sum_{j=1}^{N}k_{ij}\left(\dot{r}_{ij}-\dot{r}_{ij}^{*}\right)\label{eq:Calcul_dgi}\\
s_{i} & = & \dot{q}_{i}+k_{p}g_{i}\label{eq:Calcul_si}
\end{eqnarray}
where $g_{i}$ and $\dot{g}_{i}$ characterize the evolution of the
discrepancy between the current and target formations and $k_{p}$
is a positive scalar design parameter.

{\color{black}{ Note that since $k_{ij}=0$ if $j\notin \mathcal{N}_i$, one has $g_{i} = \sum_{j=1}^{N}k_{ij}\left(r_{ij}-r_{ij}^{*}\right) = \sum_{j \in \mathcal{N}_i}k_{ij}\left(r_{ij}-r_{ij}^{*}\right)$. Consequently, Agent~$i$ can evalutate $g_i$ and $s_i$ using only information from its neighbors. }}

\subsection{Distributed control}
\label{subsec:Command-control}

The control law proposed in \cite{Qingkai2015} is defined as $\tau_{i}=\tau_{i}(q_{i},\dot{q}_{i},q)$
and aims at reducing $P\left(q,\,t\right)$, thus making the MAS converge
to the target formation in case of permanent communication. In this
approach, each agent evaluates its control input using the state vectors
of its neighbours obtained via permanent communication. Here, in a
distributed context with limited communications between agents, agents
cannot have permanent access to $q$. Thus, one introduces the estimate
$\hat{q}_{j}^{i}$ of $q_{j}$ performed by Agent~$i$ to replace
the missing information in the control law. The MAS configuration
estimated by Agent~$i$ is denoted as $\hat{q}^{i}=\left[\begin{array}{ccc}
\hat{q}_{1}^{iT} & \ldots & \hat{q}_{N}^{iT}\end{array}\right]^{T}\in\mathbb{R}^{N.n}$. The way $\hat{q}_{j}^{i}$ is evaluated is described in Section~\ref{subsec:Estimator-dynamic-and}. 

In a distributed context with limited communications, with the help
of $\hat{q}^{i}$, Agent~$i$ is able to evaluate
\begin{eqnarray}
\bar{g}_{i} & = & \sum_{j=1}^{N}k_{ij}\left(\bar{r}_{ij}-r_{ij}^{*}\right) = \sum_{j \in \mathcal{N}_i}k_{ij}\left(\bar{r}_{ij}-r_{ij}^{*}\right)\label{eq:Calcul_bar(gi)}\\
\bar{s}_{i} & = & \dot{q}_{i}+k_{p}\bar{g}_{i}\label{eq:Calcul_bar(si)}
\end{eqnarray}
with $\bar{r}_{ij}=q_{i}-\hat{q}_{j}^{i}$ and $\dot{\bar{r}}_{ij}=\dot{q}_{i}-\dot{\hat{q}}_{j}^{i}$.
Using $\bar{g}_{i}$ and $\bar{s}_{i}$, Agent~$i$ is able to evaluate
the following adaptive distributed control input to be used in (\ref{eq: System dynamique Lagrange})
\begin{eqnarray}
\tau_{i}\left(q_{i},\,\dot{q}_{i},\,\hat{q}^{i},\,\dot{\hat{q}}^{i}\right) & = & -k_{s}\bar{s}_{i}-k_{g}\bar{g}_{i}+G-Y_{i}\left(q_{i},\,\dot{q}_{i},\,k_{p}\dot{\bar{g}}_{i},\,k_{p}\bar{g}_{i}\right)\bar{\theta}_{i},\label{eq:Control input with estimation-1}\\
\dot{\bar{\theta}}_{i} & = & \Gamma_{i}Y_{i}\left(q_{i},\,\dot{q}_{i},\,k_{p}\dot{\bar{g}}_{i},\,k_{p}\bar{g}_{i}\right)^{T}\bar{s}_{i}\label{eq:Calcul dot(Theta)}
\end{eqnarray}
where $k_{g}>0$, $k_{s}\geq1+k_{p}\left(k_{M}+1\right)$ are design
parameters and $\Gamma_{i}$ is an arbitrary symmetric positive definite
matrix.

Section~\ref{subsec:Estimator-dynamic-and} details the estimator
$\hat{q}_{j}^{i}$ of $q_{j}$ needed in (\ref{eq:Control input with estimation-1}).

\section{Time-varying formation and tracking}

\label{sec:Tracking problem}

In this section, the MAS has to follow some reference trajectory $q_{1}^{*}\left(t\right)$,
while remaining in a desired formation. Agent~$1$, taken as the
reference agent, aims at following $q_{1}^{*}\left(t\right)$. It
is assumed that all agents have access to $q_{1}^{*}\left(t\right)$.
Moreover, assume that the target formation can be time-varying and
is represented by the relative configuration vector $r^{*}\left(t\right)$.
Therefore the reference trajectory of each agent can be expressed
as $q_{i}^{*}\left(t\right)=q_{1}^{*}\left(t\right)+r_{i1}^{*}\left(t\right)$.
\begin{defn}
\textit{\textcolor{black}{\label{Def-tracking}}}The MAS reaches
its tracking objective iff there exists $\varepsilon_{1}>0$ and $\varepsilon_{2}>0$
such that (\ref{eq:Def Bounded convergence}) is satisfied and 
\begin{equation}
\lim_{t\rightarrow\infty}\left\Vert q_{1}\left(t\right)-q_{1}^{*}\left(t\right)\right\Vert \leqslant\varepsilon_{2},\label{eq:Def Bounded convergence-1}
\end{equation}
\emph{i.e.}, iff the reference agent asymptotically converges to the
reference trajectory, and the MAS asymptotically converges to the
target formation with bounded errors.
\end{defn}
A distributed control law is designed to satisfy this target. Introduce
the trajectory error terms 
\begin{eqnarray*}
\varepsilon_{i} & = & q_{i}-q_{i}^{*}\\
\hat{\varepsilon}_{i}^{j} & = & \hat{q}_{i}^{j}-q_{i}^{*}.
\end{eqnarray*}
The terms $g_{i}$, $\bar{g}_{i}$, $\hat{g}_{i}^{j}$, $\bar{s}_{i}$
and $\hat{s}_{i}^{j}$ introduced in Sections~\ref{sec:Formation problem}
are now redefined as follows to address the trajectory tracking problem
\begin{eqnarray}
g_{i} & = & \sum_{j=1}^{N}k_{ij}\left(r_{ij}-r_{ij}^{*}\right)+k_{0}\varepsilon_{i}\label{eq:gi}\\
\bar{g}_{i} & = & \sum_{j=1}^{N}k_{ij}\left(\bar{r}_{ij}-r_{ij}^{*}\right)+k_{0}\varepsilon_{i}\label{eq:gibar}\\
\hat{g}_{i}^{j} & = & \sum_{j=1}^{N}k_{ij}\left(\hat{r}_{ij}^{j}-r_{ij}^{*}\right)+k_{0}\hat{\varepsilon}_{i}^{j}\label{eq:gihat}\\
s_{i} & = & \dot{q}_{i}-\dot{q}_{i}^{*}+k_{p}g_{i}\label{eq:Si}\\
\bar{s}_{i} & = & \dot{q}_{i}-\dot{q}_{i}^{*}+k_{p}\bar{g}_{i}\label{eq:Sibar}\\
\hat{s}_{i}^{j} & = & \dot{\hat{q}}_{i}^{j}-\dot{q}_{i}^{*}+k_{p}\hat{g}_{i}^{j}\label{eq:Sihat}
\end{eqnarray}
where $k_{0}\geq0$ is a positive design parameter which may be used
to control the tracking error with respect to the reference trajectory.
When no reference trajectory is considered, $k_{0}=0$. 

From these terms, a new distributed control input to be used in (\ref{eq: System dynamique Lagrange})
is defined for Agent~$i$ as 
\begin{eqnarray}
\tau_{i} & = & -k_{s}\bar{s}_{i}-k_{g}\bar{g}_{i}+G-Y_{i}\left(q_{i},\,\dot{q}_{i},\,\dot{\bar{p}}_{i},\,\bar{p}_{i}\right)\bar{\theta}_{i}\label{eq:Control input with tracking}\\
\dot{\bar{\theta}}_{i} & = & \Gamma_{i}Y_{i}\left(q_{i},\,\dot{q}_{i},\,\dot{\bar{p}}_{i},\,\bar{p}_{i}\right)^{T}\bar{s}_{i}\label{eq:Calcul dot(Theta)-1}
\end{eqnarray}
where $\bar{p}_{i}=k_{p}\bar{g}_{i}-\dot{q}_{i}^{*}$ and $\dot{\bar{p}}_{i}=k_{p}\dot{\bar{g}}_{i}-\ddot{q}_{i}^{*}$. 

\subsection{\textcolor{black}{Communication protocol and estimator dynamics }}

\label{subsec:Estimator-dynamic-and}

\subsubsection{Communication protocol\textcolor{black}{\label{subsec:Communication-protocol}}}

In what follows, the time instant at which the $k$-th message is
sent by Agent $i$ is denoted $t_{i,k}$. Let $t_{i,k}^{j}$ be the
time at which the $k$-th message sent by Agent~$i$ is received
by Agent~$j$. \textcolor{black}{According to Assumption A6, $t_{j,k}^{i}=t_{j,k}$ for
all $i\in\mathcal{N}_{j}$.} When a communication is triggered at
$t_{i,k}$ by Agent~$i$, it transmits a message containing $t_{i,k}$,
$q_{i}\left(t_{i,k}\right)$, $\dot{q}_{i}\left(t_{i,k}\right)$ and
$\bar{\theta}_{i}\left(t_{i,k}\right)$. Upon reception of this message,
the neighbours of Agent~$i$ update their estimate of the state of
Agent~$i$ using this information.

\begin{figure}[h]
\begin{centering}
\begin{tabular}{c}
\includegraphics[bb=19bp 22bp 477bp 205bp,scale=0.6]{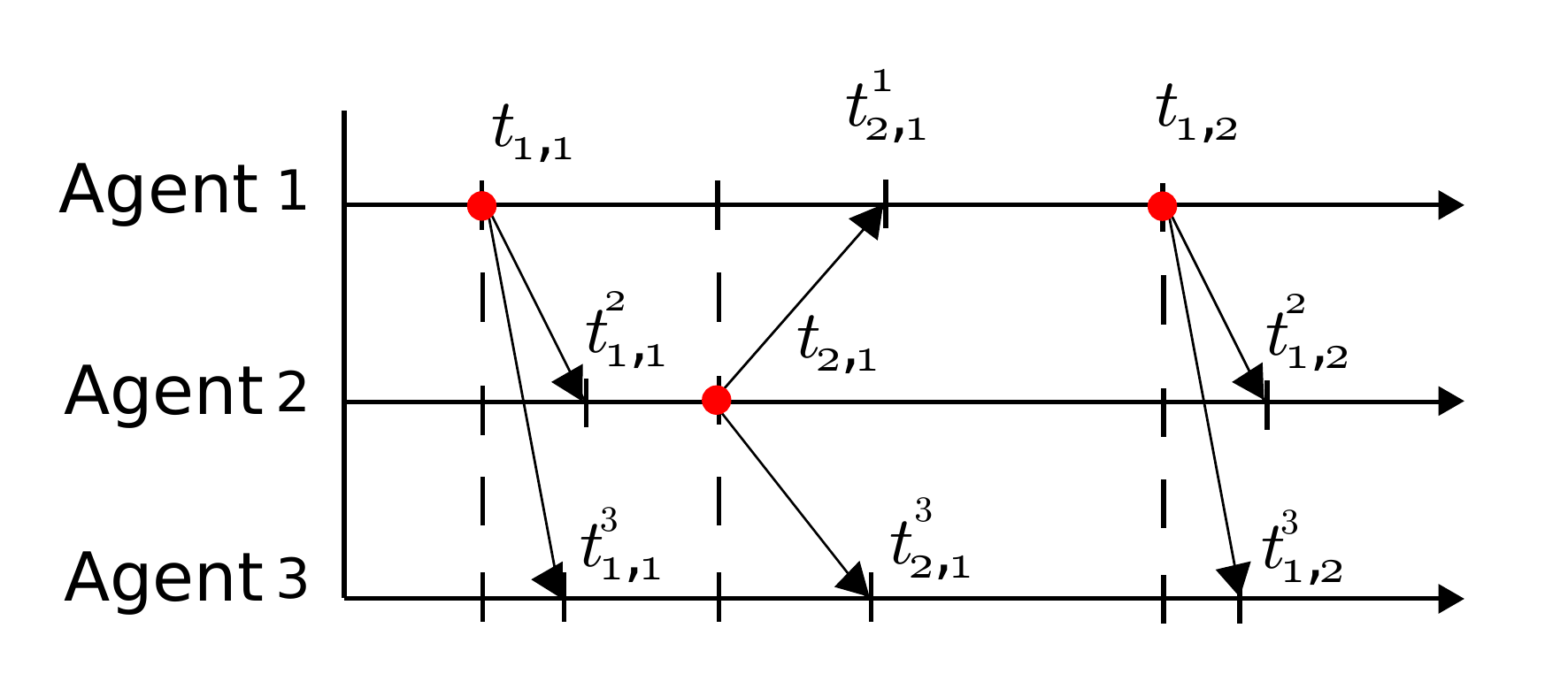}\tabularnewline
\end{tabular}
\par\end{centering}
\caption{Example of transmission times $t_{i,k}$ by Agent~$i$ of $k$-th
message and reception times $t_{i,k}^{j}$ of $k$-th message by Agent~$j$. }
\end{figure}

\subsubsection{Estimator dynamics}

\textcolor{black}{Agent~$i$ evaluates the estimate $\hat{q}_{j}^{i}$
of $q_{j}$ for all its neighbors $j\in\mathcal{N}_{i}$} as
\begin{eqnarray}
\hat{M}_{j}^{i}\left(\hat{q}_{j}^{i}\right)\ddot{\hat{q}}_{j}^{i}+\hat{C}_{j}^{i}\left(\hat{q}_{j}^{i},\,\dot{\hat{q}}_{j}^{i}\right)\dot{\hat{q}}_{j}^{i}+G & = & \hat{\tau}_{j}^{i},\mbox{ }\forall t\in\left[t_{j,k}^{i},\,t_{j,k+1}^{i}\right[\label{eq:estimation des coordonnes des voisins- avance}\\
\hat{q}_{j}^{i}\left(t_{j,k}^{i}\right) & = & q_{j}\left(t_{j,k}^{i}\right)\label{eq: Init_hat(q)_ij}\\
\dot{\hat{q}}_{j}^{i}\left(t_{j,k}^{i}\right) & = & \dot{q}_{j}\left(t_{j,k}^{i}\right),\label{eq: Init_hat(dq)_ij}
\end{eqnarray}
where $\hat{M}_{j}^{i}\left(\hat{q}_{j}^{i}\right)$ and $\hat{C}_{j}^{i}\left(\hat{q}_{j}^{i},\,\dot{\hat{q}}_{j}^{i}\right)$
are estimates of $M_{j}$ and $C_{j}$ computed from $Y_{j}\left(\hat{q}_{j}^{i},\,\dot{\hat{q}}_{j}^{i},\,x,\,y\right)$
and $\bar{\theta}_{j}\left(t_{j,k}^{i}\right)$ using
\begin{equation}
\hat{M}_{j}^{i}\left(\hat{q}_{j}^{i}\right)x+\hat{C}_{j}^{i}\left(\hat{q}_{j}^{i},\,\dot{\hat{q}}_{j}^{i}\right)y=Y_{j}\left(\hat{q}_{j}^{i},\,\dot{\hat{q}}_{j}^{i},\,x,\,y\right)\bar{\theta}_{j}\left(t_{j,k}^{i}\right).\label{eq:ModeleLineaire}
\end{equation}
The estimator (\ref{eq:estimation des coordonnes des voisins- avance})
managed by Agent~$i$ requires an estimate $\hat{\tau}_{j}^{i}$
of the control input $\tau_{j}$ evaluated by Agent~$j$. This estimate, used by Agent~$i$,
is evaluated as\textcolor{black}{
\begin{eqnarray}
\hat{\tau}_{j}^{i} & = & -k_{s}\left(\dot{\hat{\varepsilon}}_{j}^{i}+k_{p}k_{0}\hat{\varepsilon}_{j}^{i}\right)-k_{g}k_{0}\hat{\varepsilon}_{j}^{i}+G-Y_{j}\left(\hat{q}_{j}^{i},\,\dot{\hat{q}}_{j}^{i},\,\dot{\hat{m}}_{j}^{i},\,\hat{m}_{j}^{i}\right)\hat{\theta}_{j}^{i}\label{eq:Accurate estim. command-1}\\
\dot{\hat{\theta}}_{j}^{i} & = & \Gamma_{j}Y_{j}\left(\hat{q}_{j}^{i},\,\dot{\hat{q}}_{j}^{i},\,\dot{\hat{m}}_{j}^{i},\,\hat{m}_{j}^{i}\right)^{T}\left(\dot{\hat{\varepsilon}}_{j}^{i}+k_{p}k_{0}\hat{\varepsilon}_{j}^{i}\right)\\
\hat{\theta}_{j}^{i}\left(t_{j,k}^{i}\right) & = & \bar{\theta}_{j}\left(t_{j,k}^{i}\right)\label{eq:Mise =0000E0 jour theta}
\end{eqnarray}
where $\hat{\theta}_{j}^{i}$ is the estimate of $\bar{\theta}_{j}$,}
\textcolor{black}{$\hat{\varepsilon}_{j}^{i}=\hat{q}_{j}^{i}-q_{j}^{*}$, and
$\hat{m}_{j}^{i}=k_{p}k_{0}\hat{\varepsilon}_{j}^{i}-\dot{q}_{j}^{*}$ if $k_{0}>0$,
$i.e.$, in the case of a reference trajectory to be tracked and $\hat{m}_{j}^{i}=0$
else. Note that if $k_{0}=0$, $\dot{q}_{j}^{*}=0$. The estimator~\eqref{eq:estimation des coordonnes des voisins- avance}-\eqref{eq: Init_hat(dq)_ij} only requires that Agent~$i$ receives messages from Agent~$j$ to evaluate $\hat{q}_{j}^{i}$ and \eqref{eq:Accurate estim. command-1}-\eqref{eq:Mise =0000E0 jour theta}.}

\begin{figure}[h]
\begin{centering}
\includegraphics[bb=2cm 17cm 18cm 842bp,clip,scale=0.5]{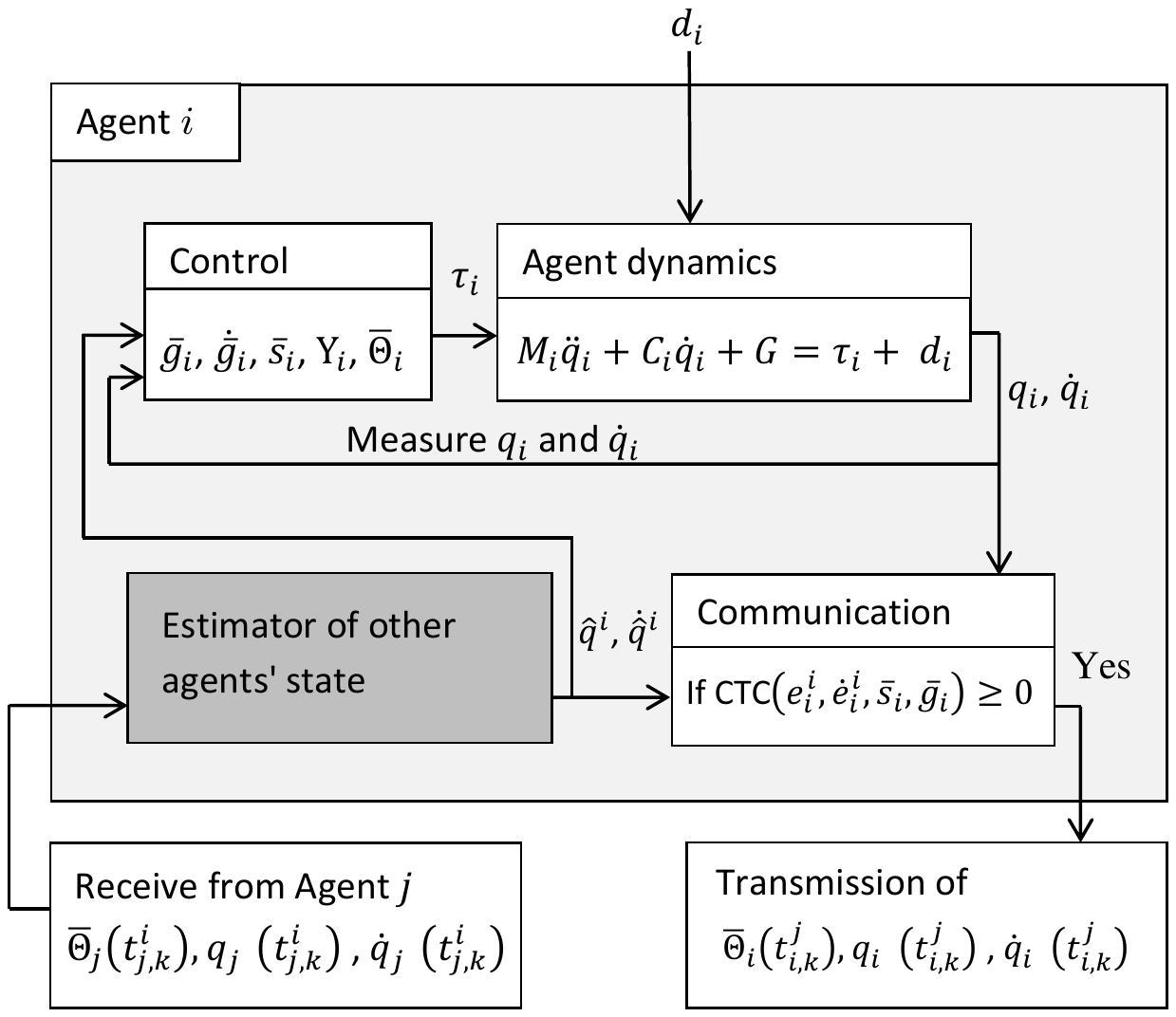}
\par\end{centering}
\caption{Formation control system architecture\label{fig:Formation-control-system}}
\end{figure}

Errors appear between $q_{i}$ and its estimate $\hat{q}_{i}^{j}$
obtained by any other Agent~$j\in\mathcal{N}_i$ due to the presence of state perturbations,
the non-permanent communication, and the mismatch between $\theta_{i}$,
$\bar{\ensuremath{\theta}}_{i}$, and $\hat{\theta}_{i}$. The errors
for the estimates performed by Agent~$j$ are expressed as
\begin{eqnarray}
e_{i}^{j} & = & \hat{q}_{i}^{j}-q_{i},\mbox{ }j\in\mathcal{N}\label{eq:EstimError}\\
e^{j} & = & \hat{q}^{j}-q.\label{eq:EstimError_Vect}
\end{eqnarray}
These errors are used in Section~\ref{sec:Event-triggered-communication}
to trigger communications when $e_{i}^{i}$ and $\dot{e}_{i}^{i}$
become too large. Figure~\ref{fig:Formation-control-system} summarizes
the overall structure of the estimator and controller. 

\textcolor{black}{Using Assumption A6 and considering the structure of the estimator~(\ref{eq:estimation des coordonnes des voisins- avance})-(\ref{eq: Init_hat(dq)_ij}),
one has $\hat{q}_{i}^{i}\left(t\right)=\hat{q}_{i}^{j}\left(t\right)$
for all $\forall i\in\mathcal{N}$ and $j\in\mathcal{N}_{i}$. This
simplifies the stability analysis in Appendix~\ref{subsec:Proof-of-convergence}.}

\section{Event-triggered communications}

\label{sec:Event-triggered-communication}

Theorem~\ref{Th event-triggered-part4} introduces a CTC used to
trigger communications to ensure a bounded asymptotic convergence
of the MAS to the target formation. Each agent knows the initial values
of the state of its neighbors. In practice, this condition can be satisfied
by triggering a communication at time $t=0$.

Let $k_{\max}=\max_{\begin{array}{c}
\ell=1\ldots N\\
j=1\ldots N
\end{array}}\left(k_{\ell j}\right)$ and $k_{\min}=\min_{\begin{array}{c}
\ell=1\ldots N\\
j=1\ldots N
\end{array}}\left(k_{\ell j}\neq0\right)$ , $\alpha_{i}=\sum_{j=1}^{N}k_{ij}$, $\alpha_{\min}=\min_{i=1,\dots,N}\alpha_{i}$
and $\alpha_{\text{M}}=\max_{i=1,\dots,N}\alpha_{i}$. {\color{black}Using Assumption~A4,} define also for
$\bar{\theta}_{i}\in\mathbb{R}^{p}$ and $\bar{\theta}_{i}=\left[\bar{\theta}_{i,1},\,\ldots,\bar{\theta}_{i,p}\right]^{T}$
\begin{equation}
\Delta\theta_{i,\max}=\left[\begin{array}{c}
\max\left\{ \left|\bar{\theta}_{i,1}-\theta_{\min,i,1}\right|,\left|\bar{\theta}_{i,1}-\theta_{\max,i,1}\right|\right\} \\
\vdots\\
\max\left\{ \left|\bar{\theta}_{i,p}-\theta_{\min,i,p}\right|,\left|\bar{\theta}_{i,p}-\theta_{\max,i,p}\right|\right\} 
\end{array}\right]\label{eq:Deltatheta}
\end{equation}
and $\Delta\theta_{i}=\bar{\theta}_{i}-\theta_{i}$. 
\begin{thm}
\label{Th event-triggered-part4} Consider a MAS with agent dynamics
given by (\ref{eq: System dynamique Lagrange}) and the control law~(\ref{eq:Control input with tracking}).
Consider some design parameters $\eta\geq0$, $\eta_{2}>0$, $0<b_{i}<\frac{k_{s}}{k_{s}k_{p}+k_{g}}$
, 
\[
c_{3}=\frac{\min\left\{ 1,k_{1},k_{p},k_{0},2k_{0}\left(2k_{0}+\frac{\alpha_{\min}k_{\min}}{k_{\max}}\right)\right\} }{\max\left\{ 1,k_{M}\right\} }
\]
and $k_{1}=k_{s}-\left(1+k_{p}\left(k_{M}+1\right)\right)$. In absence
of communication delays, the system (\ref{eq: System dynamique Lagrange})
is input-to-state practically stable (ISpS),\textcolor{black}{see \cite{jiang1996}} \textcolor{black}{or Appendix~\ref{def-ISPS}},
and the agents can be driven to some target formation such that
\begin{equation}
\lim_{t\to\infty}\sum_{i=1}^{N}k_{0}\left\Vert \varepsilon_{i}\right\Vert ^{2}+\frac{1}{2}P\left(q,t\right)\leq\xi\label{eq:Convergence formation}
\end{equation}
with 
\begin{equation}
\xi=\frac{N}{k_{g}c_{3}}\left[D_{\max}^{2}+\eta+c_{3}\Delta_{\max}\right]\label{eq:Borne eta2}
\end{equation}
where $\Delta_{\max}=\max_{i=1:N}\left(\sup_{t>0}\left(\Delta\theta_{i}^{T}\Gamma_{i}^{-1}\Delta\theta_{i}\right)\right)$,\textcolor{black}{{}
}if the communications are triggered when one of the following conditions
is satisfied
\begin{align}
k_{s}\bar{s}_{i}^{T}\bar{s}_{i}+k_{p}k_{g}\bar{g}_{i}^{T}\bar{g}_{i}+\eta & \leq\alpha_{\text{M}}^{2}\left(k_{e}e_{i}^{iT}e_{i}^{i}+k_{p}k_{M}\dot{e}_{i}^{iT}\dot{e}_{i}^{i}\right)\nonumber \\
 & +\alpha_{\text{M}}k_{C}^{2}k_{p}\left\Vert e_{i}^{i}\right\Vert ^{2}\sum_{j=1}^{N}k_{ji}\left[\left\Vert \dot{\hat{q}}_{j}^{i}\right\Vert +\eta_{2}\right]^{2}+k_{g}b_{i}\left\Vert \dot{q}_{i}-\dot{q}_{i}^{*}\right\Vert ^{2}\nonumber \\
 & +k_{p}\left\Vert e_{i}^{i}\right\Vert \left[\alpha_{\text{M}}^{2}\left(1+\left\Vert \left|Y_{i}\right|\Delta\theta_{i,\max}\right\Vert ^{2}\right)+\frac{\left\Vert \left|Y_{i}\right|\Delta\theta_{i,\max}\right\Vert ^{2}}{\left(1+\left\Vert \left|Y_{i}\right|\Delta\theta_{i,\max}\right\Vert ^{2}\right)}\right]\label{eq:condition event-triggered ei}
\end{align}
\begin{eqnarray}
\left\Vert \dot{q}_{i}\right\Vert  & \geq & \left\Vert \dot{\hat{q}}_{i}^{i}\right\Vert +\eta_{2}\label{eq:Second CTC}
\end{eqnarray}
with $k_{e}=k_{s}k_{p}^{2}+k_{g}k_{p}+\frac{k_{g}}{b_{i}}$, and $Y_{i}=Y_{i}\left(q_{i},\,\dot{q}_{i},\,\dot{\bar{p}}_{i},\,\bar{p}_{i}\right)$.

Moreover, consecutive communication triggering time instants satisfy $t_{i,k+1}>t_{i,k}$. \hfill $\Box$
\end{thm}
The proof of Theorem~\ref{Th event-triggered-part4} is given in
Appendix~\ref{subsec:Proof-of-convergence} and the proof of $t_{i,k+1}-t_{i,k}>0$ in Appendix~\ref{subsec:Proof-of-absence-Zeno}. 

The CTCs proposed in Theorem~\ref{Th event-triggered-part4} are
analyzed assuming that the estimators of the state of the agents and
the communication protocol is such that $\forall\left(i,j\right)\in\mathcal{N}\times\mathcal{N}$,
\begin{align}
\hat{x}_{i}^{i}\left(t\right)= & \hat{x}_{i}^{j}\left(t\right)\label{eq:synchronization estim}\\
\hat{x}_{i}^{i}\left(t_{i,k}\right)= & x_{i}^{i}\left(t_{i,k}\right),\label{eq:reset estimator}
\end{align}
\textcolor{black}{These properties are actually satisfied if the communication protocol
described in Section$~\ref{subsec:Estimator-dynamic-and}$ and the state estimator $(\ref{eq:estimation des coordonnes des voisins- avance})-(\ref{eq: Init_hat(dq)_ij})$ are employed.}
Theorem~\ref{Th event-triggered-part4} is valid independently of
the way the estimate $\hat{x}_{i}^{i}$ of $x_{i}$ is evaluated provided
that (\ref{eq:synchronization estim}) and (\ref{eq:reset estimator})
are satisfied.

From (\ref{eq:Convergence formation}) and (\ref{eq:condition event-triggered ei}),
one sees that $\eta$ can be used to adjust the trade-off between
the bound $\xi$ on the formation and tracking errors and the amount
of triggered communications. If $\eta=0$, there is no perturbation
and $\theta_{i}$ is perfectly known, the system converges asymptotically.

The CTC (\ref{eq:Second CTC}) is related to the discrepancy between
$\dot{q}_{i}$ and $\dot{\hat{q}}_{i}^{i}$. Choosing a small value
of $\eta_{2}$ may lead to frequent communications. On the contrary,
when $\eta_{2}$ is large, (\ref{eq:condition event-triggered ei})
is more likely to be satisfied. A value of $\eta_{2}$ that corresponds
to a trade-off between the two CTCs (\ref{eq:condition event-triggered ei})
and (\ref{eq:Second CTC}) has thus to be found to minimize the amount
of communications.

The CTCs (\ref{eq:condition event-triggered ei}) and (\ref{eq:Second CTC})
mainly depend on $e_{i}^{i}$ and $\dot{e}_{i}^{i}$. A communication
is triggered by Agent~$i$ when the state estimate $\hat{x}_{i}^{i}$
of its own state vector $x_{i}$ is not satisfying, $i.e.$, when
$e_{i}^{i}$ and $\dot{e}_{i}^{i}$ becomes large. To reduce the number
of triggered communications, one has to keep $e_{i}^{i}$ and $\dot{e}_{i}^{i}$
as small as possible. This may be achieved by increasing the accuracy
of the estimator, as proposed \textcolor{black}{in \cite{viel2017},
but possibly at the price of a more complex structure for the estimator
or the number of connection in the communication graph.}

The perturbations have a direct impact on $e_{i}^{i}$ and $\dot{e}_{i}^{i}$,
and, as a consequence, on the frequency of communications. (\ref{eq:Borne eta2})
shows the impact of $D_{\max}$ and $\eta$ on the formation and tracking
errors: in presence of perturbations, the formation and tracking errors
cannot reach a value below a minimum value due to the perturbations.
At the cost of a larger formation and tracking errors, $\eta$ can
reduce the number of triggered communications and so can reduce the
influence of perturbations on the CTC (\ref{eq:condition event-triggered ei}).

The discrepancy between the actual values of $M_{i}$ and $C_{i}$
and of their estimates $\hat{M}_{i}^{i}$ and $\hat{C}_{i}^{i}$ determines
the accuracy of $\bar{\theta}_{i}$, so $\Delta\theta_{i,\max}$,
and the estimation errors. Even in absence of state perturbations,
due to the linear parametrization, it is likely that $\hat{M}_{i}^{i}\neq M_{i}$,
$\hat{C}_{i}^{i}\neq C_{i}$ and $\Delta\theta_{i,\max}>0$, which
leads to the satisfaction of the CTCs at some time instants. Thus,
the CTC (\ref{eq:condition event-triggered ei}) leads to more communications when the model of the agent dynamics is not accurate, requiring thus
more frequent updates of the estimate of the states of agents.

The choice of the parameters $\alpha_{\text{M}}$, $k_{g}$, $k_{p}$
and $b_{i}$ also determines the number of broadcast messages. Choosing
the spring coefficients $k_{ij}$ such that $\alpha_{i}=\sum_{j=1}^{N}k_{ij}$
is small leads to a reduction in the number of communication triggered
due to the satisfaction of (\ref{eq:condition event-triggered ei}).

\section{Simulation results }

\label{sec:Example}

The performance of the proposed algorithm is evaluated considering
a set of $N=6$ agents. Two models will be considered to describe
the dynamics of the agents.

\subsection{Models of the agent dynamics and estimator}

\subsubsection{Double integrator with Coriolis term (DI)}

The first model consists in the dynamical system
\begin{eqnarray*}
M_{i}\left(q_{i}\right)\ddot{q}_{i}+C_{i}\left(q_{i},\,\dot{q}_{i}\right)\dot{q}_{i} & = & \tau_{i}+d_{i}
\end{eqnarray*}
with $q_{i} = [x_i, y_i]\in\mathbb{R}^{2}$ and where 
\begin{equation}
M_{i}=\left[\begin{array}{cc}
1 & 0\\
0 & 1
\end{array}\right]\,C_{i}\left(\dot{q}_{i}\right)=\left[\begin{array}{cc}
0.1 & 0\\
0 & 0.1
\end{array}\right]\left\Vert \dot{q}_{i}\right\Vert .\label{eq:Model simple integrateur-1}
\end{equation}

Then the vectors $\bar{\theta}_{i}\left(0\right)=\hat{\theta}_{i}^{j}\left(0\right)$,
$i=1,\dots,N$ are obtained using (\ref{eq:Prop. Yi}). In place of
the estimator in Section~\ref{subsec:Estimator-dynamic-and} a first
less accurate estimate of $x_{j}$ made by Agent~$i$, is evaluated
as
\begin{align}
\hat{q}_{j}^{i}\left(t\right) & =q_{j}\left(t_{j,k}^{i}\right)\label{eq:Basic_estimator_formation_1}\\
\dot{\hat{q}}_{j}^{i}\left(t\right) & =\dot{q}_{j}\left(t_{j,k}^{i}\right).\label{eq:Basic_estimator_formation_2}
\end{align}
This estimator allows one to better observe the tradeoff between the
potential energy of the formation and the communication requirements.

For this dynamical model, the parameters of the control law (\ref{eq:Control input with tracking})
and the CTC (\ref{eq:condition event-triggered ei}) have been selected
as: $k_{M}=\left\Vert M_{i}\right\Vert =1$, $k_{C}=\left\Vert C_{i}\right\Vert =0.1$,
$k_{p}=1$, $k_{g}=15$, $k_{s}=1+k_{p}\left(k_{M}+1\right)$, $b_{i}=\frac{1}{k_{g}}$,
and $k_{0}=2$.

\subsubsection{Surface ship (SS)}

The second model considers surface ships with coordinate vectors $q_{i}=\left[\begin{array}{ccc}
x_{i} & y_{i} & \psi_{i}\end{array}\right]^{T}\in\mathbb{R}^{3}$, $i=1\dots N$, in a local earth-fixed frame. For Agent~$i$, $\left(x_{i},y_{i}\right)$
represents its position and $\psi_{i}$ its heading angle. The dynamics
of the agents is described by the surface ship dynamical model taken
from \cite{Kyrkjeb2007}, assumed identical for all agents, and expressed
in the body frame as
\begin{equation}
M_{\text{b},i}\dot{\mathrm{v}_{i}}+C_{\text{b},i}\left(\mathrm{v}_{i}\right)\mathrm{v}_{i}+D_{\text{b},i}\mathrm{v}_{i}=\tau_{\text{b},i}+d_{\text{b},i},\label{eq:dynamics ship}
\end{equation}
where $\mathrm{v}_{i}=\left[\begin{array}{ccc}
u_{i} & v_{i} & r_{i}\end{array}\right]^{T}$ is the velocity vector in the body frame, $\tau_{\text{b},i}$ is
the control input, $d_{\text{b},i}$ is the perturbation, and
\begin{eqnarray*}
M_{\text{b},i} & = & \left[\begin{array}{ccc}
25.8 & 0 & 0\\
0 & 33.8 & 1.0115\\
0 & 1.0115 & 2.76
\end{array}\right]\\
C_{\text{b},i}\left(\mathrm{v}_{i}\right) & = & \left[\begin{array}{ccc}
0 & 0 & -33.8v_{i}-1.0115r_{i}\\
0 & 0 & 25.8u_{i}\\
33.8v_{i}+1.0115r_{i} & -25.8u_{i} & 0
\end{array}\right]\\
D_{\text{b},i} & = & \left[\begin{array}{ccc}
0.72 & 0 & 0\\
0 & 0.86 & -0.11\\
0 & -0.11 & -0.5
\end{array}\right].
\end{eqnarray*}

At $t=0$, one assumes that Agent~$i$ has access to estimates $\hat{M}_{\text{b},i}^{i}$
of $M_{\text{b},i}$, $\hat{C}_{\text{b},i}^{i}$ of $C_{\text{b},i}$,
and $\hat{D}_{\text{b},i}^{i}$ of $D_{\text{b},i}$ described as
\begin{align*}
\hat{M}_{\text{b},i}^{i} & =\left(1_{3\times3}+0.1\Xi_{i}^{\text{M}}\right)\odot M_{\text{b},i}\\
\hat{C}_{\text{b},i}^{i} & =\left(1_{3\times3}+0.1\Xi_{i}^{\text{C}}\right)\odot C_{\text{b},i}\\
\hat{D}_{\text{b},i}^{i} & =\left(1_{3\times3}+0.1\Xi_{i}^{\text{D}}\right)\odot D_{\text{b},i},
\end{align*}
where $1_{3\times3}$ is the $3\times3$ matrix of ones, $\Xi_{i}^{\text{M}}$,
$\Xi_{i}^{\text{C}}$, and $\Xi_{i}^{D}$ are matrices which components
are independent and identically Bernoulli random variables with values
in $\left\{ -1,1\right\} $, and $\odot$ is the Hadamard product.
These estimates are transmitted at $t=0$ to all other agents. As
a consequence, the estimates of $M_{\text{b},i}$ and $C_{\text{b},i}$
made by all agents at $t=0$ are all identical.

The model (\ref{eq:dynamics ship}) is expressed with the coordinate
vectors $q_{i}$ in the local earth-fixed frame using the transform
\begin{eqnarray*}
\dot{q}_{i} & = & J_{i}\left(\psi_{i}\right)\mathrm{v_{i}}\\
J_{i}\left(\psi_{i}\right) & = & \left[\begin{array}{ccc}
\cos\psi_{i} & -\sin\psi_{i} & 0\\
\sin\psi_{i} & \cos\psi_{i} & 0\\
0 & 0 & 1
\end{array}\right]
\end{eqnarray*}
where $J_{i}\left(\psi_{i}\right)$ is a simple rotation around the
$z$-axis in the earth-fixed coordinate. Define $J_{i}^{-T}=\left(J_{i}^{-1}\right)^{T}$.
Then, (\ref{eq:dynamics ship}) can be rewritten as
\begin{eqnarray*}
J_{i}^{-T}M_{\text{b},i}J_{i}^{-1}\ddot{q}_{i}+J_{i}^{-T}\left[C_{\text{b,i}}\left(\mathrm{v}\right)-M_{\text{b,i}}J_{i}^{-1}\dot{J}_{i}+D_{\text{b},i}\right]J_{i}^{-1}\dot{q}_{i} & = & J_{i}^{-T}\tau_{\text{b}}+J_{i}^{-T}d_{\text{b},i}
\end{eqnarray*}
and so

\begin{eqnarray*}
M_{i}\left(q_{i}\right)\ddot{q}_{i}+C_{i}\left(q_{i},\,\dot{q}_{i}\right)\dot{q}_{i} & = & \tau_{i}+d_{i}
\end{eqnarray*}
where 
\[
M_{i}\left(q_{i}\right)=J^{-T}M_{\text{b}}J^{-1},
\]
\[
C_{i}\left(q_{i},\,\dot{q}_{i}\right)=J_{i}^{-T}\left[C_{\text{b},i}\left(J_{i}^{-1}\dot{q}_{i}\right)-M_{\text{b},i}J_{i}^{-1}\dot{J}_{i}+D_{\text{b},i}\right]J^{-1},
\]
and $\tau_{i}$ is the control input in earth-fixed coordinates as
defined in (\ref{eq:Control input with tracking}).

The vectors $\bar{\theta}_{i}\left(0\right)=\hat{\theta}_{i}^{j}\left(0\right)$,
$i=1,\dots,N$ are obtained using (\ref{eq:Prop. Yi}). The estimator
described in Section~\ref{subsec:Estimator-dynamic-and} is employed.

For this dynamical model, the parameters of the control law (\ref{eq:Control input with tracking})
and the CTC (\ref{eq:condition event-triggered ei}) have been selected
as: $k_{M}=\left\Vert M_{i}\right\Vert =33.8$, $k_{C}=\left\Vert C_{v}\left(1_{N}\right)\right\Vert =43.96$,
$k_{p}=6$, $k_{g}=20$, $k_{s}=1+k_{p}\left(k_{M}+1\right)$, $b_{i}=\frac{1}{k_{g}}$,
and $k_{0}=1.5$. 

\subsubsection{Simulation parameters}

%
{\color{black}
The initial value are $q\left(0\right)= [
x\left(0\right)^{T}, y\left(0\right)^{T}]^T$ with $\dot{q}(0) = 0_{2N\times1} $ for the DI  and $q\left(0\right)= [
x\left(0\right)^{T}, y\left(0\right)^{T}, \psi\left(0\right)^{T}]^{T}$ with $\dot{q}\left(0\right)=0_{3N\times1}$ for the SS, where

\begin{eqnarray*}
x\left(0\right) & = & \left[
-0.35, 4.59, 4.72, 0.64, 3.53, -1.26\right]\\
y\left(0\right) & = & \left[
-1.11, -4.59, 2.42, 1.36, 1.56, 3.36\right]\\
\psi\left(0\right) & = & 0_{N} 
\end{eqnarray*} 
 
}

An hexagonal target formation
is considered with {\color{black}{$r^{*}\left(0\right)=[\begin{array}{ccc}
r_{(1)}^{*}\left(0\right)^{T} & r_{(2)}^{*}\left(0\right)^{T} \end{array}]^{T}$
for DI}} and 

$r^{*}\left(0\right)=[\begin{array}{ccc}
r_{(1)}^{*}\left(0\right)^{T} & r_{(2)}^{*}\left(0\right)^{T} & r_{(3)}^{*}\left(0\right)^{T}\end{array}]^{T}$ for SS where 
 \begin{eqnarray*} 
r_{(1)}^{*}\left(0\right) & = & \left[
0, 2, 3, 2, 0, -1\right]\\
r_{(2)}^{*}\left(0\right) & = & \left[
0, 0, \sqrt{3}, 2\sqrt{3}, 2\sqrt{3}, \sqrt{3}\right]\\
r_{(3)}^{*}\left(0\right) & = & 0_{N}
\end{eqnarray*}


\textcolor{black}{Introduce the communicqtion graphe $\mathcal{G}$
such each agent~$i$ can communicate with agents~$i+1$, $i+3$
and $i-1$.} Using the approach developed in \cite{Qingkai2015},
the following matrix $K=\left[k_{ij}\right]_{\begin{array}{c}
i=1\ldots N\\
j=1\ldots N
\end{array}}$ can be computed from $r^{*}$

\[
K=0.1\left[\begin{array}{cccccc}
0 & 1.85 & 0 & 0.926 & 0 & 1.85\\
1.85 & 0 & 1.85 & 0 & 0.926 & 0\\
0 & 1.85 & 0 & 1.85 & 0 & 0.926\\
0.926 & 0 & 1.85 & 0 & 1.85 & 0\\
0 & 0.926 & 0 & 1.85 & 0 & 1.85\\
1.85 & 0 & 0.926 & 0 & 1.85 & 0
\end{array}\right]
\]
and $\alpha_{i}=\sum_{j=1}^{N}k_{ij}=0.463$, for all $i=1,\dots,N$
and $\alpha_{\text{M}}=0.463$.

A fully-connected communication graph is considered. The simulation
duration is $T=2\:s$. Matlab's ode45 integrator is used with a step
size $\Delta t=0.01$~s. Since time has been discretized, the minimum
delay between the transmission of two messages by the same agent is
set to $\Delta t$. The perturbation $d_{i}\left(t\right)$ is assumed
of constant value over each interval of the form $\left[k\Delta t,\,\left(k+1\right)\Delta t\right[$.
The components of $d_{i}\left(t\right)$ are independent realizations
of zero-mean uniformly distributed noise $U\left(-\frac{D_{\max}}{\sqrt{3}},\,\frac{D_{\max}}{\sqrt{3}}\right)$
and are thus such that $\left\Vert d_{i}\left(t\right)\right\Vert \leq D_{\max}$.
Let $N_{\text{m}}$ be the total number of messages broadcast during
a simulation. The performance of the proposed approach is evaluated
comparing $N_{\text{m}}$ to the maximum number of messages that can
be broadcast $\overline{N}_{\text{m}}=NT/\Delta t\geq N_{\text{m}}$.
The percentage of residual communications is defined as $R_{\text{com}}=100\frac{N_{\text{m}}}{\overline{N}_{\text{m}}}$.
$R_{\text{com}}$ indicates the percentage of time slots during which
a communication has been triggered.

When a tracking has to be performed, one considers the target trajectory
of the first agent
\[
\dot{q}_{1}^{*}\left(t\right)=\left[\begin{array}{c}
4\sin\left(0.4t\right)\\
4\cos\left(0.4t\right)\\
0.4t
\end{array}\right],
\]
the other agents having to remain in formation. Define the tracking
error $\varepsilon_{0}=q_{1}-q_{1}^{*}$.

\subsection{Formation control with DI\label{subsec:Formation-control-with-simple-integrator}}

\begin{figure}[H]
\centering{}\subfloat[]{\begin{centering}
\begin{tabular}{c}
\includegraphics[width=0.3\textwidth]{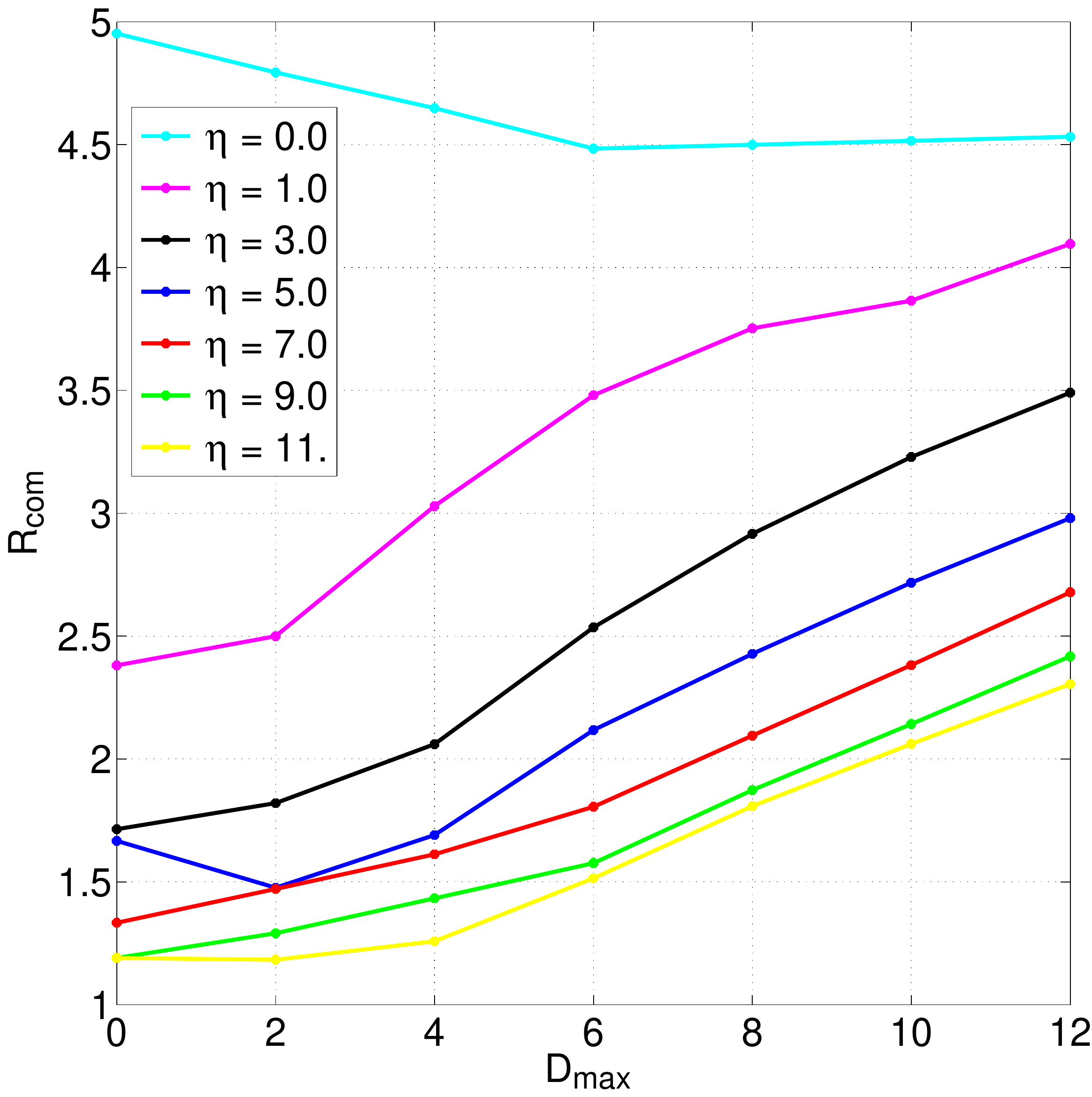}\tabularnewline
\end{tabular}
\par\end{centering}
}\subfloat[]{\begin{centering}
\begin{tabular}{c}
\includegraphics[width=0.3\textwidth]{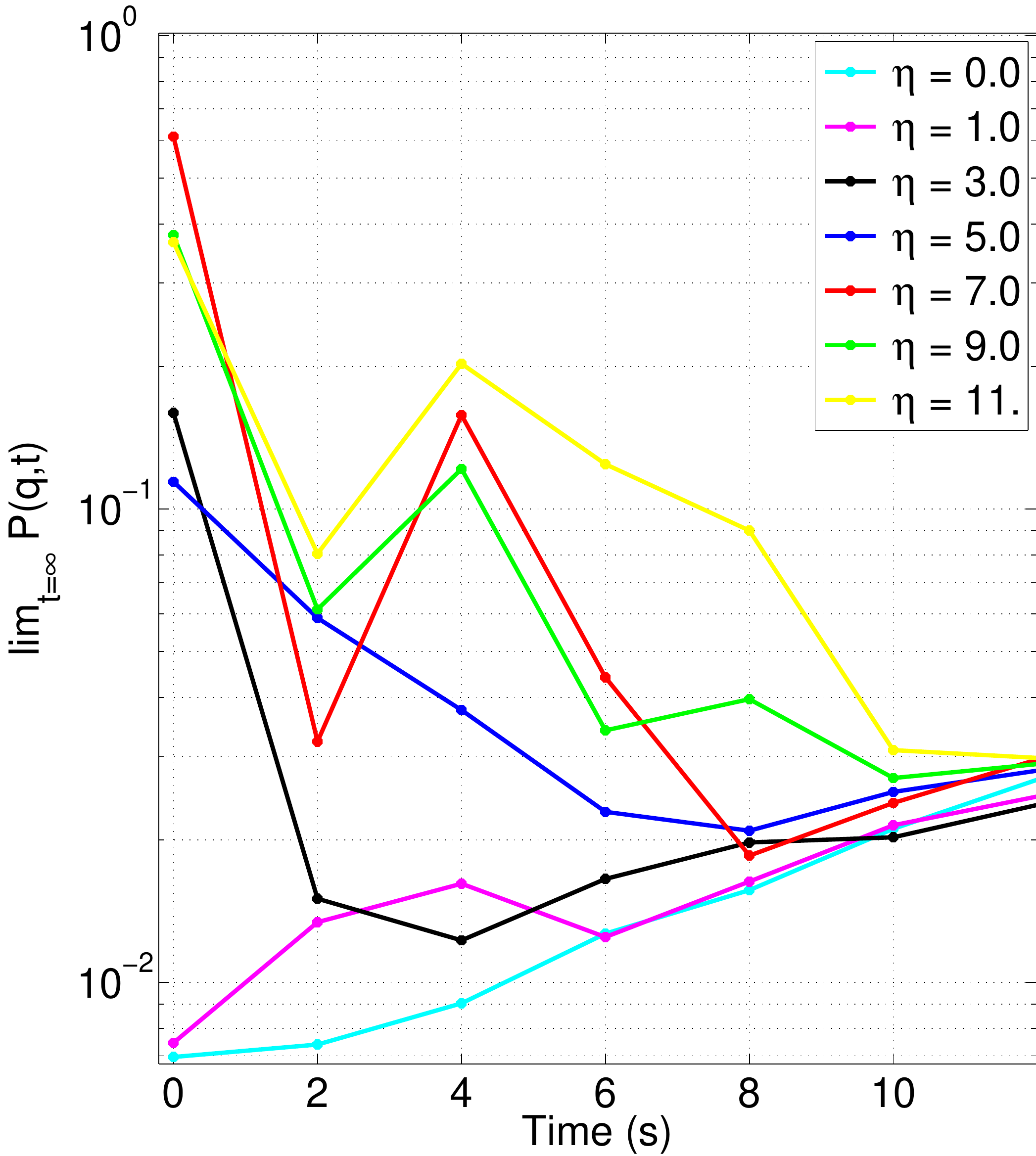}\tabularnewline
\end{tabular}
\par\end{centering}
}\subfloat[]{\begin{centering}
\begin{tabular}{c}
\includegraphics[width=0.3\textwidth]{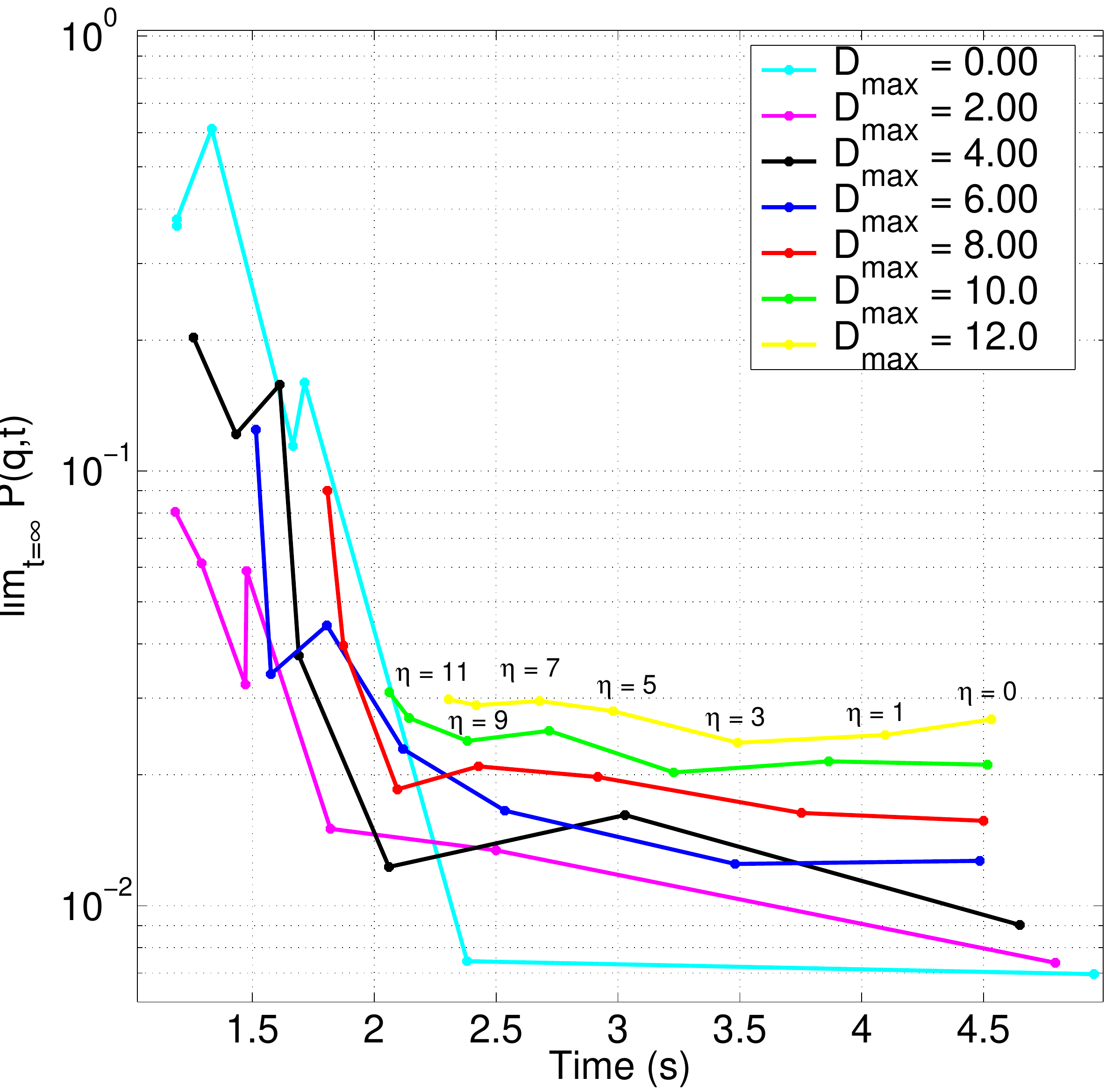}\tabularnewline
\end{tabular}
\par\end{centering}
}\caption{\label{fig:Simple integrateur estim constant}Evolution of $R_{\text{com}}$
and $P\left(q,t\right)$ for different values of $D_{\max}\in\left\{ \protect\begin{array}{ccccccc}
0, & 2, & 4, & 6, & 8, & 10, & 12\protect\end{array}\right\} $, $\eta\in\left\{ \protect\begin{array}{ccccccc}
0, & 1, & 3, & 5, & 7, & 9, & 11\protect\end{array}\right\} $, and $\eta_{2}=7.5$. The DI model and the simple estimator~(\ref{eq:Basic_estimator_formation_1})-(\ref{eq:Basic_estimator_formation_2})
are considered. }
\end{figure}

Figure~\ref{fig:Simple integrateur estim constant} shows the evolution
of the communication ratio $R_{\textrm{com}}$ and of the potential
energy at $t=T$. For all simulations, one has $P\left(q,T\right)\leq\xi$
for the different values of $D_{\max}$ and $\eta$. 

In Figure~\ref{fig:Simple integrateur estim constant}~(a), the
number of communications obtained once the system has converged increases
as the level of perturbations becomes more important, as expected.
Increasing $\eta$ in the CTC \ref{eq:condition event-triggered ei}
helps reducing $R_{\textrm{com}}$ . Nevertheless, increasing $\eta$
also increases the potential energy $P\left(q,T\right)$ of the formation,
as can be seen in Figure~\ref{fig:Simple integrateur estim constant}~(b).
In Figure~\ref{fig:Simple integrateur estim constant}~(b), when
$\eta\geq3$, one observes that the potential energy starts to decrease
with the level of perturbation $D_{\max}$ to increase again when
$D_{\max}$ gets large. To explain this surprising behavior, Figure~\ref{fig:Simple integrateur estim constant}~(c)
shows that there exists a threshold $\overline{R}_{\textrm{com}}=2.25$
below which the potential energy significantly increases to ensure
proper convergence. Therefore $\eta$ should be chosen such that $R_{\textrm{com}}$
remains above this threshold.  Even large values of $D_{\max}$ can
be tolerated provided that $\eta$ is chosen large enough to provide
a sufficient amount of communications.

\subsection{Formation control with ship dynamical model}

\begin{figure}
\centering{}\subfloat[Accurate estimator (\ref{eq:estimation des coordonnes des voisins- avance}).]{\begin{centering}
\begin{tabular}{c}
\includegraphics[width=0.4\textwidth]{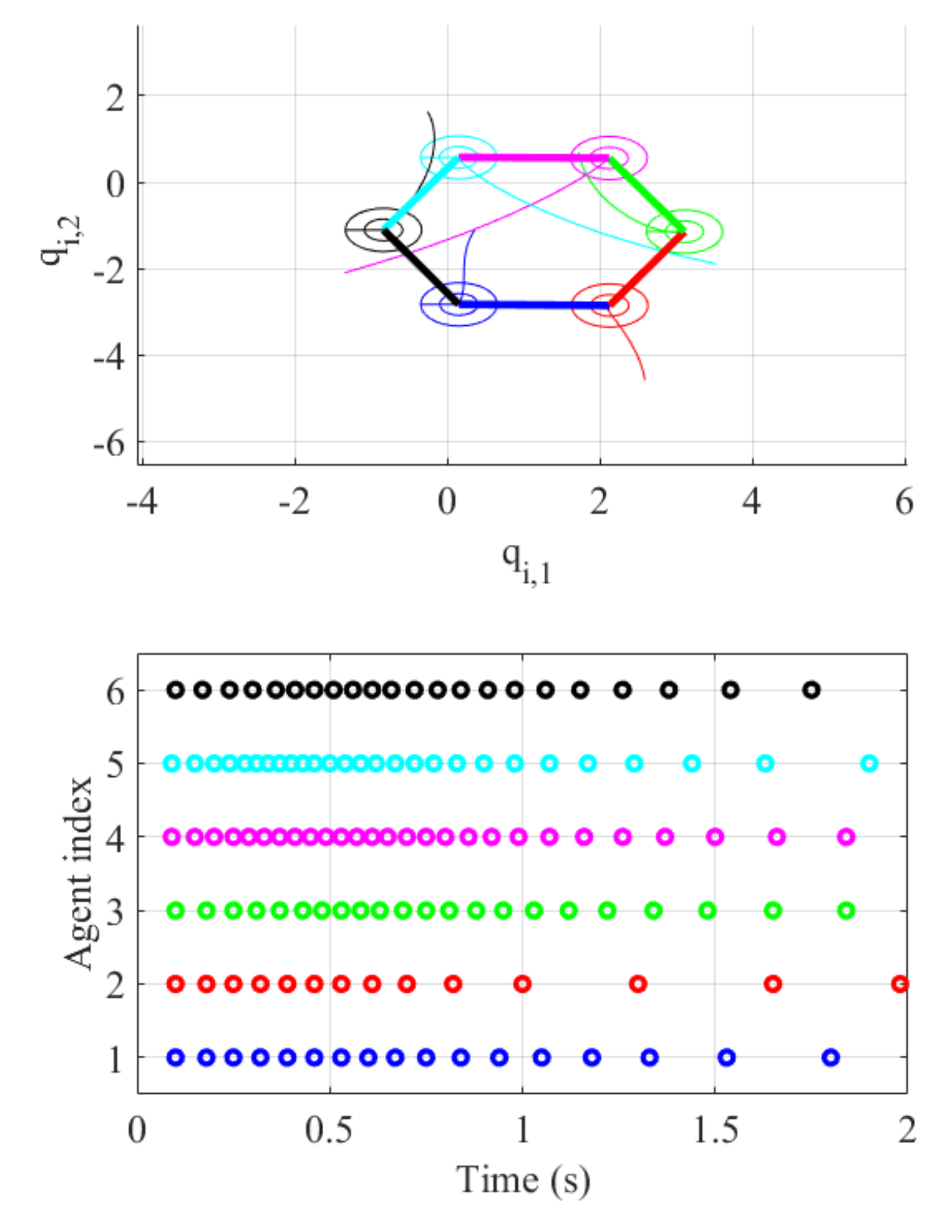}\tabularnewline
\end{tabular}
\par\end{centering}
}\subfloat[Constant estimator (\ref{eq:Basic_estimator_formation_1}).]{\begin{centering}
\begin{tabular}{c}
\includegraphics[width=0.4\textwidth]{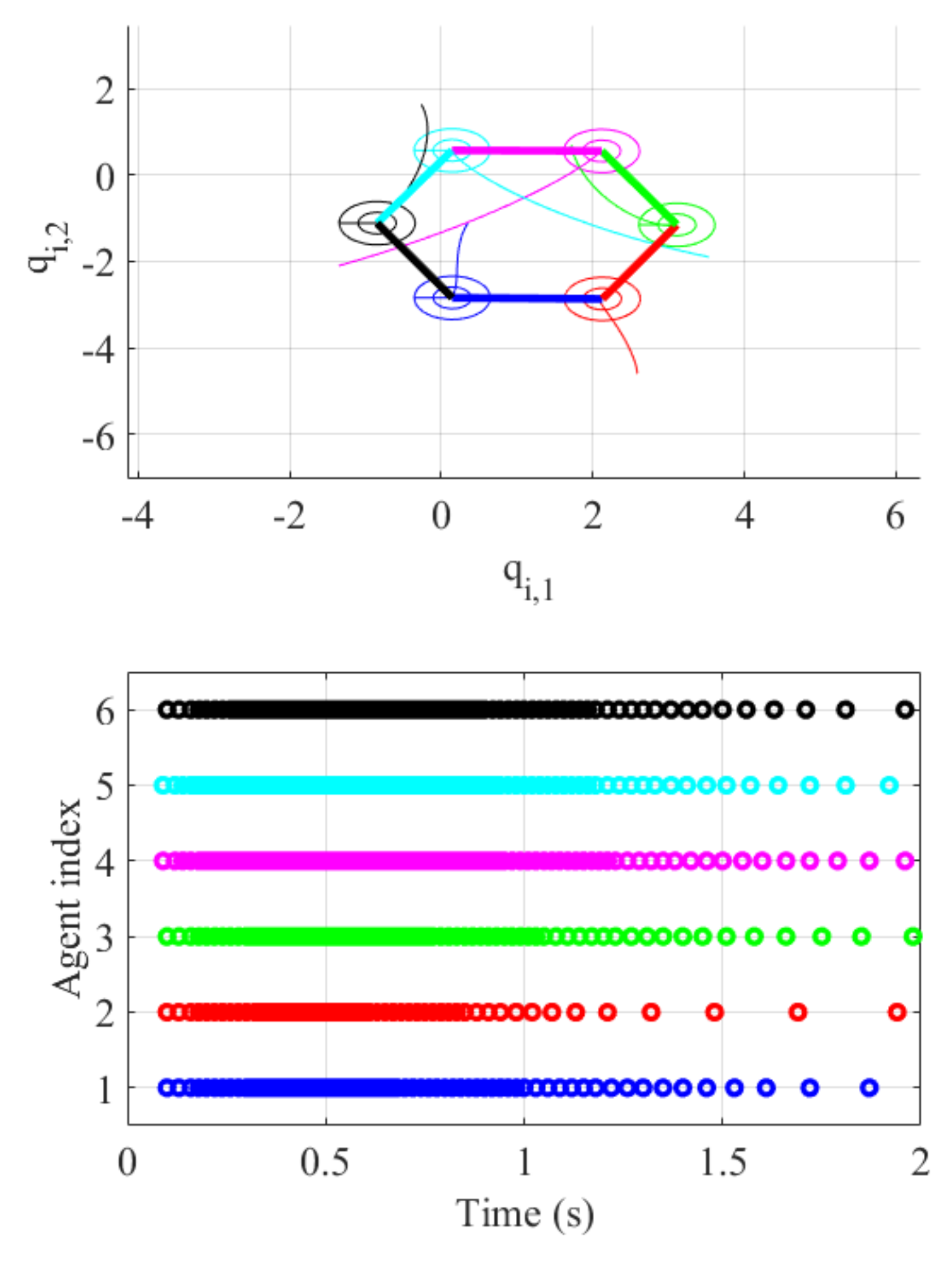}\tabularnewline
\end{tabular}
\par\end{centering}
}\caption{\label{fig:figure hexagone et trajectoire}Hexagonal formation with
$D_{\max}=20$, $\eta=20$ and $\eta_{2}=7.5$. Agents are represented
by circles. In (a), $R_{\text{com}}=10.75\%$ and $P\left(q,T\right)=0.001$.
In (b) $R_{\text{com}}=40.25\%$ and $P\left(q,T\right)=0.001$. $T=2$~s.}
\end{figure}

Figure~\ref{fig:figure hexagone et trajectoire} shows the trajectories
of the agents when the control (\ref{eq:Control input with tracking})
is applied and the communications are triggered according to the CTC
of Theorem~\ref{Th event-triggered-part4}. Figure~\ref{fig:figure hexagone et trajectoire}~(a)
illustrates the results obtained using the accurate estimator (\ref{eq:estimation des coordonnes des voisins- avance}),
Figure~\ref{fig:figure hexagone et trajectoire}(b) illustrates results
obtained using the simple estimator (\ref{eq:Basic_estimator_formation_1}).
The agents converge to the desired formation with a limited number
of communications, even in presence of perturbations.

\subsection{Tracking control with DI}

The simulation duration is $T=3.5\:s$.

\begin{figure}[H] \hspace*{-0.5cm}
\centering{}\subfloat[]{\begin{centering}
\begin{tabular}{c}
\includegraphics[width=0.30\textwidth]{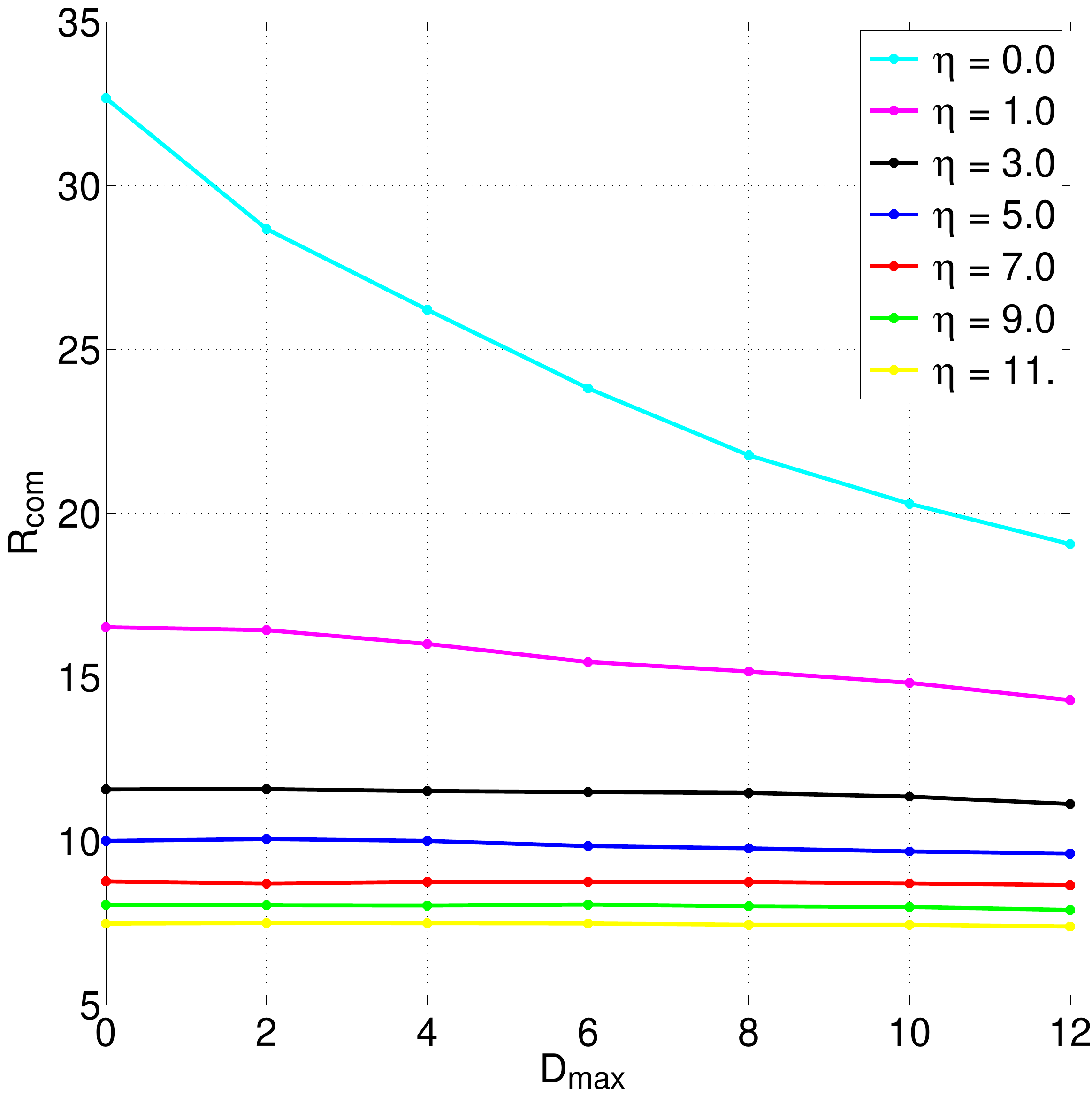}\tabularnewline
\end{tabular}
\par\end{centering}
}\subfloat[]{\begin{centering}
\begin{tabular}{c}
\includegraphics[width=0.30\textwidth]{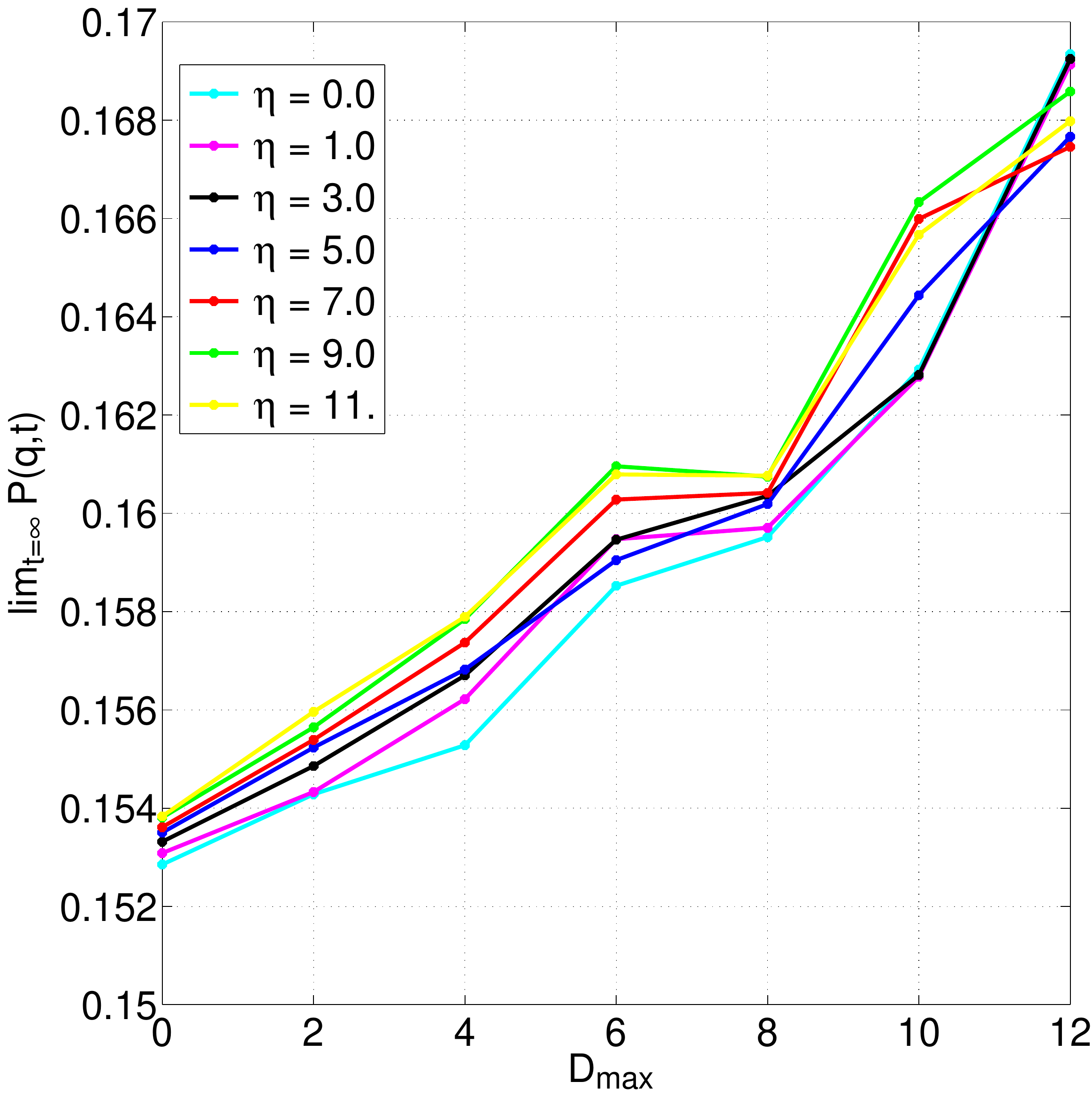}\tabularnewline
\end{tabular}
\par\end{centering}
}\subfloat[]{\begin{centering}
\begin{tabular}{c}
\includegraphics[width=0.30\textwidth]{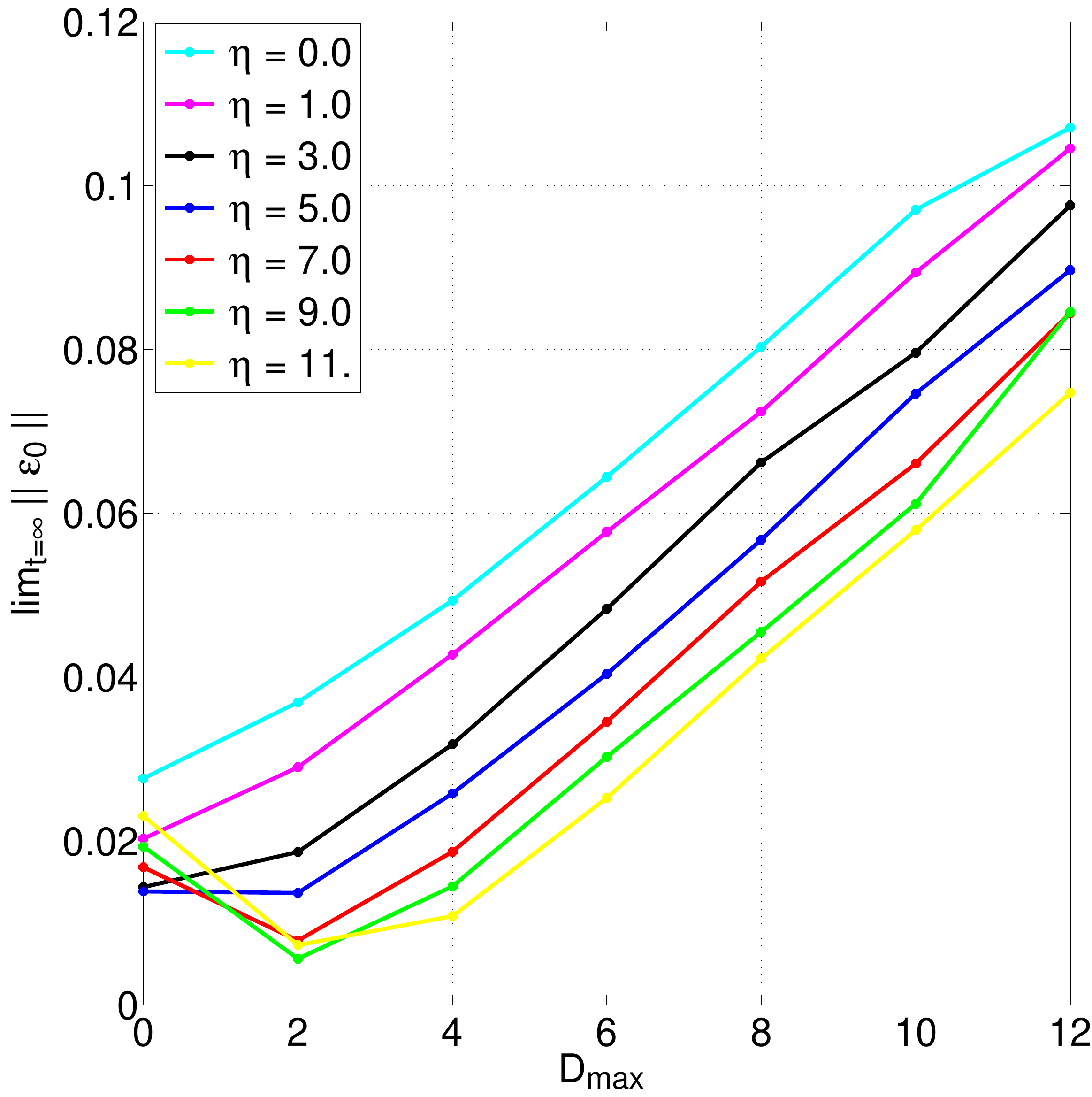}\tabularnewline
\end{tabular}
\par\end{centering}
}\caption{\label{fig:Simple integrateur estim constant-tracking-1}Evolution
of $R_{\text{com}}$, $P\left(q,t\right)$ and $\varepsilon_{0}$
for different values of $D_{\max}\in\left\{ \protect\begin{array}{ccccccc}
0, & 2, & 4, & 6, & 8, & 10, & 12\protect\end{array}\right\} $, $\eta\in\left\{ \protect\begin{array}{ccccccc}
0, & 1, & 3, & 5, & 7, & 9, & 11\protect\end{array}\right\} $ and $\eta_{2}=7.5$. Model (\ref{eq:Model simple integrateur-1})
and constant estimator~(\ref{eq:Basic_estimator_formation_1})-(\ref{eq:Basic_estimator_formation_2})
are considerate. }
\end{figure}

\begin{figure}[H]
\centering{}\subfloat[]{\begin{centering}
\begin{tabular}{c}
\includegraphics[width=0.42\textwidth]{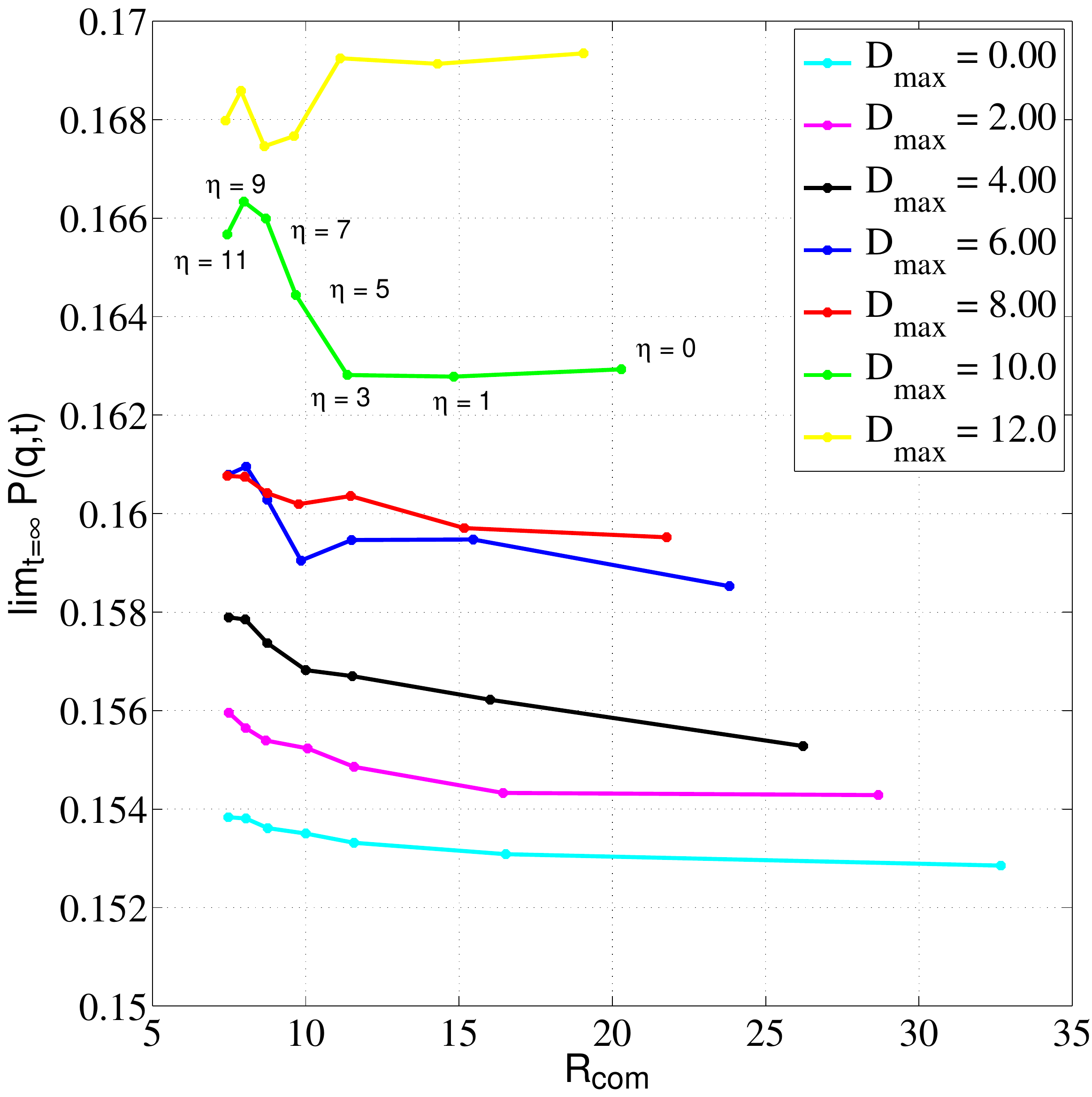}\tabularnewline
\end{tabular}
\par\end{centering}
}\subfloat[]{\begin{centering}
\begin{tabular}{c}
\includegraphics[width=0.42\textwidth]{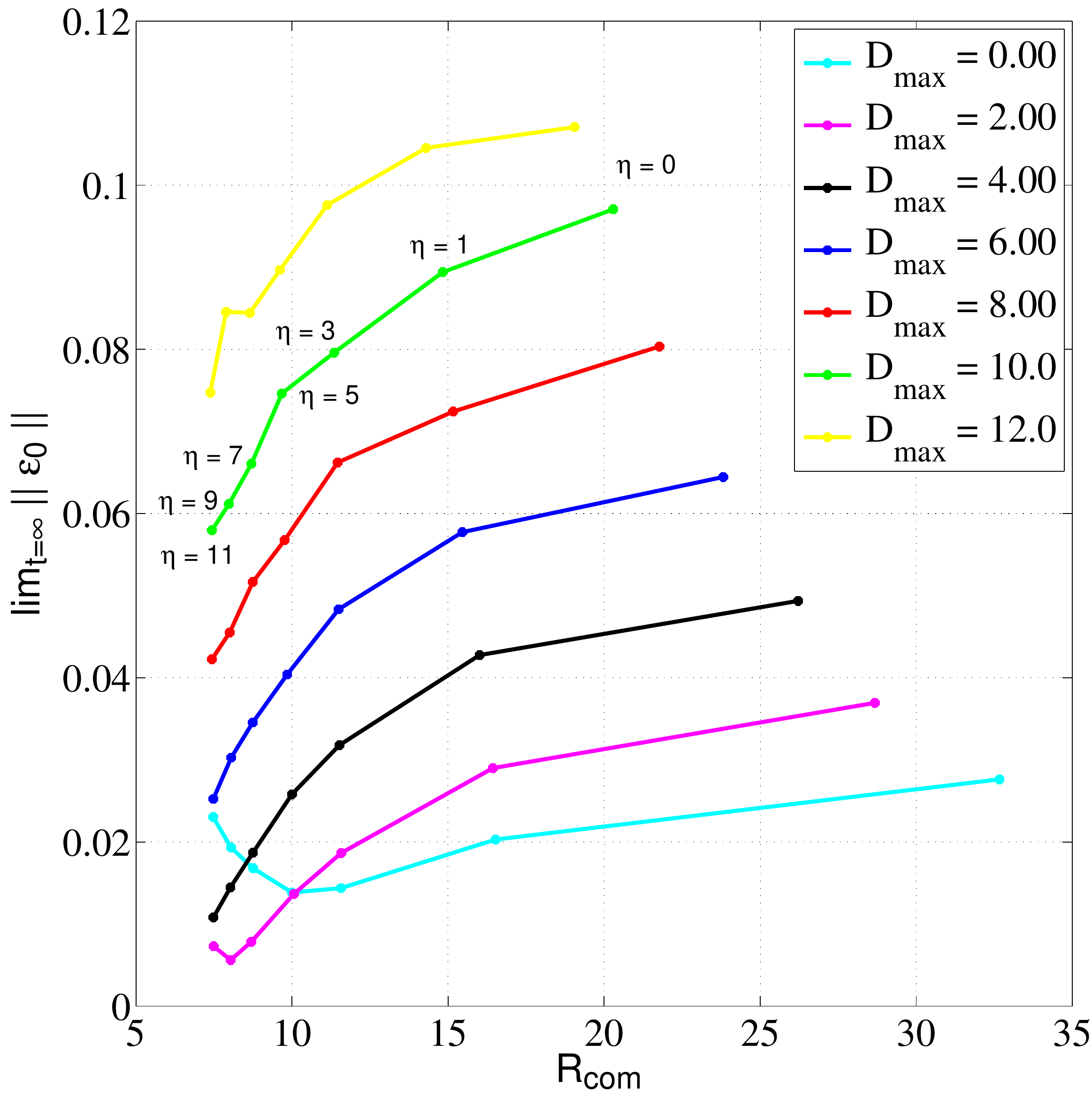}\tabularnewline
\end{tabular}
\par\end{centering}
}\caption{\label{fig:Simple integrateur estim constant-tracking-2}Evolution
of $R_{\text{com}}$, $P\left(q,t\right)$ and $\varepsilon_{0}$
for different values of $D_{\max}\in\left\{ \protect\begin{array}{ccccccc}
0, & 2, & 4, & 6, & 8, & 10, & 12\protect\end{array}\right\} $, $\eta\in\left\{ \protect\begin{array}{ccccccc}
0, & 1, & 3, & 5, & 7, & 9, & 11\protect\end{array}\right\} $ and $\eta_{2}=7.5$. Model (\ref{eq:Model simple integrateur-1})
and constant estimator~(\ref{eq:Basic_estimator_formation_1})-(\ref{eq:Basic_estimator_formation_2})
are considerate. }
\end{figure}

Figures~\ref{fig:Simple integrateur estim constant-tracking-1} and
\ref{fig:Simple integrateur estim constant-tracking-2} show the evolution
of the communication ratio $R_{\textrm{com}}$, the potential energy
and the tracking error at $t=T$. 

In Figure~\ref{fig:Simple integrateur estim constant} (a), the number
of communications obtained once the system has converged decreases
as the level of perturbation becomes more important, especially when
$\eta$ is small, which was not excepted. Such behavior is not observed
with the accurate estimator~(\ref{eq:estimation des coordonnes des voisins- avance}),
where $R_{\text{com}}$ increases when the perturbations become more
important, as illustrated in Figure~\ref{fig:Avance_model-tracking-3courbes}
(a) with the ship model. This behavior can be explained by the fact
a large $D_{\max}$ makes $\left\Vert \bar{g}_{i}\right\Vert $ and
$\left\Vert \bar{s}_{i}\right\Vert $ larger, which reduces the number
of times the CTC (\ref{eq:condition event-triggered ei}) is satisfied,
even if the error $\left\Vert e_{i}^{i}\right\Vert $ is also affected.
Difference with accurate estimator is the error $e_{i}^{i}$ is keeping
small by the estimator, so the influence of perturbations is more
significant on $e_{i}^{i}$ than on $\left\Vert \bar{g}_{i}\right\Vert $
or $\left\Vert \bar{s}_{i}\right\Vert $, which leads to a larger
number of communications triggered.

Figure~\ref{fig:Simple integrateur estim constant} (a) illustrates
that the parameter $\eta$ in the CTC (\ref{eq:condition event-triggered ei})
can help reducing $R_{\textrm{com}}$ . It can be seen that there
exists for $R_{\textrm{com}}$ a threshold ($R_{\textrm{com}}=7)$
which $R_{\textrm{com}}$ cannot reach : we can deduce a minimal number
of communications is required for system converge with the constant
estimator~(\ref{eq:Basic_estimator_formation_1})-(\ref{eq:Basic_estimator_formation_2}). 

Figures~\ref{fig:Simple integrateur estim constant} (b) and (c)
show that the potential energy of the formation $P\left(q,t\right)$
and the tracking error $\varepsilon_{0}$ increase when the perturbation
level increases. The influence of parameter $\eta$ is also illustrated:
Figure~\ref{fig:Simple integrateur estim constant} (b) shows that
a larger value of $\eta$ leads to an increase of $P\left(q,t\right)$,
but reduces $\varepsilon_{0}$. Indeed, the less communications, the
more difficult it is for some Agent~$i$ to be synchronized with
the others agents to reach the target formation. However, be less
synchronized with the other agents allows Agent~$i$ to be more synchronized
with its target trajectory $q_{i}^{*}$ , inducing a small tracking
error $\varepsilon_{0}$. Thus, a trade off between the $P\left(q,t\right)$
and $\varepsilon_{0}$ has to be reached.

\subsection{Tracking with surface ship model}

The simulation duration is $T=2.5\:s$.

\begin{figure}[H]
\centering{}%
\begin{tabular}{c}
\includegraphics[width=0.6\textwidth]{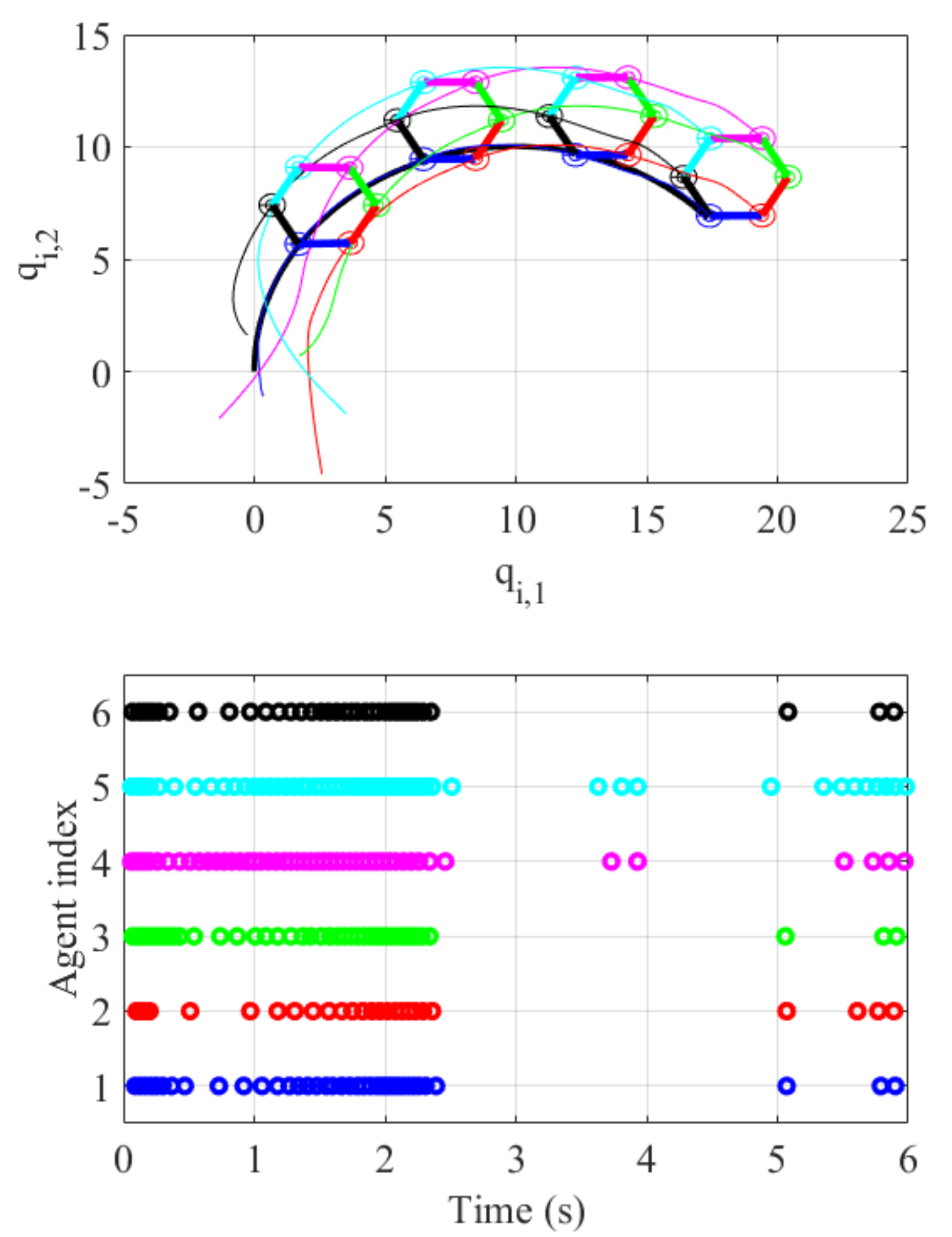}\tabularnewline
\end{tabular}\caption{\label{fig:figure hexagone et tracking}Hexagonal formation and tracking
problem with $D_{\max}=50$, $\eta=50$, and $\eta_{2}=7.5$. Circles
represents agents (top figure) and communication events (bottom figure).
$R_{\text{com}}=7.63\%$, $P\left(q,T\right)=0.001$ and $\left\Vert \varepsilon_{0}\right\Vert =0.1$.
$T=6\,$s.}
\end{figure}

\begin{figure}[ph]
\centering{}\subfloat[]{\begin{centering}
\begin{tabular}{c}
\includegraphics[bb=0bp 3cm 595bp 26cm,clip,width=0.33\textwidth]{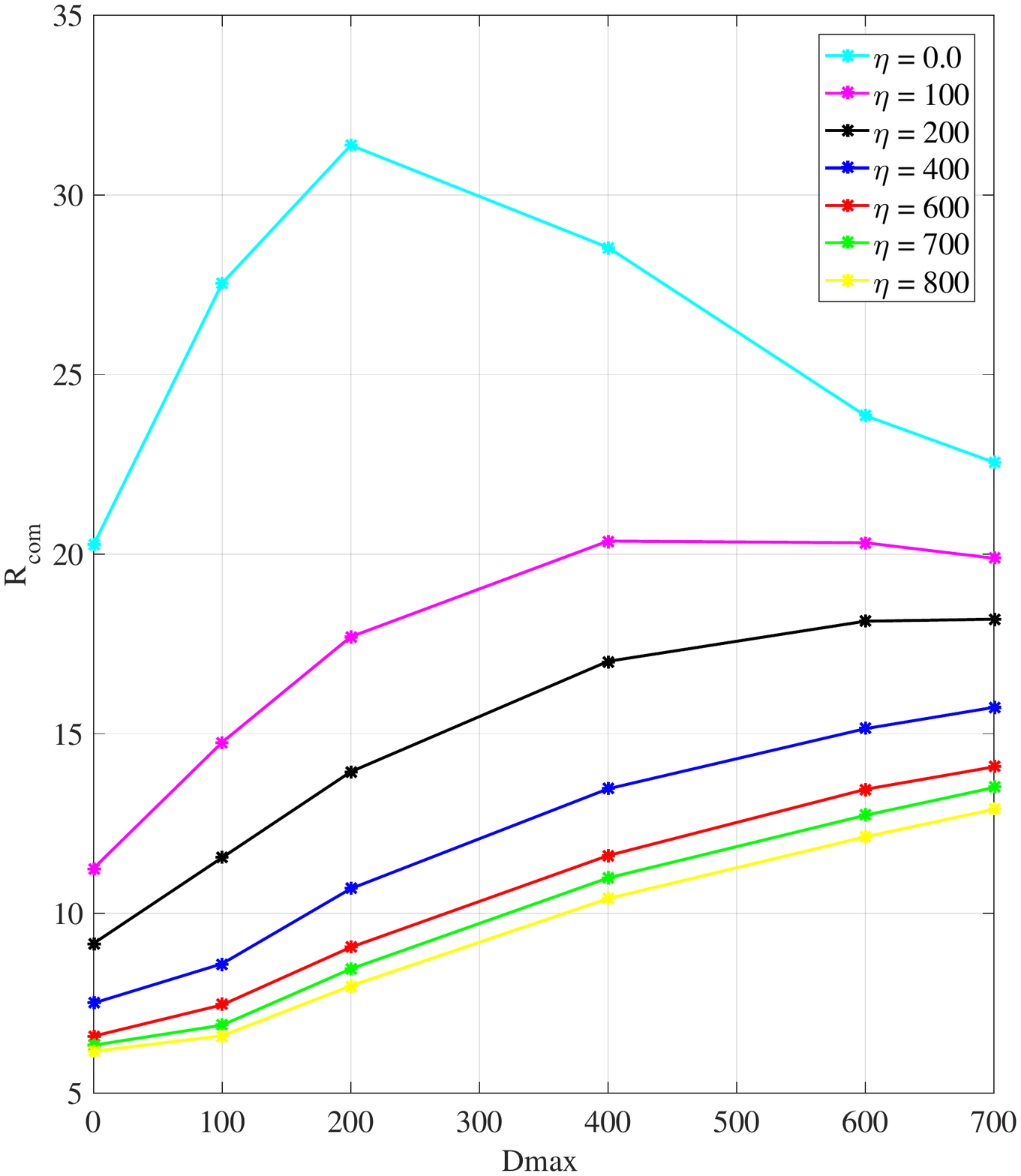}\tabularnewline
\end{tabular}
\par\end{centering}
}\subfloat[]{\begin{centering}
\begin{tabular}{c}
\includegraphics[bb=0bp 3cm 595bp 26cm,clip,width=0.33\textwidth]{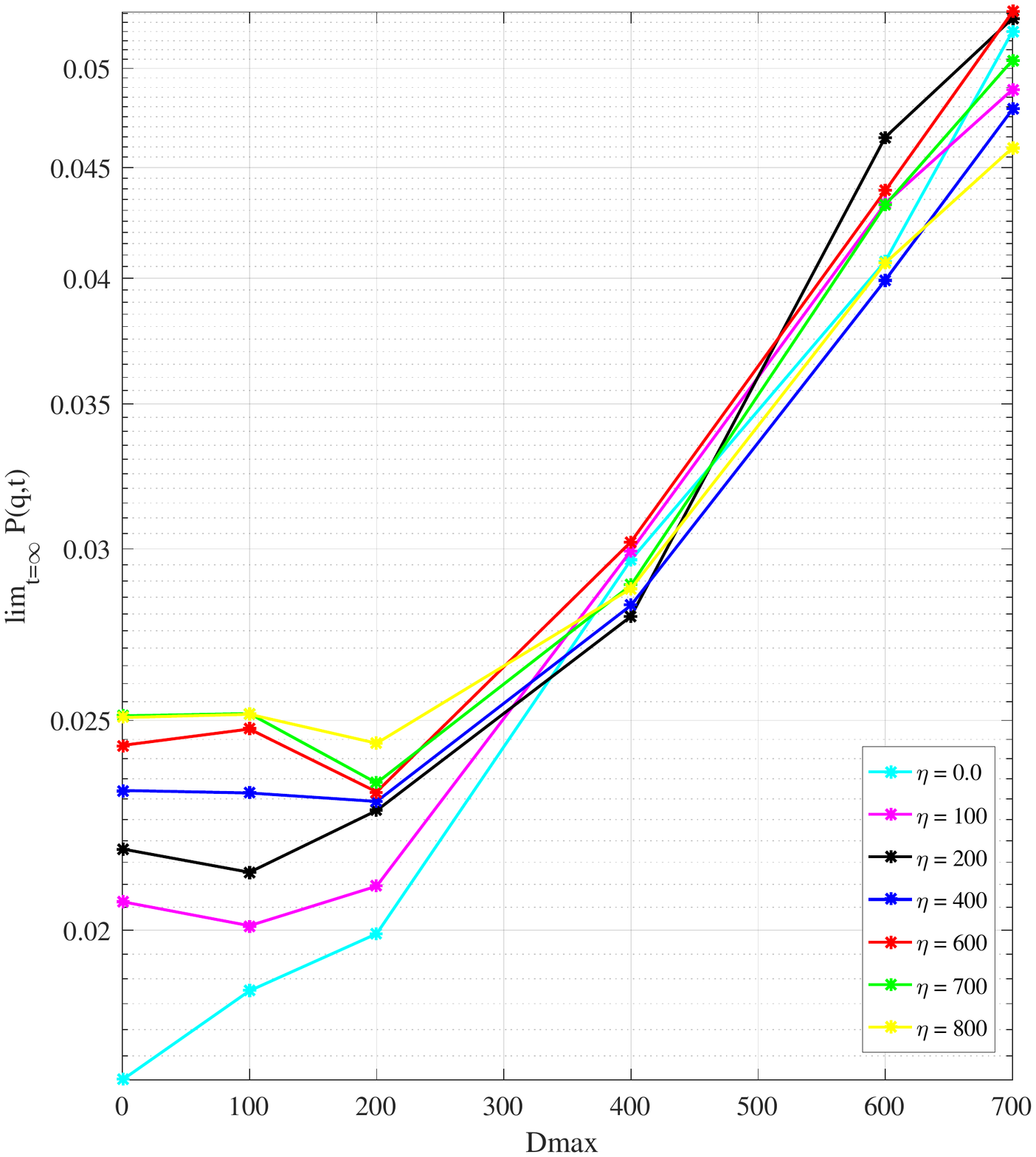}\tabularnewline
\end{tabular}
\par\end{centering}
}\subfloat[]{\begin{centering}
\begin{tabular}{c}
\includegraphics[bb=0bp 3cm 595bp 26cm,clip,width=0.33\textwidth]{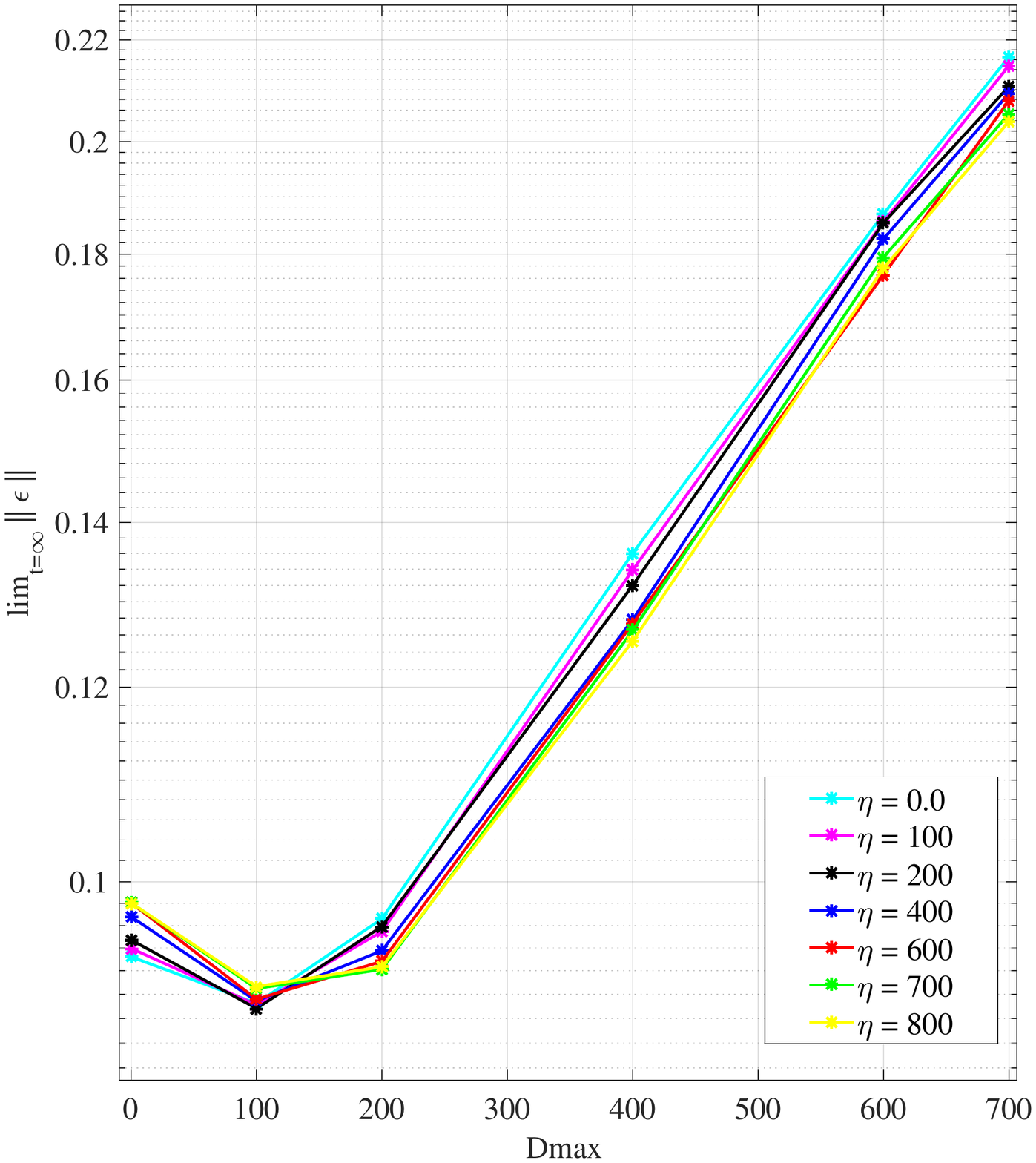}\tabularnewline
\end{tabular}
\par\end{centering}
}\caption{\label{fig:Avance_model-tracking-3courbes}Evolution of $R_{\text{com}}$,
$P\left(q,t\right)$ and $\varepsilon_{0}$ for different values of
$D_{\max}\in\left\{ \protect\begin{array}{ccccc}
0, & 100, & 200, & \ldots & 700\protect\end{array}\right\} $, $\eta\in\left\{ \protect\begin{array}{ccccc}
0, & 100, & 200, & \ldots & 800\protect\end{array}\right\} $and $\eta_{2}=7.5$. The SS model (\ref{eq:dynamics ship}) and accurate
estimator~ (\ref{eq:estimation des coordonnes des voisins- avance})
are considered. }
\end{figure}

\begin{figure}[h]
\centering{}\subfloat[]{\begin{centering}
\begin{tabular}{c}
\includegraphics[bb=0cm 3cm 595bp 26cm,clip,width=0.45\textwidth]{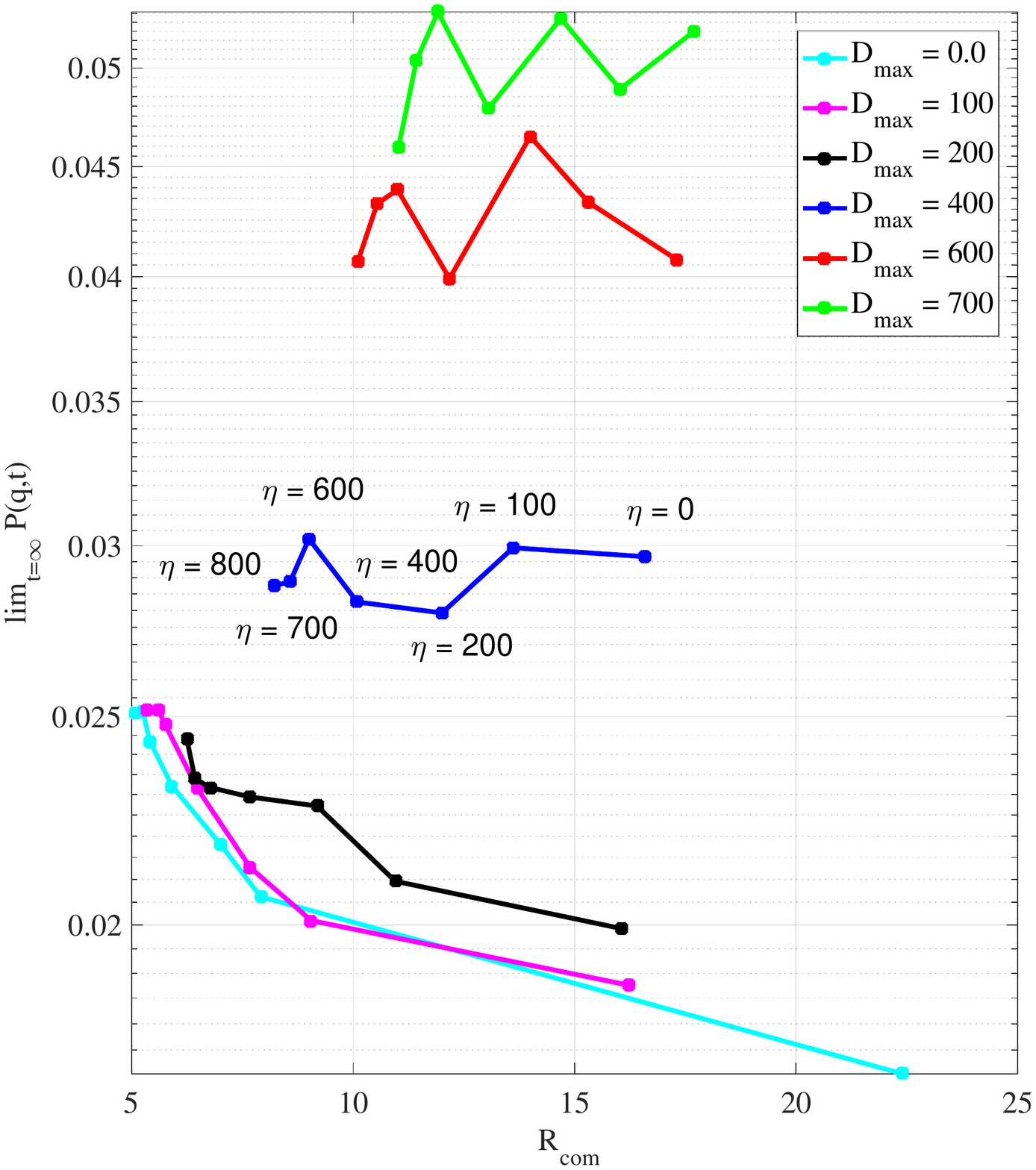}\tabularnewline
\end{tabular}
\par\end{centering}
}\subfloat[]{\begin{centering}
\begin{tabular}{c}
\includegraphics[bb=0bp 3cm 595bp 26cm,clip,width=0.45\textwidth]{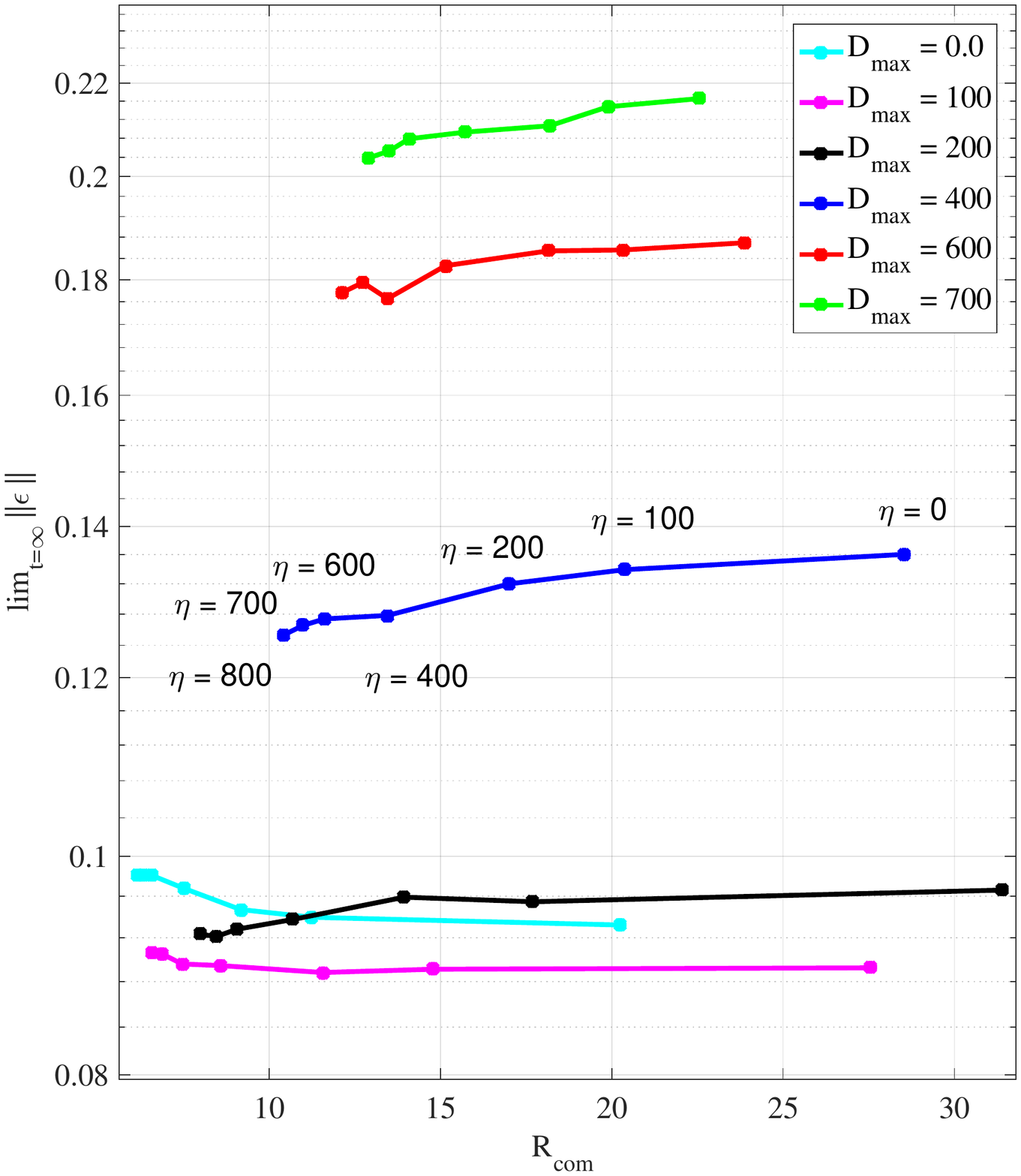}\tabularnewline
\end{tabular}
\par\end{centering}
}\caption{\label{fig:Avanced-model-tracking}Evolution of $R_{\text{com}}$,
$P\left(q,t\right)$ and $\varepsilon_{0}$ for different values of
$D_{\max}\in\left\{ \protect\begin{array}{ccccc}
0, & 100, & 200, & \ldots & 700\protect\end{array}\right\} $, $\eta\in\left\{ \protect\begin{array}{ccccc}
0, & 100, & 200, & \ldots & 800\protect\end{array}\right\} $ and $\eta_{2}=7.5$. The SS model (\ref{eq:dynamics ship}) and accurate
estimator~ (\ref{eq:estimation des coordonnes des voisins- avance})
are considered. }
\end{figure}

Figures~\ref{fig:Avance_model-tracking-3courbes} and \ref{fig:Avanced-model-tracking}
show the evolution of the communication ratio $R_{\textrm{com}}$,
the potential energy and the tracking error at $t=T$.

In Figure~\ref{fig:Avance_model-tracking-3courbes}~(a), the number
of communications obtained once the system has converged increases
as the level of perturbations becomes more important. The parameter
$\eta$ in the CTC \ref{eq:condition event-triggered ei} can help
to reduce $R_{\textrm{com}}$ . Figure~\ref{fig:Avance_model-tracking-3courbes}~(b)
and (c) show that the potential energy of the formation $P\left(q,t\right)$
and the tracking error $\varepsilon_{0}$ also increase when the perturbation
level increases. Influence of parameter $\eta$ is also illustrated
: Figure~\ref{fig:Avance_model-tracking-3courbes} (c) shows that
increasing $\eta$ results in make $\varepsilon_{0}$ decrease when
$D_{\max}>200$. Influence of $\eta$ on $P\left(q,t\right)$ is less
clearly detectable than in the case of the DI model.

In Figure~\ref{fig:Avanced-model-tracking}, it can be observed that
$R_{\text{com}}$ cannot be reduced below the value of $1$: a minimum
number of communications is indeed required to converge with the accurate
estimator~(\ref{eq:estimation des coordonnes des voisins- avance}). 

\section{Conclusion}

\label{sec:Conclusion}

This paper presents an adaptive control and event-triggered communication
strategy to reach a target formation for multi-agent systems with
perturbed Euler-Lagrange dynamics. From estimate information of agents
dynamics, an estimator has been proposed to provide the missing information
required by the control. \textcolor{black}{Each agent only require to
maintain an estimate of the state of its neighbour and communicate
with them. }Convergence to a desired formation and influence of state
perturbations on the convergence and on the amount of required communications
have been studied. Tracking control to follow an desire trajectory
has been considerate and added to the formation control. A distributed
event-triggered condition to converge to a desired formation and follow
the reference trajectory while reduce the number of communications
have been studied. Simulations have shown the effectiveness of the
proposed method in presence of state perturbations when their level
remains moderate. The time interval between two consecutive communications
by the same agent has been proved.

In future work, the considered problem will be extended to communication
delay and package drop.

\section*{Acknowledgments}

We thanks Direction Generale de l'Armement (DGA) and ICODE for a financial
support to this study.

\section{Appendix}

{\color{black}
\subsection{Characterization of a system Input-to-State practically Stable (ISpS)}
\label{def-ISPS}

Notions exposed in the section has been previously exposed in \cite{jiang1996}.

Consider the following controlled dynamical system
\begin{equation}
\dot{x}=f\left(x,u\right)\label{eq:dot(x)=00003Df(x,u)}
\end{equation}
where $x\in\mathbb{R}^{n}$, $u\in\mathbb{R}^{m}$, and $f:\mathbb{R}^{n}\times\mathbb{R}^{m}\rightarrow\mathbb{R}^{n}$
is locally Lipschitz map. 
\begin{defn}
The system (\ref{eq:dot(x)=00003Df(x,u)}) is said to be input-to-state
practically stable (ISpS) if there exist a function $\beta$ of class
$\mathcal{KL}$, a function $\mu$ of class $\mathcal{K}$ and
a non-negatie constant $\gamma$ such that, for each intial condition $x\left(0\right)$
and each measurable essentially bounded control $u\left(.\right)$
defined on $\left[0,\,\infty\right)$, the solution $x\left(.\right)$
of system (\ref{eq:dot(x)=00003Df(x,u)}) exists on $\left[0,\,\infty\right)$
and satisfies:

\begin{equation}
\left\Vert x\left(t\right)\right\Vert =\beta\left(\left\Vert x\left(0\right),t\right\Vert \right)+\mu\left(\left\Vert u\right\Vert \right)+\gamma\quad\forall t\geq0\label{eq:cond-ISpS}
\end{equation}
\end{defn}

When (\ref{eq:cond-ISpS}) is satisfied with $d=0$, the system (\ref{eq:dot(x)=00003Df(x,u)})is
said to be input-to-stable (ISS).
\begin{thm} \label{defn-ISPS}
The system (\ref{eq:dot(x)=00003Df(x,u)}) is ISpS if and only if it has an ISpS Lyapunov function $V$ for system (\ref{eq:dot(x)=00003Df(x,u)})
such that
\begin{itemize}
\item there exists functions $\psi_{1}$,
$\psi_{2}$ of class $\mathcal{K}_{\infty}$ with
\begin{equation}
\psi_{1}\left(\left\Vert x\right\Vert \right)\leq V\left(x\right)\leq\psi_{2}\left(\left\Vert x\right\Vert \right),\quad\forall x\in\mathbb{R}^{n} \label{cond1-thm-ISPS}
\end{equation}
\item there exists positive-definite functions $\Phi \in\mathcal{K}_{\infty}$ and
$\Omega\in\mathcal{K}$, and some constant $\gamma \geq 0$ with
\begin{equation}
\frac{dV\left(x\right)}{dx}f\left(x,u\right)\leq-\Phi\left(V\left(x\right)\right)+\Omega\left(\left\Vert u\right\Vert \right)+\gamma
\label{cond2-thm-ISPS}
\end{equation}
\end{itemize}
\end{thm}
These definitions will be used to prove the stability of the MAS.
}

\subsection{Proof of Theorem~\ref{Th event-triggered-part4}}

\label{subsec:Proof-of-convergence}

Consider a given value of $D_{\max}$ and $\eta$, one shows first
that the MAS is input-to-state practically stable. One then evaluates
the influence of $D_{\max}$ and $\eta$ on the behavior of the MAS. {\color{black}{For that, define a Lyapunov function and show it respects conditions defined in Theorem~\ref{defn-ISPS}}}.

\subsubsection{Proof of the input-to-state practical stability of the MAS}

Consider the continuous positive-definite candidate Lyapunov function
\begin{equation}
V=\frac{1}{2}\sum_{i=1}^{N}\left(s_{i}^{T}M_{i}s_{i}+\Delta\theta_{i}^{T}\Gamma_{i}^{-1}\Delta\theta_{i}\right)+\frac{k_{g}}{2}\left[\frac{1}{2}P\left(q,\,t\right)+\sum_{i=1}^{N}k_{0}\left\Vert q_{i}-q_{i}^{*}\right\Vert ^{2}\right]\label{eq:Lyapunov V}
\end{equation}
where $\Delta\theta_{i}=\bar{\theta}_{i}-\theta_{i}$ \textcolor{green}{.}
The time derivative of $V$ is
\begin{eqnarray}
\dot{V} & = & \sum_{i=1}^{N}\left[\frac{1}{2}s_{i}^{T}\dot{M}_{i}s_{i}+s_{i}^{T}M_{i}\dot{s}_{i}+\Delta\theta_{i}^{T}\Gamma_{i}^{-1}\dot{\bar{\theta}}_{i}\right]+\frac{k_{g}}{2}\frac{d}{dt}\left[\frac{1}{2}P\left(q,\,t\right)+\sum_{i=1}^{N}k_{0}\left\Vert q_{i}-q_{i}^{*}\right\Vert ^{2}\right]\label{eq:dot(V) - 0}
\end{eqnarray}
where, from (\ref{eq:Si}), one has $\dot{s}_{i}=\ddot{q}_{i}-\ddot{q}_{i}^{*}+k_{p}\dot{g}_{i}$.
Injecting (\ref{eq:Calcul dot(Theta)}) in (\ref{eq:dot(V) - 0})
one obtains
\begin{eqnarray}
\dot{V} & = & \sum_{i=1}^{N}\left[\frac{1}{2}s_{i}^{T}\dot{M}_{i}s_{i}+s_{i}^{T}M_{i}\dot{s}_{i}+\Delta\theta_{i}^{T}Y_{i}^T\left(q_{i},\,\dot{q}_{i},\,\dot{\bar{p}}_{i},\,\bar{p}_{i}\right)\bar{s}_{i}\right]\nonumber \\
 &  & +\frac{k_{g}}{2}\frac{d}{dt}\left[\frac{1}{2}P\left(q,\,t\right)+\sum_{i=1}^{N}k_{0}\left\Vert q_{i}-q_{i}^{*}\right\Vert ^{2}\right].\label{eq:dot(V) - 1}
\end{eqnarray}
The last term in (\ref{eq:dot(V) - 1}) may be written as
\begin{eqnarray}
 &  & \frac{1}{2}\frac{d}{dt}\left[\frac{1}{2}P\left(q,\,t\right)+\sum_{i=1}^{N}k_{0}\left\Vert q_{i}-q_{i}^{*}\right\Vert ^{2}\right]\nonumber \\
 & = & \frac{1}{4}\frac{d}{dt}\sum_{i=1}^{N}\sum_{j=1}^{N}k_{ij}\left\Vert r_{ij}-r_{ij}^{*}\right\Vert ^{2}+\frac{1}{2}\frac{d}{dt}\sum_{i=1}^{N}k_{0}\left\Vert q_{i}-q_{i}^{*}\right\Vert ^{2}\nonumber \\
 & = & \sum_{i=1}^{N}\left[\frac{1}{2}\sum_{j=1}^{N}k_{ij}\left(\dot{r}_{ij}-\dot{r}_{ij}^{*}\right)^{T}\left(r_{ij}-r_{ij}^{*}\right)+k_{0}\left(\dot{q}_{i}-\dot{q}_{i}^{*}\right)^{T}\left(q_{i}-q_{i}^{*}\right)\right]\nonumber \\
 & = & \sum_{i=1}^{N}\left[\frac{1}{2}\sum_{j=1}^{N}k_{ij}\left[\left(\dot{q}_{i}-\dot{q}_{i}^{*}\right)^{T}\left(r_{ij}-r_{ij}^{*}\right)-\left(\dot{q}_{j}-\dot{q}_{j}^{*}\right)^{T}\left(r_{ij}-r_{ij}^{*}\right)\right]\right.\nonumber \\
 &  & \left.+k_{0}\left(\dot{q}_{i}-\dot{q}_{i}^{*}\right)^{T}\left(q_{i}-q_{i}^{*}\right)\right]\nonumber \\
 & = & \sum_{i=1}^{N}\left[\frac{1}{2}\sum_{j=1}^{N}k_{ij}\left[\left(\dot{q}_{i}-\dot{q}_{i}^{*}\right)^{T}\left(r_{ij}-r_{ij}^{*}\right)-\left(\dot{q}_{i}-\dot{q}_{i}^{*}\right)^{T}\left(r_{ji}-r_{ji}^{*}\right)\right]\right.\nonumber \\
 &  & \left.+k_{0}\left(\dot{q}_{i}-\dot{q}_{i}^{*}\right)^{T}\varepsilon_{i}\right].\label{eq:deriv Pq}
\end{eqnarray}
Since $r_{ji}=-r_{ij}$, one gets, using (\ref{eq:gi})
\begin{eqnarray}
\frac{1}{2}\frac{d}{dt}\left[\frac{1}{2}P\left(q,\,t\right)+\sum_{i=1}^{N}k_{0}\left\Vert q_{i}-q_{i}^{*}\right\Vert ^{2}\right] & = & \sum_{i=1}^{N}\left(\dot{q}_{i}-\dot{q}_{i}^{*}\right)^{T}\left[\sum_{j=1}^{N}k_{ij}\left(r_{ij}-r_{ij}^{*}\right)+k_{0}\varepsilon_{i}\right]\nonumber \\
 & = & \sum_{i=1}^{N}\left(\dot{q}_{i}-\dot{q}_{i}^{*}\right)^{T}g_{i}.\label{eq:dot(P) - Final}
\end{eqnarray}
Combining (\ref{eq:dot(V) - 1}) and (\ref{eq:dot(P) - Final}), one
obtains
\begin{equation}
\dot{V}=\sum_{i=1}^{N}\left[\frac{1}{2}s_{i}^{T}\dot{M}_{i}s_{i}+s_{i}^{T}M_{i}\dot{s}_{i}+\Delta\theta_{i}^{T}Y_{i}\left(q_{i},\,\dot{q}_{i},\,\dot{\bar{p}}_{i},\,\bar{p}_{i}\right)^T\bar{s}_{i}+k_{g}\left(\dot{q}_{i}-\dot{q}_{i}^{*}\right)^{T}g_{i}\right].\label{eq:dot(V) - 2}
\end{equation}

One focuses now on the term $M_{i}\dot{s}_{i}$. Using again (\ref{eq:Si}),
one may write
\begin{equation}
M_{i}\dot{s}_{i}+C_{i}s_{i}=M_{i}\left(\ddot{q}_{i}-\ddot{q}_{i}^{*}+k_{p}\dot{g}_{i}\right)+C_{i}\left(\dot{q}_{i}-\dot{q}_{i}^{*}+k_{p}g_{i}\right)\label{eq:Transf. Ms+Cs-0}
\end{equation}
Using (\ref{eq: System dynamique Lagrange}), one gets 
\begin{eqnarray}
M_{i}\dot{s}_{i}+C_{i}s_{i} & = & \tau_{i}+d_{i}-G+M_{i}\left(k_{p}\dot{g}_{i}-\ddot{q}_{i}^{*}\right)+C_{i}\left(k_{p}g_{i}-\dot{q}_{i}^{*}\right).\label{eq:Transf. Ms+Cs-1}
\end{eqnarray}
 Now, introducing
(\ref{eq:Control input with tracking}), one gets
\begin{eqnarray}
M_{i}\dot{s}_{i}+C_{i}s_{i} & = & -k_{s}\bar{s}_{i}-k_{g}\bar{g}_{i}-Y_{i}\left(q_{i},\,\dot{q}_{i},\,k_{p}\dot{\bar{g}}_{i}-\ddot{q}_{i}^{*},k_{p}\bar{g}_{i}-\dot{q}_{i}^{*}\right)\bar{\theta}_{i}\nonumber \\
 &  & +M_{i}\left(k_{p}\dot{g}_{i}-\ddot{q}_{i}^{*}\right)+C_{i}\left(k_{p}g_{i}-\dot{q}_{i}^{*}\right)+d_{i}\label{eq:Transf. Ms+Cs}
\end{eqnarray}

In what follows, one uses $Y_{i}$ in place of $Y_{i}\left(q_{i},\,\dot{q}_{i},\,k_{p}\dot{\bar{g}}_{i}-\ddot{q}_{i}^{*},k_{p}\bar{g}_{i}-\dot{q}_{i}^{*}\right)$
to lighten notations. Since $\Delta\theta_{i}=\bar{\theta}_{i}-\theta_{i}$,
one obtains
\begin{eqnarray}
s_{i}^{T}M_{i}\dot{s}_{i} & = & -k_{s}s_{i}^{T}\bar{s}_{i}-k_{g}s_{i}^{T}\bar{g}_{i}-s_{i}^{T}C_{i}s_{i}+s_{i}^{T}\left(M_{i}\left(k_{p}\dot{g}_{i}-\ddot{q}_{i}^{*}\right)+C_{i}\left(k_{p}g_{i}-\dot{q}_{i}^{*}\right)\right)\nonumber \\
 &  & -s_{i}^{T}Y_{i}\left(\theta_{i}+\Delta\theta_{i}\right)+s_{i}^{T}d_{i}.\label{eq:Equiv. dot(si)Msi}
\end{eqnarray}
Using Assumption A3 in (\ref{eq:Equiv. dot(si)Msi})
leads to
\begin{eqnarray}
-s_{i}^{T}Y_{i}\left(\theta_{i}+\Delta\theta_{i}\right) & = & -s_{i}^{T}Y_{i}\Delta\theta_{i}-s_{i}^{T}\left(M_{i}\left(k_{p}\dot{\bar{g}}_{i}-\ddot{q}_{i}^{*}\right)+C_{i}\left(k_{p}\bar{g}_{i}-\dot{q}_{i}^{*}\right)\right).\label{eq:transf. Y}
\end{eqnarray}

Considering (\ref{eq:Prop. Yi}) and (\ref{eq:Equiv. dot(si)Msi})
in (\ref{eq:dot(V) - 2}), one gets

\begin{align}
\dot{V} & =\sum_{i=1}^{N}\left[\frac{1}{2}s_{i}^{T}\dot{M}_{i}s_{i}-k_{s}s_{i}^{T}\bar{s}_{i}-k_{g}s_{i}^{T}\bar{g}_{i}-s_{i}^{T}C_{i}s_{i}+s_{i}^{T}\left(M_{i}\left(k_{p}\dot{g}_{i}-\ddot{q}_{i}^{*}\right)+C_{i}\left(k_{p}g_{i}-\dot{q}_{i}^{*}\right)\right)\right.\nonumber \\
 & -s_{i}^{T}\left(M_{i}\left(k_{p}\dot{\bar{g}}_{i}-\ddot{q}_{i}^{*}\right)+C_{i}\left(k_{p}\bar{g}_{i}-\dot{q}_{i}^{*}\right)\right)-s_{i}^{T}Y_{i}\Delta\theta_{i}+\bar{s}_{i}^{T}Y_{i}\Delta\theta_{i}\nonumber \\
 & \left.+k_{g}\left(\dot{q}_{i}-\dot{q}_{i}^{*}\right)^{T}g_{i}+s_{i}^{T}d_{i}\right].\label{eq:VdotInterm}
\end{align}

Now, introduce (\ref{eq:gi}) in (\ref{eq:Si}) to get
\begin{equation}
s_{i}=\dot{q}_{i}-\dot{q}_{i}^{*}+k_{p}\left[\sum_{j=1}^{N}k_{ij}\left(q_{i}-q_{j}-r_{ij}^{*}\right)+k_{0}\varepsilon_{i}\right].\label{eq:Decompo Si}
\end{equation}
Since $e_{j}^{i}=\hat{q}_{j}^{i}-q_{j}$, one gets
\begin{eqnarray}
s_{i} & = & \dot{q}_{i}-\dot{q}_{i}^{*}+k_{p}\left[\sum_{j=1}^{N}k_{ij}\left(q_{i}-\hat{q}_{j}^{i}+e_{j}^{i}-r_{ij}^{*}\right)+k_{0}\varepsilon_{i}\right]\nonumber \\
 & = & \dot{q}_{i}-\dot{q}_{i}^{*}+k_{p}\left[\sum_{j=1}^{N}k_{ij}\left(\bar{r}_{ij}-r_{ij}^{*}\right)+k_{0}\varepsilon_{i}\right]+k_{p}\sum_{\begin{array}{c}
j=1\\
j\neq i
\end{array}}^{N}k_{ij}e_{j}^{i}\nonumber \\
 & = & \bar{s}_{i}+k_{p}E_{j}^{i},\label{eq:Transf. si}
\end{eqnarray}
where
\begin{equation}
E_{}^{i}=\sum_{j=1}^{N}k_{ij}e_{j}^{i}.\label{eq:Eij}
\end{equation}
Using similar derivations, one may show that 
\begin{equation}
g_{i}=\bar{g}_{i}+E_{}^{i}.\label{eq:transf.gi}
\end{equation}

Replacing (\ref{eq:Transf. si}) and (\ref{eq:transf.gi}) in (\ref{eq:VdotInterm}),
one gets

\begin{eqnarray}
\dot{V} & = & \sum_{i=1}^{N}\left[s_{i}^{T}\left[\frac{1}{2}\dot{M}_{i}-C_{i}\right]s_{i}-k_{s}s_{i}^{T}\bar{s}_{i}-k_{g}\left(\dot{q}_{i}-\dot{q}_{i}^{*}+k_{p}g_{i}\right)^{T}\bar{g}_{i}\right.\nonumber \\
 &  & \left.+k_{p}s_{i}^{T}\left(M_{i}\dot{E}_{}^{i}+C_{i}E_{}^{i}\right)-k_{p}E_{}^{iT}Y_{i}\Delta\theta_{i}+k_{g}\left(\dot{q}_{i}-\dot{q}_{i}^{*}\right)^{T}g_{i}+s_{i}^{T}d_{i}\right].\label{eq:V - 1}
\end{eqnarray}

Let 
\[
\dot{V}_{1}=\sum_{i=1}^{N}2k_{p}s_{i}^{T}\left(M_{i}\dot{E}_{}^{i}+C_{i}E_{}^{i}\right)
\]
 and 
\[
\dot{V}_{2}=-2k_{p}\sum_{i=1}^{N}E_{}^{iT}Y_{i}\Delta\theta_{i}.
\]
Using Assumption A2, $\frac{1}{2}\dot{M}_{i}-C_{i}$ is skew symmetric or definite
negative thus $s_{i}^{T}\left[\frac{1}{2}\dot{M}_{i}-C_{i}\right]s_{i}\leq0$\textcolor{blue}{.}
For all $b>0$ and all vectors $x$ and $y$ of similar size, one
has 
\begin{equation}
x^{T}y\leq\frac{1}{2}\left(bx^{T}x+\frac{1}{b}y^{T}y\right).\label{eq:Carre}
\end{equation}
Using (\ref{eq:Carre}) with $b=1$, one deduces that $d_{i}^{T}s_{i}\leq\frac{1}{2}\left(D_{\max}^{2}+s_{i}^{T}s_{i}\right)$
and (\ref{eq:gi}) that
\begin{eqnarray}
\dot{V} & \leq & \sum_{i=1}^{N}\left[-k_{s}s_{i}^{T}\bar{s}_{i}-k_{g}k_{p}g_{i}^{T}\bar{g}_{i}+\frac{1}{2}s_{i}^{T}s_{i}+\frac{1}{2}D_{\max}^{2}\right.\nonumber \\
 &  & \left.+k_{g}\left(\dot{q}_{i}-\dot{q}_{i}^{*}\right)^{T}\left(g_{i}-\bar{g}_{i}\right)\right]+\frac{1}{2}\left(\dot{V}_{1}+\dot{V}_{2}\right)\label{eq:V - 2}
\end{eqnarray}

One notices that $r_{ij}=q_{i}-q_{j}=q_{i}-\hat{q}_{j}^{i}+e_{j}^{i}=\bar{r}_{ij}+e_{j}^{i}$, using (\ref{eq:Eij})
\begin{eqnarray}
\left\Vert s_{i}-\bar{s}_{i}\right\Vert ^{2} & = & s_{i}^{T}s_{i}-2s_{i}^{T}\bar{s}_{i}+\bar{s}_{i}^{T}\bar{s}_{i}\nonumber \\
\left\Vert k_{p}E_{}^{i}\right\Vert ^{2} & = & s_{i}^{T}s_{i}-2s_{i}^{T}\bar{s}_{i}+\bar{s}_{i}^{T}\bar{s}_{i} 
\end{eqnarray}
thus
\begin{eqnarray}
s_{i}^{T}\bar{s}_{i} & = & -\frac{1}{2}\left\Vert k_{p}E_{}^{i}\right\Vert ^{2}+\frac{1}{2}s_{i}^{T}s_{i}+\frac{1}{2}\bar{s}_{i}^{T}\bar{s}_{i}\label{eq:Si*Sibar}
\end{eqnarray}

Similarly, using (\ref{eq:Si*Sibar}), one shows that
\begin{equation}
g_{i}^{T}\bar{g}_{i}=-\frac{1}{2}\left\Vert E_{}^{i}\right\Vert ^{2}+\frac{1}{2}g_{i}^{T}g_{i}+\frac{1}{2}\bar{g}_{i}^{T}\bar{g}_{i}.\label{eq:GiGib}
\end{equation}
Injecting (\ref{eq:GiGib}) in (\ref{eq:V - 2}),
\begin{eqnarray}
\dot{V} & \leq & \sum_{i=1}^{N}\left[\frac{k_{s}}{2}\left(k_{p}^{2}\left\Vert E_{}^{i}\right\Vert ^{2}-s_{i}^{T}s_{i}-\bar{s}_{i}^{T}\bar{s}_{i}\right)+k_{p}k_{g}\frac{1}{2}\left(\left\Vert E_{}^{i}\right\Vert ^{2}-g_{i}^{T}g_{i}-\bar{g}_{i}^{T}\bar{g}_{i}\right)+\frac{1}{2}s_{i}^{T}s_{i}+\frac{1}{2}D_{\max}^{2}\right.\nonumber \\
 &  & \left.+k_{g}\left(\dot{q}_{i}-\dot{q}_{i}^{*}\right)^{T}\left(g_{i}-\bar{g}_{i}\right)\right]+\frac{1}{2}\left(\dot{V}_{1}+\dot{V}_{2}\right)\nonumber \\
 & \leq & \sum_{i=1}^{N}\left[-\frac{\left(k_{s}-1\right)}{2}s_{i}^{T}s_{i}-\frac{k_{s}}{2}\bar{s}_{i}^{T}\bar{s}_{i}+\frac{k_{s}k_{p}^{2}+k_{g}k_{p}}{2}\left\Vert E_{}^{i}\right\Vert ^{2}-\frac{1}{2}k_{p}k_{g}\left(g_{i}^{T}g_{i}+\bar{g}_{i}^{T}\bar{g}_{i}\right)+\frac{1}{2}D_{\max}^{2}\right.\nonumber \\
 &  & \left.+k_{g}\left(\dot{q}_{i}-\dot{q}_{i}^{*}\right)^{T}\left(g_{i}-\bar{g}_{i}\right)\right]+\frac{1}{2}\left(\dot{V}_{1}+\dot{V}_{2}\right).\label{eq:V - 3}
\end{eqnarray}
Using (\ref{eq:Carre}) with $b=b_{i}>0$, one shows that $2(\dot{q}_{i} - \dot{q}_{i}^{*})^{T}\left(g_{i}-\bar{g}_{i}\right)\leq\left(b_{i}\left\Vert \dot{q}_{i} - \dot{q}_{i}^{*}\right\Vert ^{2}+\frac{1}{b_{i}}\left\Vert E_{}^{i}\right\Vert ^{2}\right)$.
Using this result in (\ref{eq:V - 3}), one gets 
\begin{eqnarray}
\dot{V} & \leq & \frac{1}{2}\sum_{i=1}^{N}\left[-\left(k_{s}-1\right)s_{i}^{T}s_{i}-k_{s}\bar{s}_{i}^{T}\bar{s}_{i}+\left(k_{s}k_{p}^{2}+k_{g}k_{p}+\frac{k_{g}}{b_{i}}\right)\left\Vert E_{}^{i}\right\Vert ^{2}+b_{i}k_{g}\left\Vert \dot{q}_{i}-\dot{q}_{i}^{*}\right\Vert ^{2}\right.\nonumber \\
 &  & \left.-k_{p}k_{g}\left(g_{i}^{T}g_{i}+\bar{g}_{i}^{T}\bar{g}_{i}\right)+D_{\max}^{2}\right]+\frac{1}{2}\left(\dot{V}_{1}+\dot{V}_{2}\right).\label{eq:V - 4}
\end{eqnarray}

Consider now $\dot{V}_{1}$. Using (\ref{eq:Carre}) with $b=1$ and Assumption A1, one obtains
\begin{eqnarray}
\sum_{i=1}^{N}2k_{p}s_{i}^{T}\left(M_{i}\dot{E}_{}^{i}+C_{i}E_{}^{i}\right) & \leq & \sum_{i=1}^{N}k_{p}\left(s_{i}^{T}M_{i}s_{i}+s_{i}^{T}s_{i}+\left[\dot{E}_{}^{iT}M_{i}\dot{E}_{}^{i}+E_{}^{iT}C_{i}^{T}C_{i}E_{}^{i}\right]\right)\nonumber \\
 & \leq & \sum_{i=1}^{N}k_{p}\left(\left(k_{M}+1\right)s_{i}^{T}s_{i}+\left[k_{M}\dot{E}_{}^{iT}\dot{E}_{}^{i}+E_{}^{iT}C_{i}^{T}C_{i}E_{}^{i}\right]\right)\label{eq:Maj V1}
\end{eqnarray}

Focus now on the terms $E_{}^{iT}C_{i}^{T}C_{i}E_{}^{i}$. Using Assumption A2, one has
{\color{black}{

\begin{equation}
x^{T}C_{i}x	\leq x^{T}x\lambda_{\max}\left(C_{i}\right)
	\leq x^{T}x\left(k_{C}\left\Vert \dot{q}_{i}\right\Vert \right)
\end{equation}
and
\begin{equation}
	x^{T}C_{i}^{T}C_{i}x	\leq x^{T}x\left(\lambda_{\max}\left(C_{i}\right)\right)^{2}
	\leq x^{T}x\left(k_{C}\left\Vert \dot{q}_{i}\right\Vert \right)^{2}
\end{equation}

Then
\begin{eqnarray}
\sum_{i=1}^{N}\left(\sum_{j=1}^{N}k_{ij}e_{j}^{i}\right)^{T}C_{i}^{T}C_{i}\left(\sum_{\ell=1}^{N}k_{i\ell}e_{\ell}^{i}\right)	=\sum_{i=1}^{N}X^{T}C_{i}^{T}C_{i}Y
\end{eqnarray}

where $X=\left(\sum_{j=1}^{N}k_{ij}e_{j}^{i}\right)=\left(\sum_{\ell=1}^{N}k_{i\ell}e_{\ell}^{i}\right)=Y$, so
\begin{eqnarray}
\sum_{i=1}^{N}\left(\sum_{j=1}^{N}k_{ij}e_{j}^{i}\right)^{T}C_{i}^{T}C_{i}\left(\sum_{\ell=1}^{N}k_{i\ell}e_{\ell}^{i}\right)	&=& \sum_{i=1}^{N}X^{T}C_{i}^{T}C_{i}X \\
	&=& \sum_{i=1}^{N}X^{T}X\left(\lambda_{\max}\left(C_{i}\right)\right)^{2}  \\
	&\leq& \sum_{i=1}^{N}\left(\lambda_{\max}\left(C_{i}\right)\right)^{2}\left(\sum_{\ell=1}^{N}k_{i\ell}e_{\ell}^{i}\right)\left(\sum_{\ell=1}^{N}k_{i\ell}e_{\ell}^{i}\right) \\
	&\leq& \sum_{i=1}^{N}\sum_{j=1}^{N}\sum_{\ell=1}^{N}k_{i\ell}k_{ij}e_{j}^{iT}e_{\ell}^{i}\left(\lambda_{\max}\left(C_{i}\right)\right)^{2} \\
	&\leq& \sum_{i=1}^{N}\sum_{j=1}^{N}\sum_{\ell=1}^{N}k_{i\ell}k_{ij}e_{j}^{iT}e_{\ell}^{i}\left(k_{C}\left\Vert \dot{q}_{i}\right\Vert \right)^{2}
\label{eq:Terme ECCE}
\end{eqnarray}
}}
Remind $\alpha_{i}=\sum_{j=1}^{N}k_{ij}$ and $\alpha_{\text{M}}=\max_{i=1,\dots,N}\alpha_{i}$. Using (\ref{eq:Carre}) with $b=1$, one gets
\begin{eqnarray}
\sum_{i=1}^{N}E_{}^{iT}C_{i}^{T}C_{i}E_{}^{i} & \leq & \frac{1}{2}\sum_{i=1}^{N}\sum_{j=1}^{N}\sum_{\ell=1}^{N}k_{i\ell}k_{ij}k_{C}^{2}\left\Vert \dot{q}_{j}\right\Vert ^{2}\left(e_{j}^{iT}e_{j}^{i}+e_{\ell}^{iT}e_{\ell}^{i}\right)\nonumber \\
 & \leq & \sum_{i=1}^{N}\sum_{j=1}^{N}\sum_{\ell=1}^{N}k_{i\ell}k_{ij}k_{C}^{2}\left\Vert \dot{q}_{j}\right\Vert ^{2}\left(e_{j}^{iT}e_{j}^{i}\right)\nonumber \\
 & \leq & \sum_{i=1}^{N}\alpha_{i}\sum_{j=1}^{N}k_{ij}k_{C}^{2}\left\Vert \dot{q}_{j}\right\Vert ^{2}\left(e_{j}^{iT}e_{j}^{i}\right).\label{eq:Maj ECCE}
\end{eqnarray}

Since one has assumed that (\ref{eq:synchronization estim})-(\ref{eq:reset estimator})
are satisfied, one has $\forall i\in\mathcal{N}_{j}$ $\hat{q}_{j}^{i}=\hat{q}_{j}^{j}$ and $e_{j}^{i}=e_{j}^{j}$. Moreover, since $k_{ij}=k_{ji}$ and $k_{ij}=0$ if $(i,j)\in\mathcal{N}_{j}$, one may write
\begin{equation}
\sum_{i=1}^{N}\sum_{j=1}^{N}k_{ij}\left\Vert e_{j}^{i}\right\Vert ^{2}=\sum_{i=1}^{N}\sum_{j=1}^{N}k_{ij}\left\Vert e_{j}^{j}\right\Vert ^{2}=\sum_{i=1}^{N}\sum_{j=1}^{N}k_{ji}\left\Vert e_{i}^{i}\right\Vert ^{2}.\label{eq:Trans. Sum_ei-1-1}
\end{equation}
 Injecting (\ref{eq:Trans. Sum_ei-1-1}) in (\ref{eq:Maj ECCE}),
\begin{eqnarray}
\sum_{i=1}^{N}E_{}^{iT}C_{i}^{T}C_{i}E_{}^{i}
 & \leq & \sum_{i=1}^{N}\left(\alpha_{\text{M}}\sum_{j=1}^{N}\left[k_{ij}\left\Vert e_{i}^{i}\right\Vert ^{2}k_{C}^{2}\left\Vert \dot{q}_{j}\right\Vert ^{2}\right]\right).\label{eq:Maj ECCE - 2}
\end{eqnarray}
Thank to the second CTC (\ref{eq:Second CTC}), one has
\begin{eqnarray}
\sum_{i=1}^{N}E_{}^{iT}C_{i}^{T}C_{i}E_{}^{i} & \leq & \sum_{i=1}^{N}\left(\alpha_{\text{M}}k_{C}^{2}\left\Vert e_{i}^{i}\right\Vert ^{2}\sum_{j=1}^{N}k_{ij}\left(\left\Vert \dot{\hat{q}}_{j}^{i}\right\Vert +\eta_{2}\right)^{2}\right).\label{eq:Maj Eji}
\end{eqnarray}
Similarly, one shows that 
\begin{equation}
\sum_{i=1}^{N}E_{}^{iT}E_{}^{i}\leq\sum_{i=1}^{N}\alpha_{\text{M}}^{2}\left\Vert e_{i}^{i}\right\Vert ^{2} \label{Ei-maj}
\end{equation}
 and 
\begin{equation}
\sum_{i=1}^{N}\dot{E}_{}^{iT}\dot{E}_{}^{i}\leq\sum_{i=1}^{N}\alpha_{\text{M}}^{2}\left\Vert \dot{e}_{i}^{i}\right\Vert ^{2}. \label{dotEi-maj}
\end{equation}

Consider now $\dot{V}_{2}$ 
\begin{align}
\dot{V}_{2} & = -2k_{p}\sum_{i=1}^{N}E_{}^{iT}Y_{i}\Delta\theta_{i}\nonumber \\
 & = -2k_{p}\sum_{i=1}^{N}\left(\sum_{j=1}^{N}k_{ij}e_{j}^{i}\right)^{T}Y_{i}\Delta\theta_{i}.\label{eq:dot(V2)}
\end{align}
\textcolor{black}{Thank to (\ref{eq:synchronization estim}) and $k_{ij}=0$, one has $\forall j\in\mathcal{N}_{i}$, $k_{ij}e_{j}^{i}=k_{ij}e_{j}^{j}$
}. One gets
\begin{align}
\dot{V}_{2} & =-2k_{p}\sum_{i=1}^{N}\left(\sum_{j=1}^{N}k_{ij}e_{j}^{j}\right)^{T}Y_{i}\Delta\theta_{i}\nonumber \\
 & =-2k_{p}\sum_{j=1}^{N}\sum_{i=1}^{N}\left(k_{ji}e_{i}^{i}\right)^{T}Y_{j}\Delta\theta_{j}\nonumber \\
 & =-2k_{p}\sum_{i=1}^{N}e_{i}^{i}{}^{T}\sum_{j=1}^{N}k_{ji}Y_{j}\Delta\theta_{j}.\label{eq:dot(V2)-1}
\end{align}
Let $0_{n}=\left[0,\,\ldots\,0\right]^{T}\in\mathbb{R}^{n}$ be the
all-zero vector. If $e_{i}^{i}=0_{n}$, one has $2k_{p}e_{i}^{i}{}^{T}\sum_{j=1}^{N}k_{ji}Y_{j}\Delta\theta_{j}=0$.
Considering now the case $e_{i}^{i}\neq0_{n}$. Using (\ref{eq:Carre})
with $b=b_{i2}>0$, one obtains
\begin{align}
\dot{V}_{2} & =-2k_{p}\sum_{i=1}^{N}E_{}^{iT}Y_{i}\Delta\theta_{i}\label{eq:Maj V2-1}\\
 & \leq k_{p}\sum_{i=1}^{N}\left(b_{i2}E_{}^{iT}E_{}^{i}+\frac{1}{b_{i2}}\left\Vert Y_{i}\Delta\theta_{i}\right\Vert ^{2}\right).
\end{align}
Since (\ref{Ei-maj}) and $|\Delta \theta_i|\leq  \Delta\theta_{i,\max}$,
one gets
\begin{align}
\dot{V}_{2} & \leq\sum_{i=1}^{N}k_{p}\left(\alpha_{\text{M}}^{2}b_{i2}\left\Vert e_{i}^{i}\right\Vert ^{2}+\frac{1}{b_{i2}}\left\Vert \left|Y_{i}\right|\left|\Delta\theta_{i}\right|\right\Vert ^{2}\right)\nonumber \\
 & \leq\sum_{i=1}^{N}k_{p}\left(\alpha_{\text{M}}^{2}b_{i2}\left\Vert e_{i}^{i}\right\Vert ^{2}+\frac{1}{b_{i2}}\left\Vert \left|Y_{i}\right|\Delta\theta_{i,\max}\right\Vert ^{2}\right),\label{eq:dot(V2)-3}
\end{align}
where $\Delta\theta_{i,\max}$ is given by (\ref{eq:Deltatheta}) {\color{black} and A4}.

Since $e_{i}^{i}\neq0_{n}$, choosing $b_{i2}=\frac{1+\left\Vert \left|Y_{i}\right|\Delta\theta_{i,\max}\right\Vert ^{2}}{\left\Vert e_{i}^{i}\right\Vert }$,
one obtains $\dot{V}_{2}\leq\dot{V}_{3}$ with
\begin{align}
\dot{V}_{3} & =\sum_{i=1}^{N}k_{p}\left[\alpha_{\text{M}}^{2}\left(\frac{1+\left\Vert \left|Y_{i}\right|\Delta\theta_{i,\max}\right\Vert ^{2}}{\left\Vert e_{i}^{i}\right\Vert }\right)\left\Vert e_{i}^{i}\right\Vert ^{2}+\frac{\left\Vert e_{i}^{i}\right\Vert \left\Vert \left|Y_{i}\right|\Delta\theta_{i,\max}\right\Vert ^{2}}{\left(1+\left\Vert \left|Y_{i}\right|\Delta\theta_{i,\max}\right\Vert ^{2}\right)}\right]\nonumber \\
 & =\sum_{i=1}^{N}k_{p}\left\Vert e_{i}^{i}\right\Vert \left[\alpha_{\text{M}}^{2}\left(1+\left\Vert \left|Y_{i}\right|\Delta\theta_{i,\max}\right\Vert ^{2}\right)+\frac{\left\Vert \left|Y_{i}\right|\Delta\theta_{i,\max}\right\Vert ^{2}}{\left(1+\left\Vert \left|Y_{i}\right|\Delta\theta_{i,\max}\right\Vert ^{2}\right)}\right].\label{eq:V3}
\end{align}

Injecting (\ref{eq:Maj V1}), (\ref{eq:Maj Eji}), and (\ref{eq:V3})
in (\ref{eq:V - 4}), one gets
\begin{eqnarray}
\dot{V} & \leq & \frac{1}{2}\sum_{i=1}^{N}\left[-\left(k_{s}-1-k_{p}\left(k_{M}+1\right)\right)s_{i}^{T}s_{i}-k_{s}\bar{s}_{i}^{T}\bar{s}_{i}+D_{\max}^{2}\right.\nonumber \\
 &  & -k_{p}k_{g}g_{i}^{T}g_{i}-k_{p}k_{g}\bar{g}_{i}^{T}\bar{g}_{i}+k_{g}b_{i}\left\Vert \dot{q}_{i}-\dot{q}_{i}^{*}\right\Vert ^{2}+k_{p}k_{M}\alpha_{\text{M}}^{2}\left\Vert \dot{e}_{i}^{i}\right\Vert ^{2}\nonumber \\
 &  & \left.+\alpha_{\text{M}}^{2}\left(k_{s}k_{p}^{2}+k_{g}k_{p}+\frac{k_{g}}{b_{i}}\right)\left\Vert e_{i}^{i}\right\Vert ^{2}+\alpha_{\text{M}}k_{p}k_{C}^{2}\left\Vert e_{i}^{i}\right\Vert ^{2}\sum_{j=1}^{N}k_{ij}\left[\left\Vert \dot{\hat{q}}_{j}^{i}\right\Vert +\eta_{2}\right]^{2}\right]+\frac{1}{2}\dot{V}_{3}.\label{eq:dot(V)+v3}
\end{eqnarray}

The CTC (\ref{eq:condition event-triggered ei}) makes sure that
\begin{align}
\dot{V} & \leq\frac{1}{2}\sum_{i=1}^{N}\left[-\left(k_{s}-1-k_{p}\left(k_{M}+1\right)\right)s_{i}^{T}s_{i}-k_{g}k_{p}g_{i}^{T}g_{i}+D_{\max}^{2}+\eta\right]\nonumber \\
\dot{V} & \leq\frac{1}{2}\sum_{i=1}^{N}\left[-k_{1}s_{i}^{T}s_{i}-k_{g}k_{p}g_{i}^{T}g_{i}+D_{\max}^{2}+\eta\right]\label{eq:dot(V) - Final}
\end{align}
with $k_{1}=k_{s}-1-k_{p}\left(k_{M}+1\right)$. 

Following the steps given in Appendix~\ref{subsec:Differential-equation-of-V}
from (\ref{eq:equa diff - etape1}) to (\ref{eq:equa diff V - final-2}),
one shows that
\begin{equation}
\dot{V}\leq-c_{3}V+\frac{N}{2}\left[D_{\max}^{2}+\eta\right]+\frac{c_{3}}{2}\sum_{i=1}^{N}\left(\Delta\theta_{i}{}^{T}\Gamma_{i}^{-1}\Delta\theta_{i}\right),\label{eq:equa diff V - final 3}
\end{equation}
where $c_{3}>0$ is a positive constant. Introducing $\Delta_{\max}=\max_{i=1:N}\left(\sup_{t>0}\left(\Delta\theta_{i}^{T}\Gamma_{i}^{-1}\Delta\theta_{i}\right)\right)$,
one has
\begin{equation}
\dot{V}\leq-c_{3}V+\frac{N}{2}\left[c_{3}\Delta_{\max}+D_{\max}^{2}+\eta\right].\label{eq:equa diff V - final}
\end{equation}

Define the function $W$ such that $W\left(0\right)=V\left(0\right)$
and
\begin{equation}
\dot{W}=-c_{3}W+\frac{N}{2}\left[D_{\max}^{2}+\eta+c_{3}\Delta_{\max}\right].\label{eq:dot(W)}
\end{equation}
Using the initial condition $W\left(0\right)=V\left(0\right)$, the
solution of (\ref{eq:dot(W)}) is 
\begin{equation}
W\left(t\right)=\exp\left(-c_{3}t\right)V\left(0\right)+\left(1-\exp\left(-c_{3}t\right)\right)\frac{N}{2c_{3}}\left[D_{\max}^{2}+\eta+c_{3}\Delta_{\max}\right].\label{eq:W}
\end{equation}
Then, using the \cite{NonlinearSystem}, Lemma 3.4 (Comparison lemma),
one has $V\left(t\right)\leq W\left(t\right)$ and so
\begin{equation}
V\left(t\right)\leq\exp\left(-c_{3}t\right)V\left(0\right)+\left(1-\exp\left(-c_{3}t\right)\right)\frac{N}{2c_{3}}\left[D_{\max}^{2}+\eta+c_{3}\Delta_{\max}\right]\label{eq:Maj V expo}
\end{equation}

Since $M_{i}$ and $\Gamma_{i}$ are symmetric, there exists matrices
$S_{M_{i}}$ and $S_{\Gamma_{i}}$ such that $M_{i}=S_{M_{i}}^{T}S_{M_{i}}$
and $\Gamma_{i}=S_{\Gamma_{i}}^{T}S_{\Gamma_{i}}$. Introduce now
\begin{align}
y_{M} & =\left[\begin{array}{ccccc}
\left(S_{M_{1}}s_{1}\right)^{T} & \ldots & \left(S_{M_{i}}s_{i}\right)^{T} & \ldots & \left(S_{M_{N}}s_{N}\right)^{T}\end{array}\right]^{T}\label{eq:yM-part4}\\
y_{\Gamma} & =\left[\begin{array}{ccccc}
\left(S_{\Gamma_{1}}^{-1}\Delta\theta_{1}\right)^{T} & \ldots & \left(S_{\Gamma_{i}}^{-1}\Delta\theta_{i}\right)^{T} & \ldots & \left(S_{\Gamma_{N}}^{-1}\Delta\theta_{N}\right)^{T}\end{array}\right]^{T}\label{eq:yT-part4}\\
y_{q} & =\left[\begin{array}{ccccc}
\left(q_{1}-q_{1}^{*}\right)^{T} & \ldots & \left(q_{i}-q_{i}^{*}\right)^{T} & \ldots & \left(q_{N}-q_{N}^{*}\right)^{T}\end{array}\right]^{T}\label{eq:yq-part4}\\
z & =\left[\begin{array}{cccc}
y_{M}^{T} & y_{\Gamma}^{T} & \sqrt{k_{g}k_{0}}y_{q}^{T} & \sqrt{\frac{k_{g}}{2}P\left(x,\,t\right)}\end{array}\right]^{T}\label{eq:z-part4}
\end{align}

{\color{black}
Then, $V\left(t\right)$ can be written as
\begin{equation}
V\left(z\left(t\right)\right)=\frac{1}{2}z\left(t\right)^{T}z\left(t\right)
\end{equation}
of the system
\begin{equation}
\dot{z}\left(t\right)=f\left(z\left(t\right),u_{z}\right)
\end{equation}
with $u_{z}=0$ and
\begin{align}
f\left(z, uz\right)=\left[\begin{array}{cccc}
\left(\frac{d}{dt}y_{M}\right)^{T} & \left(\frac{d}{dt}y_{\Gamma}\right)^{T} & \sqrt{k_{g}k_{0}}\left(\frac{d}{dt}y_{q}\right)^{T} & \frac{d}{dt}\left(\sqrt{\frac{k_{g}}{2}P\left(x,\,t\right)}\right)^{T}\end{array}\right]^{T}
\end{align}
with
\begin{eqnarray}
\frac{d}{dt}y_{M}	&=&\left[\begin{array}{ccccc}
\left(S_{M_{1}}\dot{s}_{1}\right)^{T} & \ldots & \left(S_{M_{i}}\dot{s}_{i}\right)^{T} & \ldots & \left(S_{M_{N}}\dot{s}_{N}\right)^{T}\end{array}\right]^{T} \\
\frac{d}{dt}y_{\Gamma}	&=&\left[\begin{array}{ccccc}
\left(S_{\Gamma_{1}}^{-1}\left(\dot{\bar{\theta}}_{1}-\dot{\theta}_{1}\right)\right)^{T} & \ldots & \left(S_{\Gamma_{i}}^{-1}\left(\dot{\bar{\theta}}_{i}-\dot{\theta}_{i}\right)\right)^{T} & \ldots & \left(S_{\Gamma_{N}}^{-1}\left(\dot{\bar{\theta}}_{N}-\dot{\theta}_{N}\right)\right)^{T}\end{array}\right]^{T} \\
\frac{d}{dt}y_{q}	&=& \left[\begin{array}{ccccc}
\left(\dot{q}_{1}-\dot{q}_{1}^{*}\right)^{T} & \ldots & \left(\dot{q}_{i}-\dot{q}_{i}^{*}\right)^{T} & \ldots & \left(\dot{q}_{N}-\dot{q}_{N}^{*}\right)^{T}\end{array}\right]^{T} \\
\frac{d}{dt}\sqrt{\frac{k_{g}}{2}P\left(x,\,t\right)}	&=& \sqrt{\frac{k_{g}}{2}}\left(\frac{\sum_{j=1}^{N}k_{ij}\left(\dot{r}_{ij}-\dot{r}_{ij}^{*}\right)^{T}\left(r_{ij}-r_{ij}^{*}\right)}{\sqrt{P\left(x,\,t\right)}}\right).
\end{eqnarray}

Then, one can observe
\begin{equation}
\psi_{1}\left(\left\Vert z\left(t\right)\right\Vert \right)\leq V\left(t\right)\leq\psi_{2}\left(\left\Vert z\left(t\right)\right\Vert \right)\label{eq:cond1-ISPS}
\end{equation}
where $\psi_{1}\left(\left\Vert z\left(t\right)\right\Vert \right)=\frac{1}{4}\left\Vert z\left(t\right)\right\Vert ^{2}$
and $\psi_{2}\left(\left\Vert z\left(t\right)\right\Vert \right)=\left\Vert z\left(t\right)\right\Vert ^{2}$.
Then, one has
\begin{align}
\frac{d}{dz}V\left(t\right)f\left(z\left(t\right),u_{z}\right) & \leq-c_{3}V\left(t\right)+\frac{N}{2}\left[c_{3}\Delta_{\max}+D_{\max}^{2}+\eta\right]\nonumber \\
 & \leq-\Phi\left(\left\Vert z\left(t\right)\right\Vert \right)+\gamma\label{eq:cond2-ISPS}
\end{align}
where $\Phi\left(\left\Vert z\left(t\right)\right\Vert \right)=\frac{c_{3}}{2}\left\Vert z\left(t\right)\right\Vert ^{2}$,
$\theta\left(\left\Vert u_{z}\right\Vert \right)=u_{z}$ and $\gamma=\frac{N}{2}\left[c_{3}\Delta_{\max}+D_{\max}^{2}+\eta\right]$
. 

Consequently, (\ref{eq:cond1-ISPS})-(\ref{eq:cond2-ISPS})
satisfy (\ref{cond1-thm-ISPS})-(\ref{cond2-thm-ISPS}), which  implies $V\left(t\right)$ is an ISpS-lyapunov, thus the MAS is input-to-state
practically stable.
}

\subsubsection{Convergence of $V$}

From previous section, we have shown the system is ISpS. One has
\begin{align}
\dot{V} & \leq-c_{3}V+\frac{N}{2}\left[D_{\max}^{2}+\eta\right]+\frac{c_{3}}{2}\sum_{i=1}^{N}\left(\Delta\theta_{i}{}^{T}\Gamma_{i}^{-1}\Delta\theta_{i}\right)\label{eq:dot(V) < c3 V}
\end{align}
Then, if initially 
\begin{equation}
-c_{3}V\left(0\right)+\frac{N}{2}\left[D_{\max}^{2}+\eta\right]+\frac{c_{3}}{2}\sum_{i=1}^{N}\left(\Delta\theta_{i}{}^{T}\Gamma_{i}^{-1}\Delta\theta_{i}\right)<0\label{eq:dot(V) < c3 V(0)}
\end{equation}
one has $\dot{V}\leq0$ and $V$ is decreasing. Then, one has from
(\ref{eq:Maj V expo})
\begin{align}
\lim_{t\to\infty}V\left(t\right) & \leq\frac{N}{2c_{3}}\left[D_{\max}^{2}+\eta+c_{3}\Delta_{\max}\right]\nonumber
\end{align}
\begin{align}
\lim_{t\to\infty}\frac{1}{2}\sum_{i=1}^{N}\left(s_{i}^{T}M_{i}s_{i}+\Delta\theta_{i}{}^{T}\Gamma_{i}^{-1}\Delta\theta_{i}\right)+\frac{k_{g}}{2}\left(\sum_{i=1}^{N}k_{0}\left\Vert \varepsilon_{i}\right\Vert ^{2}+\frac{1}{2}P\left(q,t\right)\right) & \leq\frac{N}{2c_{3}}\left[D_{\max}^{2}+\eta+c_{3}\Delta_{\max}\right]\nonumber 
\end{align}
\begin{align}
\lim_{t\to\infty}\frac{k_{g}}{2}\left(\sum_{i=1}^{N}k_{0}\left\Vert \varepsilon_{i}\right\Vert ^{2}+\frac{1}{2}P\left(q,t\right)\right) & \leq\frac{N}{2c_{3}}\left[D_{\max}^{2}+\eta+c_{3}\Delta_{\max}\right]\nonumber -\lim_{t\to\infty}\frac{1}{2}\sum_{i=1}^{N}\left(s_{i}^{T}M_{i}s_{i}+\Delta\theta_{i}{}^{T}\Gamma_{i}^{-1}\Delta\theta_{i}\right)\nonumber 
 \end{align}
 \begin{align}
\lim_{t\to\infty}\sum_{i=1}^{N}k_{0}\left\Vert \varepsilon_{i}\right\Vert ^{2}+\frac{1}{2}P\left(q,t\right) & \leq\frac{N}{k_{g}c_{3}}\left[D_{\max}^{2}+\eta+c_{3}\Delta_{\max}\right].\label{eq:lim P + sum(ri)}
\end{align}
Asymptotically, the formation and tracking error are bounded.

\subsection{Showing $t_{i,k+1} - t_{i,k}$}
\label{subsec:Proof-of-absence-Zeno}

According to (\ref{eq:condition event-triggered ei}) and (\ref{eq:Second CTC}), a communication
is triggered at $t=t_{i,k}^{-}$ when
{\color{black}{
\begin{eqnarray}
\left\Vert \dot{q}_{i}\right\Vert  & \geq & \left\Vert \dot{\hat{q}}_{i}^{i}\right\Vert +\eta_{2}
\end{eqnarray}
or }}
\begin{align}
k_{s}\bar{s}_{i}^{T}\bar{s}_{i}+k_{p}k_{g}\bar{g}_{i}^{T}\bar{g}_{i}+\eta & =\alpha_{\text{M}}^{2}\left(k_{e}\left\Vert e_{i}^{i}\right\Vert ^{2}+k_{p}k_{M}\left\Vert \dot{e}_{i}^{i}\right\Vert ^{2}\right)\nonumber \\
 & +\alpha_{\text{M}}k_{C}^{2}k_{p}\left\Vert e_{i}^{i}\right\Vert ^{2}\sum_{j=1}^{N}k_{ji}\left[\left\Vert \dot{\hat{q}}_{j}^{i}\right\Vert +\eta_{2}\right]^{2}+k_{g}b_{i}\left\Vert \dot{q}_{i}-\dot{q}_{i}^{*}\right\Vert ^{2}\nonumber \\
 & +k_{p}\left\Vert e_{i}^{i}\right\Vert \left[\alpha_{\text{M}}^{2}\left(1+\left\Vert \left|Y_{i}\right|\Delta\theta_{i,\max}\right\Vert ^{2}\right)+\frac{\left\Vert \left|Y_{i}\right|\Delta\theta_{i,\max}\right\Vert ^{2}}{\left(1+\left\Vert \left|Y_{i}\right|\Delta\theta_{i,\max}\right\Vert ^{2}\right)}\right]\label{eq:preuve zeno-0-part4}
\end{align}
with $k_{e}=\left(k_{s}k_{p}^{2}+k_{g}k_{p}+\frac{k_{g}}{b_{i}}\right)$.
Then, the estimation errors $e_{i}^{i}$ and $\dot{e}_{i}^{i}$ are
reset and one has $e_{i}^{i}\left(t_{i,k}^{+}\right)=0$ and $\dot{e}_{i}^{i}\left(t_{i,k}^{+}\right)=0$.
As a consequence, 
{\color{black}{(\ref{eq:Second CTC})
in Theorem~\ref{Th event-triggered-part4} is not satisfied at $t=t_{i,k}^{+}$
iff
\begin{align}
\left\Vert \dot{q}_{i}\right\Vert  & <  \left\Vert \dot{q}_{i} + e^{i}_i(t_{i,k+1}^{+})\right\Vert +\eta_{2}  \nonumber \\
\left\Vert \dot{q}_{i}\right\Vert  & <  \left\Vert \dot{q}_{i}\right\Vert +\eta_{2} \nonumber \\
0 & <  \eta_{2}. 
\end{align}
Deduce (\ref{eq:Second CTC}) is not satisfied when $t=t_{i,k+1}^{+}$.

In the same way,}}
(\ref{eq:condition event-triggered ei})
in Theorem~\ref{Th event-triggered-part4} is not satisfied at $t=t_{i,k}^{+}$
iff
\begin{equation}
k_{s}\bar{s}_{i}^{T}\bar{s}_{i}+k_{p}k_{g}\bar{g}_{i}^{T}\bar{g}_{i}+\eta>k_{g}b_{i}\left\Vert \dot{q}_{i}-\dot{q}_{i}^{*}\right\Vert ^{2}.\label{eq:CTC apres declenchement}
\end{equation}
To prove $t_{i,k+1}>t_{i,k}$, one has to show that (\ref{eq:CTC apres declenchement})
is satisfied.

Using (\ref{eq:Carre})
for some $b_{i2}>0$, one deduces that
\begin{align}
\bar{s}_{i}^{T}\bar{s}_{i} & =k_{p}^{2}\bar{g}_{i}^{T}\bar{g}_{i}+\left\Vert \dot{q}_{i}-\dot{q}_{i}^{*}\right\Vert ^{2}+2k_{p}\bar{g}_{i}^{T}\left(\dot{q}_{i}-\dot{q}_{i}^{*}\right)\nonumber \\
 & \geq\left(k_{p}^{2}-k_{p}b_{i2}\right)\bar{g}_{i}^{T}\bar{g}_{i}+\left(1-\frac{k_{p}}{b_{i2}}\right)\left\Vert \dot{q}_{i}-\dot{q}_{i}^{*}\right\Vert ^{2}.\label{eq:Minoration norm(si)}
\end{align}

Using (\ref{eq:Minoration norm(si)}), a sufficient condition for
(\ref{eq:CTC apres declenchement}) to be satisfied is 

\begin{align}
k_{s}\left(k_{p}^{2}-k_{p}b_{i2}\right)\bar{g}_{i}^{T}\bar{g}_{i}+k_{s}\left(1-\frac{k_{p}}{b_{i2}}\right)\left\Vert \dot{q}_{i}-\dot{q}_{i}^{*}\right\Vert ^{2}+k_{p}k_{g}\bar{g}_{i}^{T}\bar{g}_{i}+\eta & >k_{g}b_{i}\left\Vert \dot{q}_{i}-\dot{q}_{i}^{*}\right\Vert ^{2}\nonumber \\
k_{s}\left(1-\frac{k_{p}}{b_{i2}}\right)\left\Vert \dot{q}_{i}-\dot{q}_{i}^{*}\right\Vert ^{2}+\left[k_{p}k_{g}+k_{s}\left(k_{p}^{2}-k_{p}b_{i2}\right)\right]\bar{g}_{i}^{T}\bar{g}_{i}+\eta & >k_{g}b_{i}\left\Vert \dot{q}_{i}-\dot{q}_{i}^{*}\right\Vert ^{2}\nonumber \\
k_{1}\bar{g}_{i}^{T}\bar{g}_{i}+\eta & >k_{2}\left\Vert \dot{q}_{i}-\dot{q}_{i}^{*}\right\Vert ^{2}\label{eq:Inegalite zeno}
\end{align}
where $k_{1}=\left[k_{p}k_{g}+k_{s}\left(k_{p}^{2}-k_{p}b_{i2}\right)\right]$
and $k_{2}=\left[k_{g}b_{i}-k_{s}\left(1-\frac{k_{p}}{b_{i2}}\right)\right]$.
To ensure that the inequality (\ref{eq:Inegalite zeno}) is satisfied
independently of the values of $\bar{g}_{i}$ and $\dot{q}_{i}-\dot{q}_{i}^{*}$,
it is sufficient to find $b_{i}$ and $b_{i2}$ such that $k_{1}>0$
and $k_{2}<0$. Consider first $k_{1}$.
\begin{align}
k_{p}k_{g}+k_{s}\left(k_{p}^{2}-k_{p}b_{i2}\right) & >0\nonumber \\
\frac{k_{g}}{k_{s}} & >\left(-k_{p}+b_{i2}\right)\nonumber \\
\frac{k_{s}k_{p}+k_{g}}{k_{s}} & >b_{i2}.\label{eq:Condition 1 sur bi2}
\end{align}
Focus now on $k_{2}$
\begin{align}
k_{g}b_{i}-k_{s}\left(1-\frac{k_{p}}{b_{i2}}\right) & <0\nonumber \\
\frac{k_{g}b_{i}}{k_{s}} & <1-\frac{k_{p}}{b_{i2}}\nonumber \\
\frac{1}{b_{i2}} & <\frac{1}{k_{p}}\left(1-\frac{k_{g}b_{i}}{k_{s}}\right).\label{eq:Condition 2 sur bi2}
\end{align}
Since $b_{i2}>0$, one has $\frac{k_{g}b_{i}}{k_{s}}<1$ and so $b_{i}<\frac{k_{s}}{k_{g}}$.
Then
\begin{align}
\frac{k_{s}k_{p}}{k_{s}-k_{g}b_{i}} & <b_{i2}.\label{eq:min(bi2)}
\end{align}
Finally, one has to find a condition on $b_{i}$ such that (\ref{eq:Condition 1 sur bi2})
and (\ref{eq:Condition 2 sur bi2}) can be satisfied simultaneously

\begin{align}
\frac{k_{s}k_{p}+k_{g}}{k_{s}} & >b_{i2}>\frac{k_{s}k_{p}}{k_{s}-k_{g}b_{i}}.\label{eq:condition bi2 full}
\end{align}
One may find $b_{i2}$ if
\begin{align}
k_{s}-k_{g}b_{i} & >\frac{k_{s}^{2}k_{p}}{k_{s}k_{p}+k_{g}}\nonumber \\
\frac{1}{k_{g}}\left(k_{s}-\frac{k_{s}^{2}k_{p}}{k_{s}k_{p}+k_{g}}\right) & >b_{i}\nonumber \\
b_{i} & <\frac{k_{s}}{k_{s}k_{p}+k_{g}}.\label{eq:maj(bi)}
\end{align}
which also ensures that $b_{i}<\frac{k_{s}}{k_{g}}$. Thus, once $b_{i}<\frac{k_{s}}{k_{s}k_{p}+k_{g}}$,
there exists some $b_{i2}$ such that (\ref{eq:condition bi2 full})
is satisfied. As a consequence $t_{i,k+1}-t_{i,k}>0$.

\subsection{Complementary proof elements}

\subsubsection{Differential inequation satisfied by $V$}

\label{subsec:Differential-equation-of-V}

From (\ref{eq:dot(V) - Final}), one gets

\begin{align}
\dot{V} & \leq\frac{1}{2}\sum_{i=1}^{N}\left[-k_{\text{m}}\left(s_{i}^{T}s_{i}-k_{g}g_{i}^{T}g_{i}\right)+D_{\max}^{2}+\eta\right]\label{eq:equa diff - etape1}
\end{align}
where $k_{\text{m}}=\min\left\{ k_{1},k_{p}\right\} $. Using (\ref{eq:gigi minoration}) in Appendix~\ref{sum-gigi},
one may write 
\begin{align}
\sum_{i=1}^{N}g_{i}^{T}g_{i} & \geq\sum_{i=1}^{N}k_{0}^{2}\left\Vert \varepsilon_{i}\right\Vert ^{2}+\left(2k_{0}+\frac{\alpha_{\min}k_{\min}}{k_{\max}}\right)P\left(q,t\right)\nonumber \\
 & \geq k_{2}\left(\sum_{i=1}^{N}k_{0}^{2}\left\Vert \varepsilon_{i}\right\Vert ^{2}+\frac{1}{2}P\left(q,t\right)\right)\label{eq:min(gigi)-part4}
\end{align}
where 
\[
k_{2}=\begin{cases}
2\left(2k_{0}+\frac{\alpha_{\min}k_{\min}}{k_{\max}}\right) & \text{if }\left(2k_{0}+\frac{\alpha_{\min}k_{\min}}{k_{\max}}\right)<\frac{1}{2}\\
1 & \text{else.}
\end{cases}
\]
 Then
\begin{align}
\sum_{i=1}^{N}g_{i}^{T}g_{i} & \geq k_{2}\left(\sum_{i=1}^{N}k_{0}^{2}\left\Vert \varepsilon_{i}\right\Vert ^{2}+\frac{1}{2}P\left(q,t\right)\right)\nonumber \\
 & \geq k_{3}\left(\sum_{i=1}^{N}k_{0}\left\Vert \varepsilon_{i}\right\Vert ^{2}+\frac{1}{2}P\left(q,t\right)\right)\label{eq:min(gigi)-2-part4}
\end{align}
where $k_{3}=k_{2}k_{0}$ if $k_{0}<1$, $k_{3}=1$ else. Then
\begin{align}
\dot{V} & \leq-\frac{1}{2}\sum_{i=1}^{N}\left(k_{\text{m}}s_{i}^{T}s_{i}\right)-\frac{k_{3}k_{g}}{2}\left(\sum_{i=1}^{N}k_{0}\left\Vert \varepsilon_{i}\right\Vert ^{2}+\frac{1}{2}P\left(q,t\right)\right)+\frac{N}{2}\left(D_{\max}^{2}+\eta\right)\nonumber \\
 & \leq-\frac{1}{k_{M}^{*}}\left[\frac{1}{2}\sum_{i=1}^{N}k_{\text{m}}\left(k_{\text{M}}s_{i}^{T}s_{i}\right)+\frac{k_{3}k_{g}}{2}\left(\sum_{i=1}^{N}k_{0}\left\Vert \varepsilon_{i}\right\Vert ^{2}+\frac{1}{2}P\left(q,t\right)\right)\right]+\frac{N}{2}\left(D_{\max}^{2}+\eta\right)\nonumber \\
 & \leq-\frac{k_{4}}{k_{M}^{*}}\left[\frac{1}{2}\sum_{i=1}^{N}\left(k_{\text{M}}s_{i}^{T}s_{i}\right)+\frac{k_{g}}{2}\left(\sum_{i=1}^{N}k_{0}\left\Vert \varepsilon_{i}\right\Vert ^{2}+\frac{1}{2}P\left(q,t\right)\right)\right]+\frac{N}{2}\left(D_{\max}^{2}+\eta\right)\label{eq:maj(dot(V))}
\end{align}
with $k_{M}^{*}=1$ if $k_{M}<1$ and $k_{M}^{*}=k_{M}$ else, and
$k_{4}=\min\left(k_{\text{m}},k_{3}\right)$. Introduce $c_{3}=\frac{k_{4}}{k_{M}^{*}}$, one gets
\begin{align}
\dot{V} & \leq-c_{3}\left[\frac{1}{2}\sum_{i=1}^{N}\left[s_{i}^{T}M_{i}s_{i}+\Delta\theta_{i}{}^{T}\Gamma_{i}^{-1}\Delta\theta_{i}\right]+\frac{k_{g}}{2}\left(\sum_{i=1}^{N}k_{0}\left\Vert \varepsilon_{i}\right\Vert ^{2}+\frac{1}{2}P\left(q,t\right)\right)\right]\nonumber \\
 & +\frac{N}{2}\left[D_{\max}^{2}+\eta\right]+\frac{c_{3}}{2}\sum_{i=1}^{N}\Delta\theta_{i}{}^{T}\Gamma_{i}^{-1}\Delta\theta_{i}\nonumber \\
\dot{V} & \leq-c_{3}V+\frac{N}{2}\left[D_{\max}^{2}+\eta\right]+\frac{c_{3}}{2}\sum_{i=1}^{N}\left(\Delta\theta_{i}{}^{T}\Gamma_{i}^{-1}\Delta\theta_{i}\right).\label{eq:equa diff V - final-2}
\end{align}
The evaluation of $c_{3}$ is described in Appendix~\ref{subsec:Evaluation-of-c3}.

\subsubsection{Upper-bound on $\sum_{i=1}^{N}g_{i}^{T}g_{i}$}\label{sum-gigi}

From (\ref{eq:gi}), one may write

\begin{align}
\sum_{i=1}^{N}g_{i}^{T}g_{i} & =\sum_{i=1}^{N}\left[\sum_{j=1}^{N}k_{ij}\left(r_{ij}-r_{ij}^{*}\right)+k_{0}\varepsilon_{i}\right]^{T}\left[\sum_{j=1}^{N}k_{ij}\left(r_{ij}-r_{ij}^{*}\right)+k_{0}\varepsilon_{i}\right]\nonumber \\
 & =\sum_{i=1}^{N}\left[\left\Vert \sum_{j=1}^{N}k_{ij}\left(r_{ij}-r_{ij}^{*}\right)\right\Vert ^{2}+\left\Vert k_{0}\varepsilon_{i}\right\Vert ^{2}+2 k_{0}\varepsilon_{i}^{T}\left(\sum_{j=1}^{N}k_{ij}\left(r_{ij}-r_{ij}^{*}\right)\right)\right].\label{eq:Minoration gi-0}
\end{align}

Let 
\begin{equation}
P_{1}=\sum_{i=1}^{N}\varepsilon_{i}^{T}\left(\sum_{j=1}^{N}k_{ij}\left(r_{ij}-r_{ij}^{*}\right)\right).\label{eq:P1Def}
\end{equation}
Since $r_{ij}-r_{ij}^{*}=q_{i}-q_{j}-\left(q_{i}^{*}-q_{j}^{*}\right)=\varepsilon_{i}-\varepsilon_{j}$,
\begin{align}
P_{1} & =\sum_{i=1}^{N}\sum_{j=1}^{N}k_{ij}\varepsilon_{i}^{T}\left(\varepsilon_{i}-\varepsilon_{j}\right)\nonumber \\
 & =\sum_{i=1}^{N}\sum_{j=1}^{N}k_{ij}\left(\varepsilon_{i}^{T}\varepsilon_{i}-\varepsilon_{i}^{T}\varepsilon_{j}\right).\label{eq:P1-1}
\end{align}
Using the fact that 
\begin{equation}
\left(a-b\right)^{T}\left(a-b\right) = a^{T}a+b^{T}b-2a^{T}b,\label{2ab}
\end{equation}
one gets
\begin{align}
P_{1} & =\sum_{i=1}^{N}\sum_{j=1}^{N}k_{ij}\left(\left\Vert \varepsilon_{i}\right\Vert ^{2}-\frac{1}{2}\left(\left\Vert \varepsilon_{i}\right\Vert ^{2}+\left\Vert \varepsilon_{j}\right\Vert ^{2}-\left\Vert \varepsilon_{i}-\varepsilon_{j}\right\Vert ^{2}\right)\right).\label{eq:P1-2}
\end{align}
Since $k_{ij}=k_{ji}$ and $\varepsilon_{i}-\varepsilon_{j}=r_{ij}-r_{ij}^{*}$
\begin{align}
P_{1} & =\frac{1}{2}\sum_{i=1}^{N}\sum_{j=1}^{N}k_{ij}\left\Vert \varepsilon_{i}\right\Vert ^{2}-\frac{1}{2}\sum_{i=1}^{N}\sum_{j=1}^{N}k_{ji}\left\Vert \varepsilon_{j}\right\Vert ^{2}+\frac{1}{2}\sum_{i=1}^{N}\sum_{j=1}^{N}k_{ij}\left\Vert r_{ij}-r_{ij}^{*}\right\Vert ^{2}\nonumber \\
 & =\frac{1}{2}\sum_{i=1}^{N}\sum_{j=1}^{N}k_{ij}\left\Vert \varepsilon_{i}\right\Vert ^{2}-\frac{1}{2}\sum_{i=1}^{N}\sum_{j=1}^{N}k_{ij}\left\Vert \varepsilon_{i}\right\Vert ^{2}+P\left(q,t\right)\nonumber \\
 & =P\left(q,t\right).\label{eq:P1-3}
\end{align}

Injecting $P_{1}$ in (\ref{eq:Minoration gi-0}), one gets
\begin{align}
\sum_{i=1}^{N}g_{i}^{T}g_{i} & =\sum_{i=1}^{N}\left[\left\Vert \sum_{j=1}^{N}k_{ij}\left(r_{ij}-r_{ij}^{*}\right)\right\Vert ^{2}+\left\Vert k_{0}\varepsilon_{i}\right\Vert ^{2}\right]+2k_{0}P\left(q,t\right)\label{eq:gigi avec P1}
\end{align}
and using (\ref{eq:bound of sum((sum(rij-rij*))^2)}) in Appendix~\ref{upper-bound-sum}, one gets

\begin{align}
\sum_{i=1}^{N}g_{i}^{T}g_{i} & \geq\sum_{i=1}^{N}k_{0}^{2}\left\Vert \varepsilon_{i}\right\Vert ^{2}+\left(2k_{0}+\frac{\alpha_{\min}k_{\min}}{k_{\max}}\right)P\left(q,t\right)\label{eq:gigi minoration}
\end{align}

\subsubsection{Upper-bound on $\sum_{i=1}^{N}\left\Vert \sum_{j=1}^{N}k_{ij}\left(r_{ij}-r_{ij}^{*}\right)\right\Vert ^{2}$} \label{upper-bound-sum}

One may write

\begin{align}
\sum_{i=1}^{N}\left\Vert \sum_{j=1}^{N}k_{ij}\left(r_{ij}-r_{ij}^{*}\right)\right\Vert ^{2} & =\sum_{i=1}^{N}\left(\sum_{j=1}^{N}k_{ij}\left(r_{ij}-r_{ij}^{*}\right)\right)^{T}\left(\sum_{\ell=1}^{N}k_{i\ell}\left(r_{i\ell}-r_{i\ell}^{*}\right)\right)\nonumber \\
 & =\sum_{i=1}^{N}\sum_{\ell=1}^{N}\sum_{j=1}^{N}k_{i\ell}k_{ij}\left(r_{ij}-r_{ij}^{*}\right)^{T}\left(r_{i\ell}-r_{i\ell}^{*}\right)\label{eq:Minoration gi-2}
\end{align}
Using (\ref{2ab}), one obtains
\begin{equation}
\sum_{i=1}^{N}\left\Vert \sum_{j=1}^{N}k_{ij}\left(r_{ij}-r_{ij}^{*}\right)\right\Vert ^{2}=\sum_{i=1}^{N}\left[\frac{1}{2}\sum_{\ell=1}^{N}\sum_{j=1}^{N}k_{i\ell}k_{ij}\left[\left\Vert r_{ij}-r_{ij}^{*}\right\Vert ^{2}+\left\Vert r_{i\ell}-r_{i\ell}^{*}\right\Vert ^{2}-\left\Vert r_{ij}-r_{ij}^{*}-\left(r_{i\ell}-r_{i\ell}^{*}\right)\right\Vert ^{2}\right]\right].\label{eq:Minoration gi-3}
\end{equation}
One has 
\begin{align*}
\left(r_{ij}-r_{ij}^{*}\right)-\left(r_{i\ell}-r_{i\ell}^{*}\right) & =\left(r_{ij}-r_{i\ell}\right)-\left(r_{ij}^{*}-r_{i\ell}^{*}\right)\\
 & =r_{\ell j}-r_{\ell j}^{*}
\end{align*}
Injecting this result in (\ref{eq:Minoration gi-3}) leads to
\begin{equation}
\sum_{i=1}^{N}\left\Vert \sum_{j=1}^{N}k_{ij}\left(r_{ij}-r_{ij}^{*}\right)\right\Vert ^{2}=\sum_{i=1}^{N}\left[\frac{1}{2}\sum_{\ell=1}^{N}\sum_{j=1}^{N}k_{i\ell}k_{ij}\left[\left\Vert r_{ij}-r_{ij}^{*}\right\Vert ^{2}+\left\Vert r_{i\ell}-r_{i\ell}^{*}\right\Vert ^{2}-\left\Vert r_{\ell j}-r_{\ell j}^{*}\right\Vert ^{2}\right]\right]\label{eq:norm(sum(sum(rij-rij*))}
\end{equation}
with $k_{\max}=\max_{\begin{array}{c}
\ell=1\ldots N\\
j=1\ldots N
\end{array}}\left(k_{\ell j}\right)$ 
\begin{align}
k_{\max}\sum_{i=1}^{N}\left\Vert \sum_{j=1}^{N}k_{ij}\left(r_{ij}-r_{ij}^{*}\right)\right\Vert ^{2} & \geq\sum_{i=1}^{N}\left[\frac{1}{2}\sum_{\ell=1}^{N}\sum_{j=1}^{N}k_{i\ell}k_{ij}k_{\ell j}\left[\left\Vert r_{ij}-r_{ij}^{*}\right\Vert ^{2}+\left\Vert r_{i\ell}-r_{i\ell}^{*}\right\Vert ^{2}-\left\Vert r_{\ell j}-r_{\ell j}^{*}\right\Vert ^{2}\right]\right]\nonumber \\
k_{\max}\sum_{i=1}^{N}\left\Vert \sum_{j=1}^{N}k_{ij}\left(r_{ij}-r_{ij}^{*}\right)\right\Vert ^{2} & \geq\frac{1}{2}\sum_{i=1}^{N}\sum_{\ell=1}^{N}\sum_{j=1}^{N}k_{i\ell}k_{ij}k_{\ell j}\left\Vert r_{ij}-r_{ij}^{*}\right\Vert ^{2}+\frac{1}{2}\sum_{i=1}^{N}\sum_{\ell=1}^{N}\sum_{j=1}^{N}k_{i\ell}k_{ij}k_{\ell j}\left\Vert r_{i\ell}-r_{i\ell}^{*}\right\Vert ^{2}\nonumber \\
 & -\frac{1}{2}\sum_{i=1}^{N}\sum_{\ell=1}^{N}\sum_{j=1}^{N}k_{i\ell}k_{ij}k_{\ell j}\left\Vert r_{\ell j}-r_{\ell j}^{*}\right\Vert ^{2}\nonumber \\
k_{\max}\sum_{i=1}^{N}\left\Vert \sum_{j=1}^{N}k_{ij}\left(r_{ij}-r_{ij}^{*}\right)\right\Vert ^{2} & \geq\frac{1}{2}\sum_{i=1}^{N}\sum_{\ell=1}^{N}\sum_{j=1}^{N}k_{i\ell}k_{ij}k_{\ell j}\left\Vert r_{ij}-r_{ij}^{*}\right\Vert ^{2}+\frac{1}{2}\sum_{i=1}^{N}\sum_{\ell=1}^{N}\sum_{j=1}^{N}k_{i\ell}k_{ij}k_{\ell j}\left\Vert r_{ij}-r_{ij}^{*}\right\Vert ^{2}\nonumber \\
 & -\frac{1}{2}\sum_{i=1}^{N}\sum_{\ell=1}^{N}\sum_{j=1}^{N}k_{i\ell}k_{ij}k_{\ell j}\left\Vert r_{ij}-r_{ij}^{*}\right\Vert ^{2}\nonumber \\
k_{\max}\sum_{i=1}^{N}\left\Vert \sum_{j=1}^{N}k_{ij}\left(r_{ij}-r_{ij}^{*}\right)\right\Vert ^{2} & \geq\frac{1}{2}\sum_{i=1}^{N}\sum_{\ell=1}^{N}\sum_{j=1}^{N}k_{i\ell}k_{ij}k_{\ell j}\left\Vert r_{ij}-r_{ij}^{*}\right\Vert ^{2}.\label{eq:Minoration gi-4}
\end{align}
 Let $k_{\min}=\min_{\begin{array}{c}
\ell=1\ldots N\\
j=1\ldots N
\end{array}}\left(k_{\ell j}\neq0\right)$ and $\alpha_{\min}=\min_{i=1,\dots,N}\alpha_{i}$. One may write
\begin{align}
\sum_{i=1}^{N}\sum_{\ell=1}^{N}\sum_{j=1}^{N}k_{i\ell}k_{ij}k_{\ell j}\left\Vert r_{ij}-r_{ij}^{*}\right\Vert ^{2} & =\sum_{i=1}^{N}\sum_{\ell=1}^{N}k_{i\ell}\sum_{j=1}^{N}k_{ij}k_{\ell j}\left\Vert r_{ij}-r_{ij}^{*}\right\Vert ^{2}\nonumber \\
 & \geq\sum_{i=1}^{N}\sum_{\ell=1}^{N}k_{i\ell}k_{\min}\sum_{j=1}^{N}k_{ij}\left\Vert r_{ij}-r_{ij}^{*}\right\Vert ^{2}\nonumber \\
 & \geq\sum_{i=1}^{N}\alpha_{i}k_{\min}\sum_{j=1}^{N}k_{ij}\left\Vert r_{ij}-r_{ij}^{*}\right\Vert ^{2}\nonumber \\
 & \geq\alpha_{\min}k_{\min}\sum_{i=1}^{N}\sum_{j=1}^{N}k_{ij}\left\Vert r_{ij}-r_{ij}^{*}\right\Vert ^{2}\nonumber \\
 & \geq2\alpha_{\min}k_{\min}P\left(q,t\right)\label{eq:minoration sum}
\end{align}

Injecting (\ref{eq:minoration sum}) in (\ref{eq:Minoration gi-4})
one gets
\begin{align}
k_{\max}\sum_{i=1}^{N}\left\Vert \sum_{j=1}^{N}k_{ij}\left(r_{ij}-r_{ij}^{*}\right)\right\Vert ^{2} & \geq\alpha_{\min}k_{\min}P\left(q,t\right)\nonumber \\
\sum_{i=1}^{N}\left\Vert \sum_{j=1}^{N}k_{ij}\left(r_{ij}-r_{ij}^{*}\right)\right\Vert ^{2} & \geq\frac{\alpha_{\min}k_{\min}}{k_{\max}}P\left(q,t\right).\label{eq:bound of sum((sum(rij-rij*))^2)}
\end{align}

\subsubsection{Evaluation of $c_{3}$\label{subsec:Evaluation-of-c3}}

One has

\begin{align}
c_{3} & =\frac{k_{4}}{k_{M}^{*}}\nonumber \\
 & =\frac{\min\left(k_{\text{m}},k_{3}\right)}{\max\left\{ 1,k_{M}\right\} }\nonumber \\
 & =\frac{\min\left\{ \min\left\{ k_{1},k_{p}\right\} ,\min\left\{ k_{2}k_{0},1\right\} \right\} }{\max\left\{ 1,k_{M}\right\} }\nonumber \\
 & =\frac{\min\left\{ k_{1},k_{p},1,k_{2}k_{0}\right\} }{\max\left\{ 1,k_{M}\right\} }\nonumber \\
 & =\frac{\min\left\{ k_{1},k_{p},1,k_{0}\min\left\{ 2\left(2k_{0}+\frac{\alpha_{\min}k_{\min}}{k_{\max}}\right),1\right\} \right\} }{\max\left\{ 1,k_{M}\right\} }\nonumber \\
 & =\frac{\min\left\{ k_{1},k_{p},1,k_{0},2k_{0}\left(2k_{0}+\frac{\alpha_{\min}k_{\min}}{k_{\max}}\right)\right\} }{\max\left\{ 1,k_{M}\right\} }\label{eq:c3-part4}
\end{align}
where $k_{1}=k_{s}-1-k_{p}\left(k_{M}+1\right)$, $\alpha_{\min}=\min_{i=1,\dots,N}\alpha_{i}$
, $k_{\max}=\max_{\begin{array}{c}
\ell=1\ldots N\\
j=1\ldots N
\end{array}}\left(k_{\ell j}\right)$ and $k_{\min}=\min_{\begin{array}{c}
\ell=1\ldots N\\
j=1\ldots N
\end{array}}\left(k_{\ell j}\neq0\right)$ .\bibliographystyle{plain}
\bibliography{biblio}

\end{document}